\documentclass[5p]{elsarticle}
\usepackage{url}
\usepackage{amsmath}
\usepackage[table,xcdraw]{xcolor}
\usepackage{subfig}
\usepackage{xcolor}
\usepackage[normalem]{ulem}
\usepackage{multirow}
\usepackage{tabu}
\usepackage{amssymb}
\usepackage{graphicx}
   \graphicspath{{../figures/}}
    \DeclareGraphicsExtensions{.pdf,.jpeg,.pdf,.eps}
\usepackage{enumitem} 
\setitemize{noitemsep,topsep=0pt,parsep=0pt,partopsep=0pt}  

\newcommand{\figref}[1]{Fig.~\ref{#1}}
\newcommand{\tabref}[1]{Tab.~\ref{#1}}
\newcommand{\abs}[1]{\left|#1\right|}
\newcommand*\rot{\rotatebox{90}}
\setlist[itemize]{leftmargin=*}
\journal{Signal Processing: Image Communication}

\begin{document}

\begin{frontmatter}

\title{Comprehensive Performance Analysis of Objective Quality Metrics for Digital Holography}

\author[add1,add2]{Ayyoub~Ahar\corref{correauthor}} 
\cortext[correauthor]{Corresponding author}
\ead{ayyoub.ahar@vub.be}

\author[add1,add2]{Tobias~Birnbaum}
\author[WUTadd]{Maksymilian~Chlipala} 
\author[WUTadd]{Weronika~Zaperty} 
\author[add1,add2]{Saeed~Mahmoudpour}
\author[WUTadd]{Tomasz~Kozacki}
\author[WUTadd]{Malgorzata~Kujawinska}
\author[add1,add2]{Peter~Schelkens}

\address[add1]{Dept. of Electronics and Informatics (ETRO), Vrije Universiteit Brussel (VUB),Pleinlaan 2, B-1050 Brussels}
\address[add2]{imec, Kapeldreef 75, B-3001 Leuven, Belgium}
\address[WUTadd]{Warsaw University of Technology, Institute of Micromechanics and Photonics, 8 Sw. A.~Boboli St., 02-525 Warsaw, Poland.}

\begin{abstract}
	Objective quality assessment of digital holograms has proven to be a challenging task. While prediction of perceptual quality of the recorded 3D content from the holographic wavefield is an open problem; perceptual quality assessment from content after rendering, requires a  time-consuming rendering step and a multitude of possible viewports. 
	In this research, we use 96 Fourier holograms of the recently released HoloDB database to evaluate the performance of well-known and state-of-the-art image quality metrics on digital holograms. We compare the reference holograms with their distorted versions: (i) before rendering on the real and imaginary parts of the quantized complex-wavefield, (ii) after converting Fourier to Fresnel holograms, (iii) after rendering, on the quantized amplitude of the reconstructed data, and (iv) after subsequently removing speckle noise using a Wiener filter. For every experimental track, the quality metric predictions are compared to the Mean Opinion Scores (MOS) gathered on a 2D screen, light field display and a holographic display. Additionally, a statistical analysis of the results and a discussion on the performance of the metrics are presented. The tests demonstrate that while for each test track a few quality metrics present a highly correlated performance compared to the multiple sets of available MOS, none of them demonstrates a consistently high-performance across all four test-tracks.
\end{abstract}

\begin{keyword}
Digital Holography, Fourier Holography, Holographic Display, Hologram Quality Assessment, Visual Quality Assessment, Visual Quality Metrics
\MSC[2020] 00-01\sep  99-00
\end{keyword}

\end{frontmatter}

\section{Introduction} 
\label{sec:Intro}
Predicting perceived visual quality for 3D media in general is highly desired. In this regard, objective quality evaluation of 3D content has been particularly pursued based on the type of the utilized technology to capture the depth information plus visual parallax. For example in \cite{li2019no,fan2018no,zhou2019dual,xu2019predictive} methods are introduced to evaluate the stereoscopic scenes. Alternatively, for Depth-Image-Based-Rendering techniques, which are essentially based on providing a depth value per regular 2D image pixel, methods like \cite{battisti2015objective,gu2019multiscale,ling2019quality,zhou2019no,sandic2019fast} are proposed for evaluation. For light field (LF) imaging, some recent efforts have proved to be efficient in their perceptual analysis, among them \cite{paudyal2019reduced,fang2018light,zhou2019tensor,kara2019key,shi2019no}. 

One of the emerging plenoptic modalities, which essentially can provide the most complete set of visual depth cues compared to the other 3D data modalities \cite{blinder_signal_2018}, is digital holography . Holography can present faithful reconstruction of the real object, with continuous parallax (within the recorded Field of View (FoV) and the freedom to refocus on unlimited number of focal distances --- just like watching the real 3D scene. 

However, perceptual quality prediction of the holographic content requires taking into account extra layers of complexity, on top of the general difficulties associated to quality prediction on plenoptic content \cite{blinder_signal_2018, schelkens_jpeg_2019}, such as stereoscopic or light field content. A major complexity added to the Visual Quality Assessment (VQA) of digital holograms, consists of predicting the visual quality of the captured 3D scene from the complex-valued holographic fringes, instead of analyzing several reconstructed view-ports of a scene involving an expensive rendering step. Even without considering the perceptual quality and visual appearance aspects, direct calculation of a mathematical error between a pair of complex-valued data is a challenge of its own (i.e. given the wrapping nature of phase and unboundedness of the magnitude, quantifying the maximum possible error or in general defining a bounded measure, which gives the relative error compared to the magnitude of the reference data, is not straightforward). Nonetheless, in \cite{ahar2017new,ahar2017new2}, the authors have recently proposed a framework which can potentially be utilized with this extent. 

The next major challenge for objective holographic VQA, is introduced by the lack of comprehensive perceptually-annotated holographic data sets. This comes from the fact that creating such data sets is on its own a challenging task. First of all, collecting appropriately diverse sets of holograms that take into account the different characteristics of the many types of holograms is critical. This diversity can be addressed via an increased complexity of the recorded scene, \emph{i.e.} via the number of objects available in the scene, the objects' depth, the distance from the recording sensor, the presence of spatial occlusions, and the object materials and surface structures, \emph{e.g} diffused, transparent, specular etc. Various types of hologram production are possible, \emph{e.g} computer generated hologram(CGH) or optically recorded hologram (ORH). But each type is limited by (virtual) sensor resolution, pixel pitch, reference wavelength, numerical aperture, sources of noise etc. All of the mentioned factors contribute to the attributes of the produced hologram. Some humble efforts to create publicly available holographic data sets can be found in \cite{BERNARDO2018193, blinder_open_2015}. Apart from data selection, conducting subjective experiments requires inventive but reproducible methodologies and scoring protocols for such plenoptic content. 

Yet another issue for holographic subjective experiments arises from the fact that holographic displays with acceptable visual attributes are still rare and mostly operate under laboratory conditions, which require advanced technical skills to be properly configured. This makes it difficult to directly evaluate the holograms. To address this issue, most of the related researches, have utilized other types of displays to conduct holographic subjective experiments. In an early effort, Lehtimaki et al. \cite{lehtimaki2009visual, lehtimaki2010evaluation} utilized a stereoscopic display to investigate applicable visual depth cues and appearance of a handful of optically recorded holograms. In \cite{Lehtimaki:16}, the perceptual quality of a limited set of holograms was even studied after  printing on glass plates. Later in \cite{ahar_subjective_2015}, a large 2D screen with high resolution was utilized to conduct subjective experiment on the central view of CGHs. In \cite{symeonidou_three-dimensional_2016, Symeonidou:18}, a light field display was proposed to render a set of CGHs. Nonetheless, the most comprehensive subjective experiment for holographic content to this date is provided in~\cite{Ahar2019SuitabilityAO} where a subjective experiment was conducted on a holographic display \cite{Kozacki2018color}, a light field display \cite{holografika}, and a regular 2D display \cite{eizo} using the same data set. A diverse set of both CGH and ORH was generated from single and multi-objects scenes carefully picked to represent various depths, recording distances and surface materials. The 96 holograms of this data set, called HoloDB, and the Mean Opinion Scores (MOS) gathered from each display setup have been made publicly available~\cite{HoloDB}. 

An additional complexity is the presence of speckle noise which is due to coherent light being used for the recording and displaying of holograms\cite{goodman2007speckle}. The speckle stems from the interference of multiple wavefronts, each of which is generated from a different illuminated point on the scene surfaces and is super-positioned with the reference light beam. Although, those interfering wavefronts are generated from $\emph{e.g}$ the reflection of the highly coherent reference beam, the slightly different distances of the points to the recording sensor introduces spatial incoherence. Constructive, as well as destructive interference show-up in the reconstruction as very dark and bright spots, randomly positioned which alter the appearance of the main recorded data. The visibility of the speckle noise may vary significantly from object to object depending on the smoothness level of the object surface or, in case of an in-line recording setup, based on the thickness of the recorded sample. In \cite{bianco2018strategies}, Bianco et al. provide a good review of the many proposed solutions for the suppression of speckle noise. More recently, in \cite{Birnbaum:19} and \cite{fonseca2019assessment}, state-of-the-art denoising methods have been evaluated both objectively and subjectively, using different sets of digital holograms and various objective metrics.

To the best of our knowledge, a full-fledged holographic perceptual quality metric is yet to be introduced due to the obstacles we reviewed above. Nevertheless, optimized compression techniques for digital holograms~\cite{blinder_signal_2018,el_rhammad_color_2018, peixeiro_holographic_2018, bernardo_holographic_2018, schelkens_jpeg_2019} and sophisticated numerical methods for CGH generation~\cite{Symeonidou:15,park_recent_2017,pan_review_2016, sugie_high-performance_2018, shimobaba_tomoyoshi_computer_2019,nishitsuji2017review} continue to advance. Consequently, lack of reliable perceptual quality predictors is being felt more than ever. As a first step towards designing such algorithms, we report in this paper a comprehensive analysis on the prediction performance of state-of-the-art Image Quality Metrics(IQMs), based on the subjective scores provided by the HoloDB data set. We also explore their strengths and weaknesses with regard to the studied holographic data and summarize their behaviour when used to compare the holograms before and after rendering. Additionally, we study a conventional speckle denoising filter and compare the performance of the IQMs before and after denoising. Furthermore, to ensure that the results of our study is not biased to the characteristics of Fourier holograms, we study the performance of IQMs after a lossless, numerical conversion to the more conventional Fresnel hologram type.  

The main objectives and novelties of this manuscript include:
\begin{itemize} 
    \item analysis and comparison of the accuracy of the predicted visual quality by the IQMs operating on:
    \begin{itemize}
        \item the hologram plane;
        \item numerically reconstructed holograms for different viewing angles and focal depth planes;
        \item numerically reconstructed holograms after speckle denoising;
    \end{itemize}
    \item assessing the impact of the hologram type, Fresnel or Fourier, on the prediction performance of the IQMs;
    \item definition of general guidelines for deployment of currently available IQMs on holographic content.
\end{itemize}
\smallskip
In section~\ref{sec:ExpPipeline}, we discuss the characteristics of the tested data, we explain the experimental pipeline and the conversion process of Fourier holograms to Fresnel holograms. The tested IQMs as well as their input requirements are also discussed here, along with the utilized tools for the statistical analysis of the results. The test results along with statistical analysis and discussion about the performance of quality metrics are provided in section~\ref{sec:Results}. Finally, Section~\ref{sec:conclusion} provides the concluding remarks.     
\section{Experimental pipeline and test methodology}
\label{sec:ExpPipeline}
In this section, we describe the technical details required for the test data and experimental setup, the IQMs under test and the deployed methods for statistical analysis.
\subsection{Description of test data}
\label{sec:Data}
\begin{figure}
	\centering
		   \captionsetup[subfigure]{labelformat=empty}
        \subfloat[CG-Ball]
	{\includegraphics[width=0.245\columnwidth]{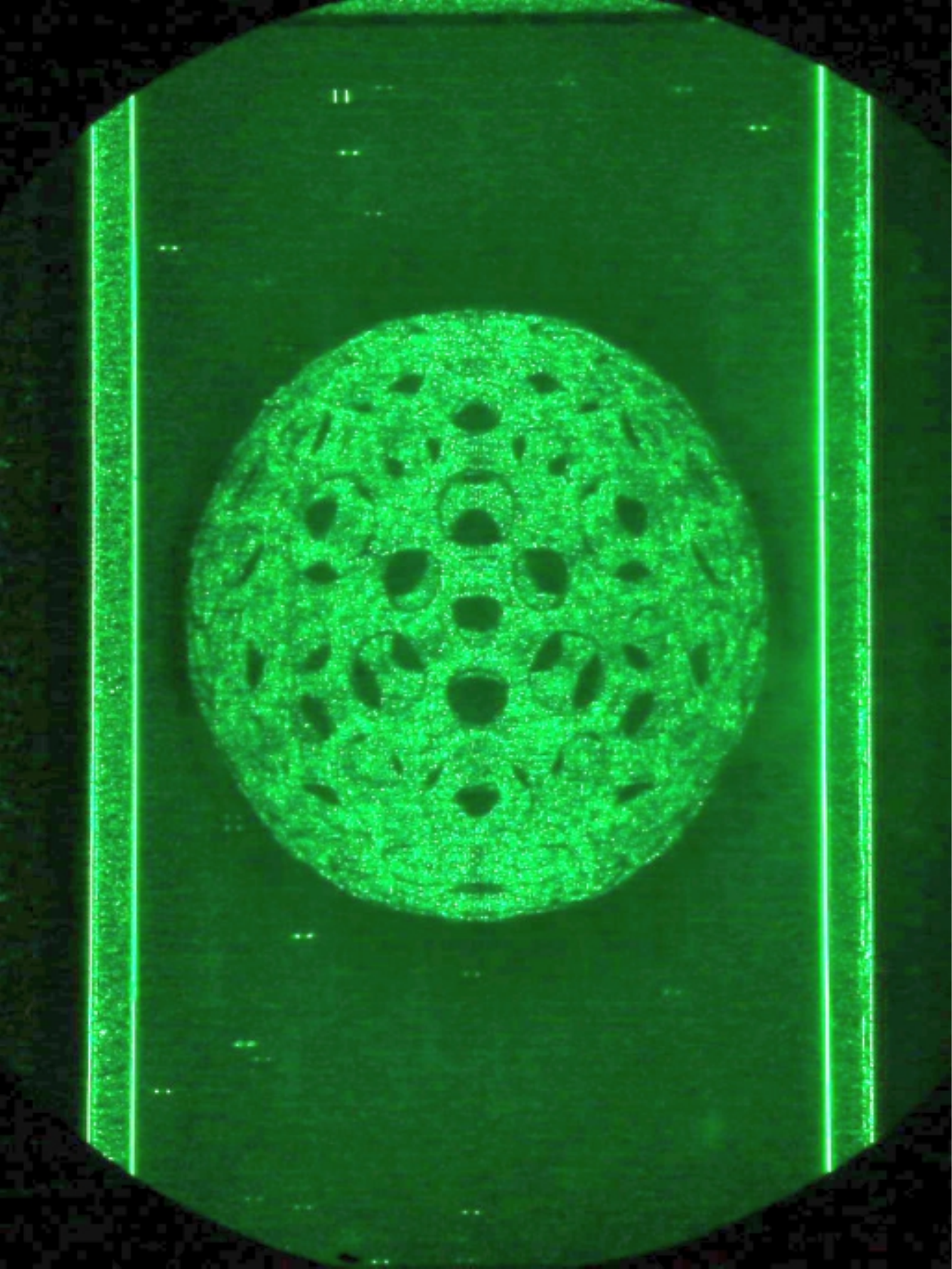}}\hfill
	    \subfloat[CG-Chess]
	{\includegraphics[width=0.245\columnwidth]{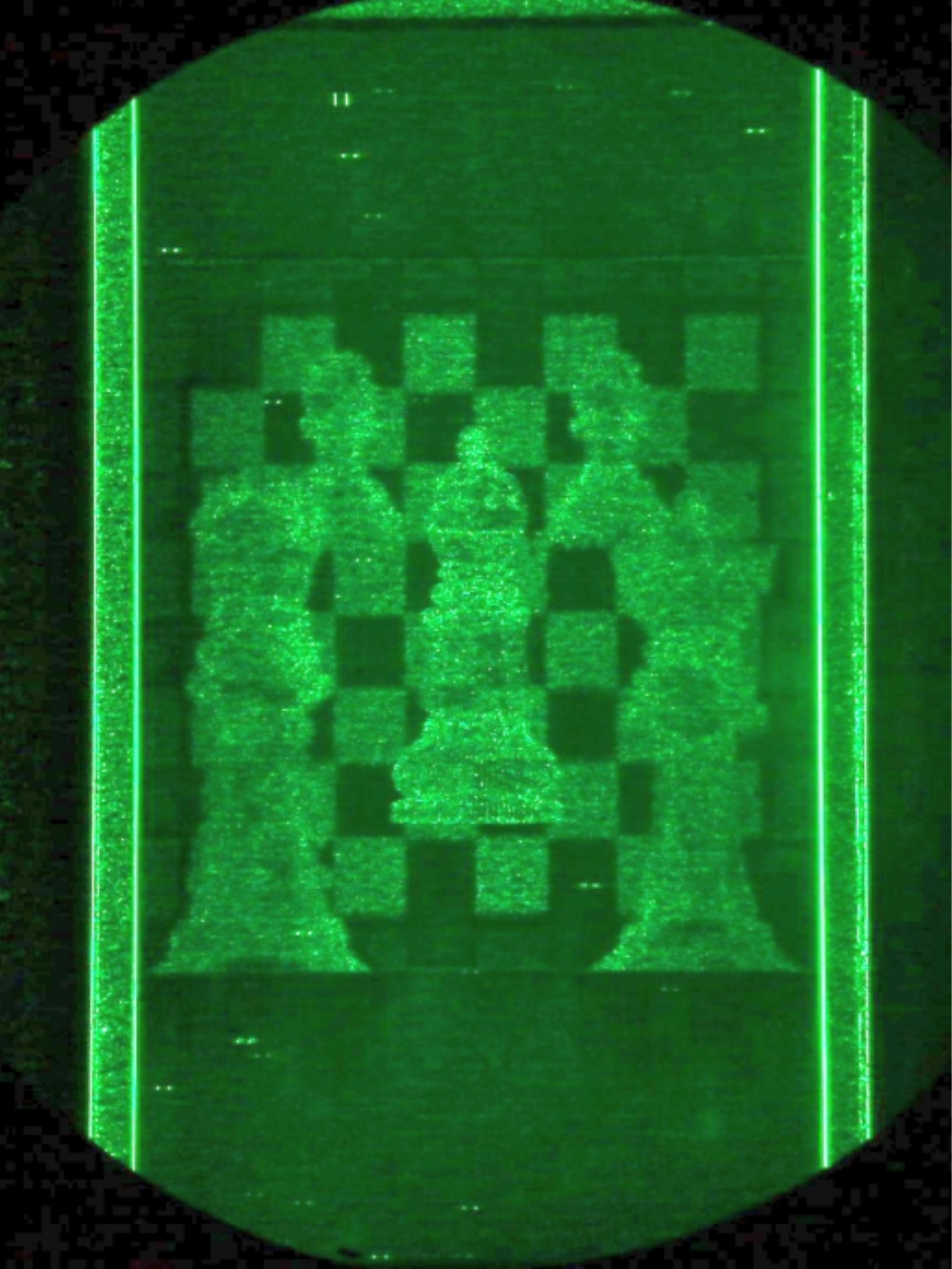}}\hfill
	    \subfloat[CG-Earth]
	{\includegraphics[width=0.245\columnwidth]{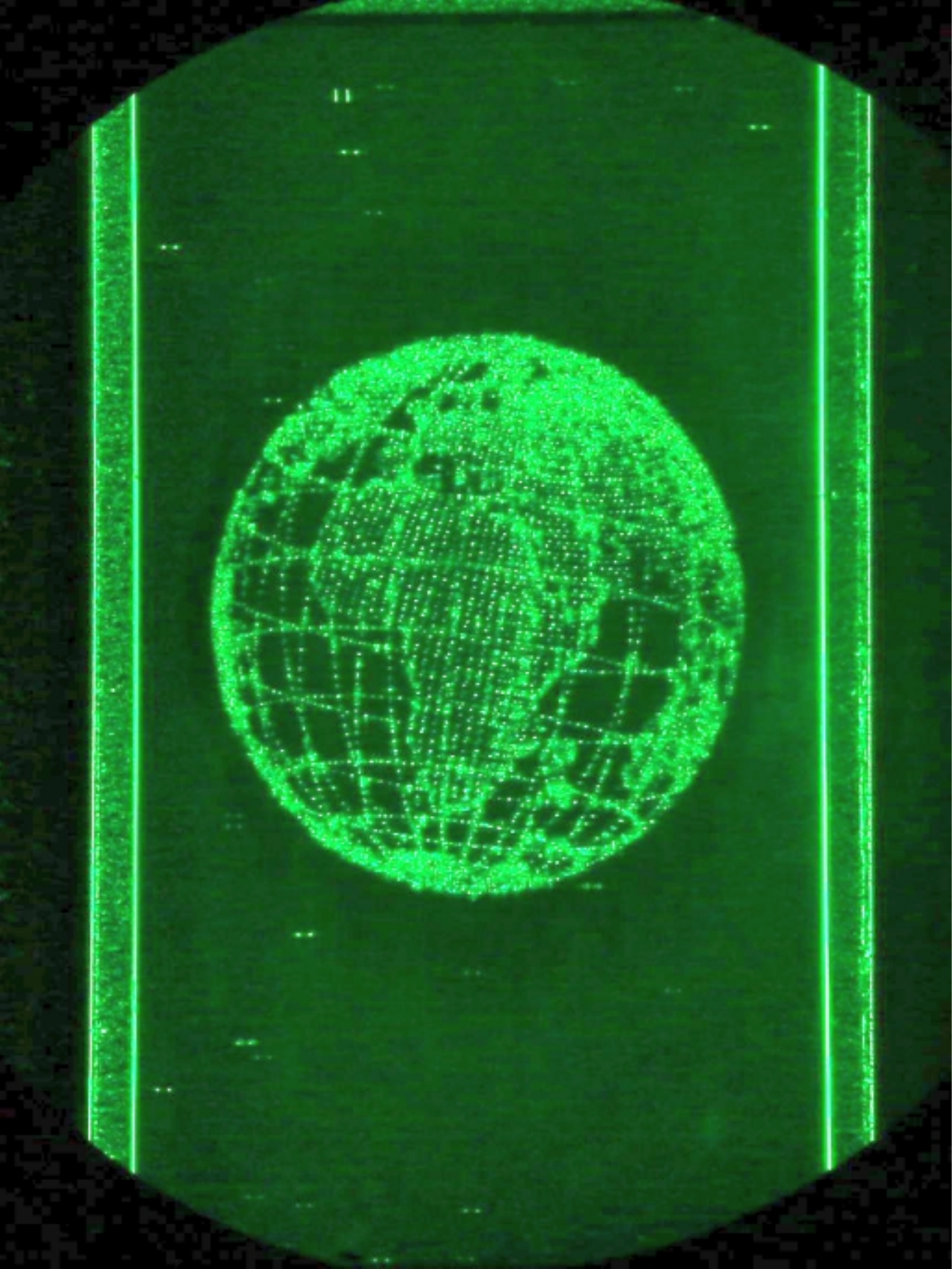}}\hfill
	    \subfloat[CG-Earth]
	{\includegraphics[width=0.245\columnwidth]{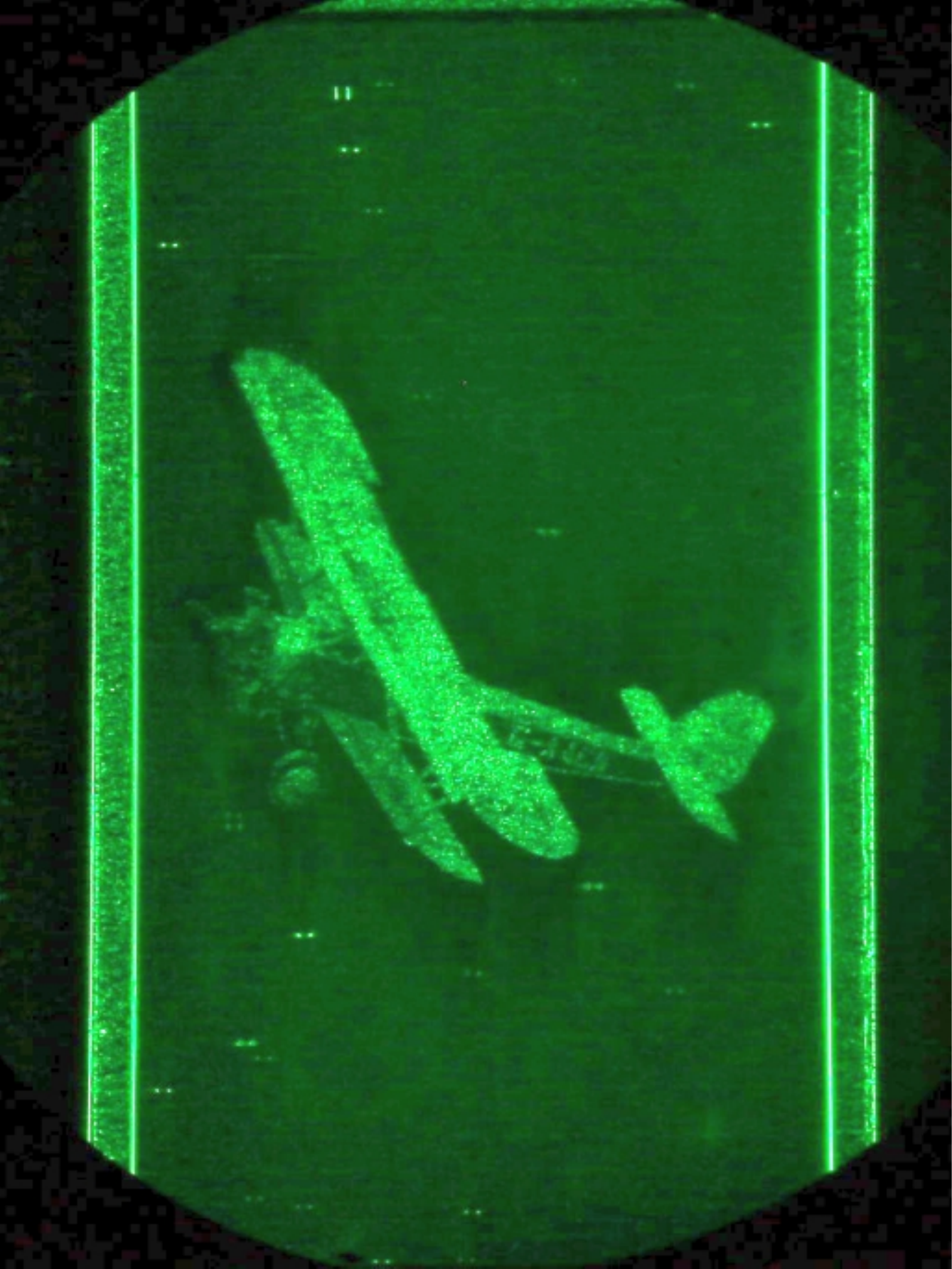}}
				
	\vspace*{-1em}
	    \subfloat[OR-Ball]
	{\includegraphics[width=0.245\columnwidth]{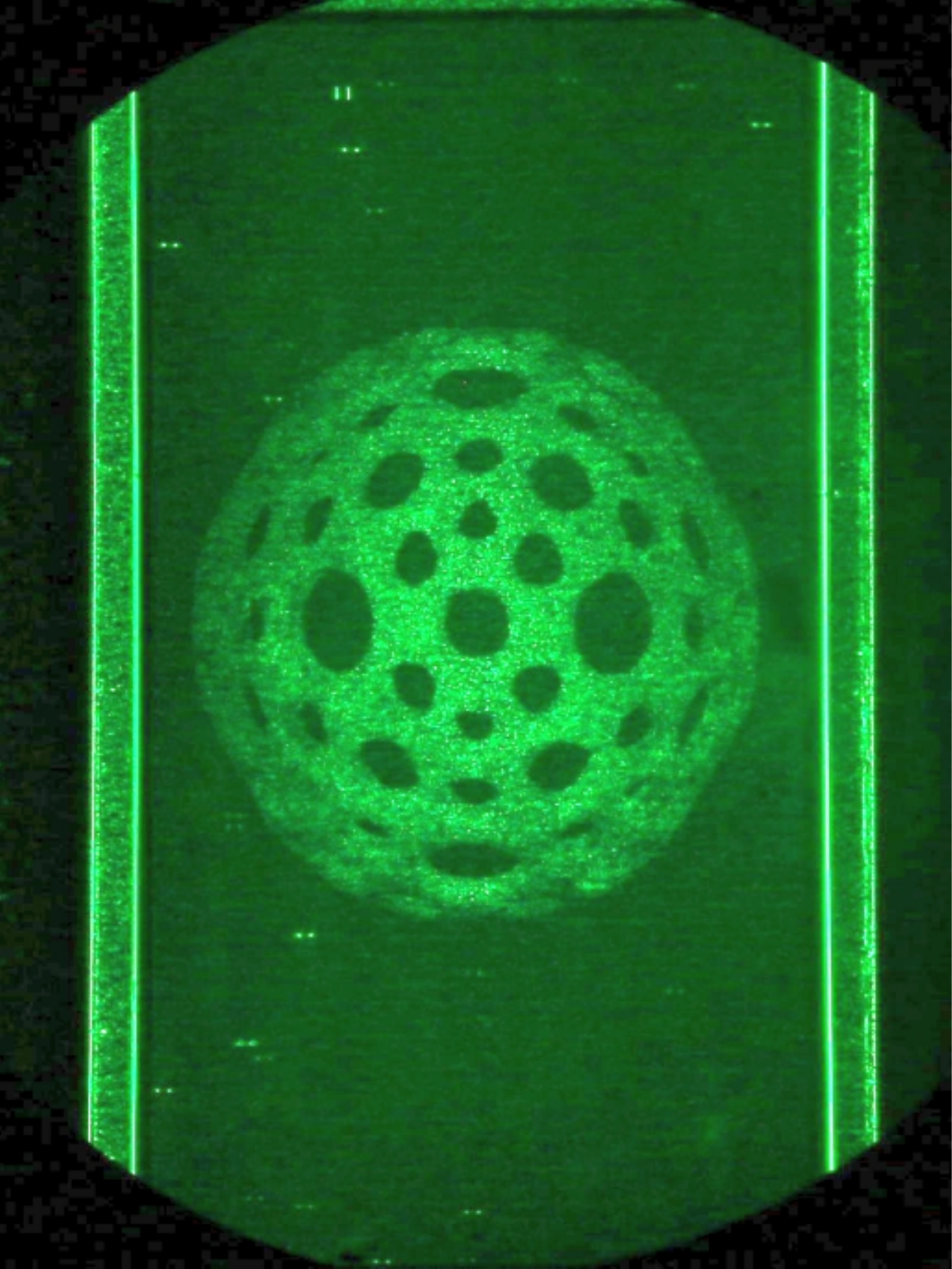}}\hfill
	    \subfloat[OR-Mermaid]
	{\includegraphics[width=0.245\columnwidth]{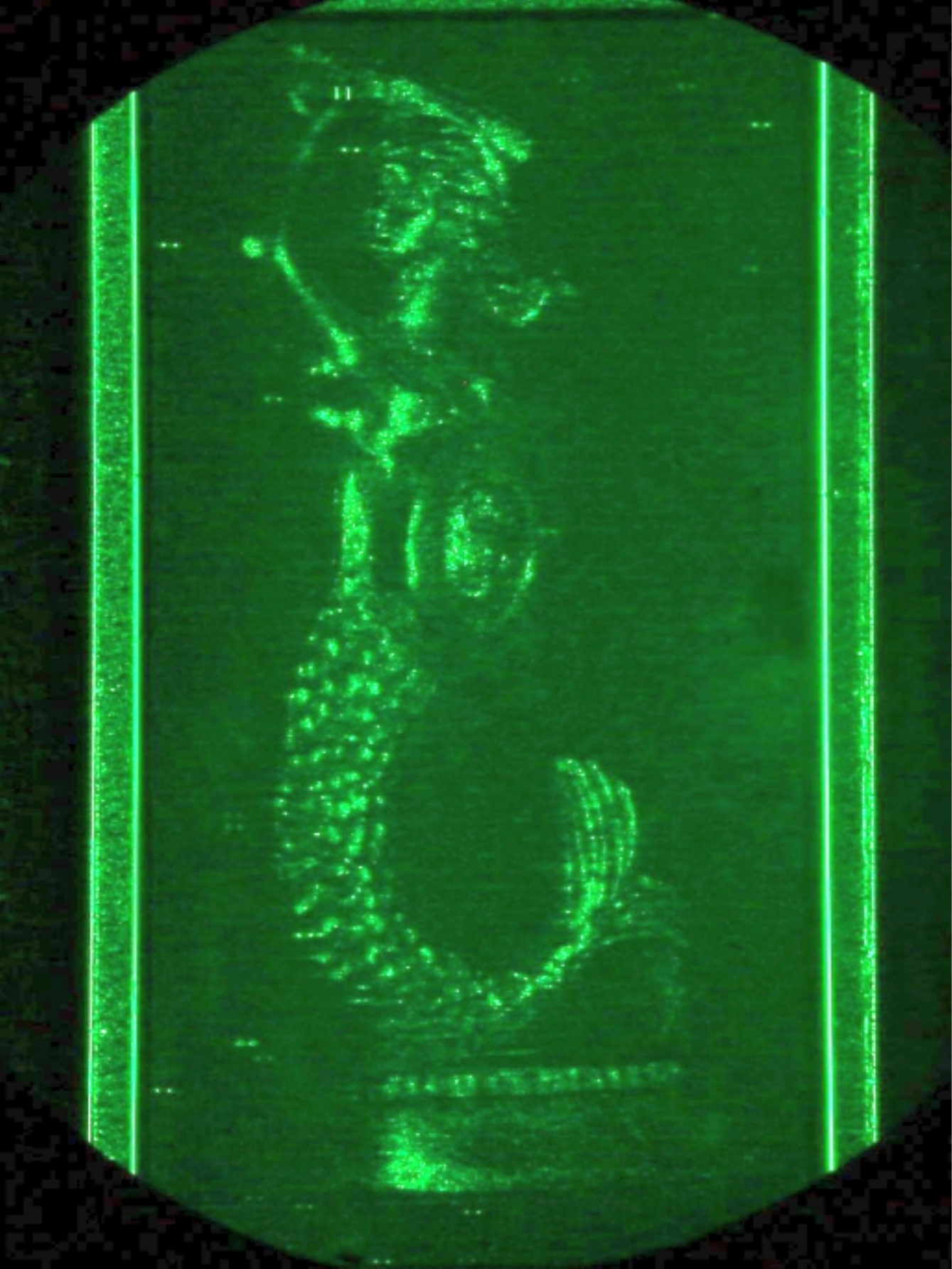}}\hfill
	    \subfloat[OR-Squirrel]
	{\includegraphics[width=0.245\columnwidth]{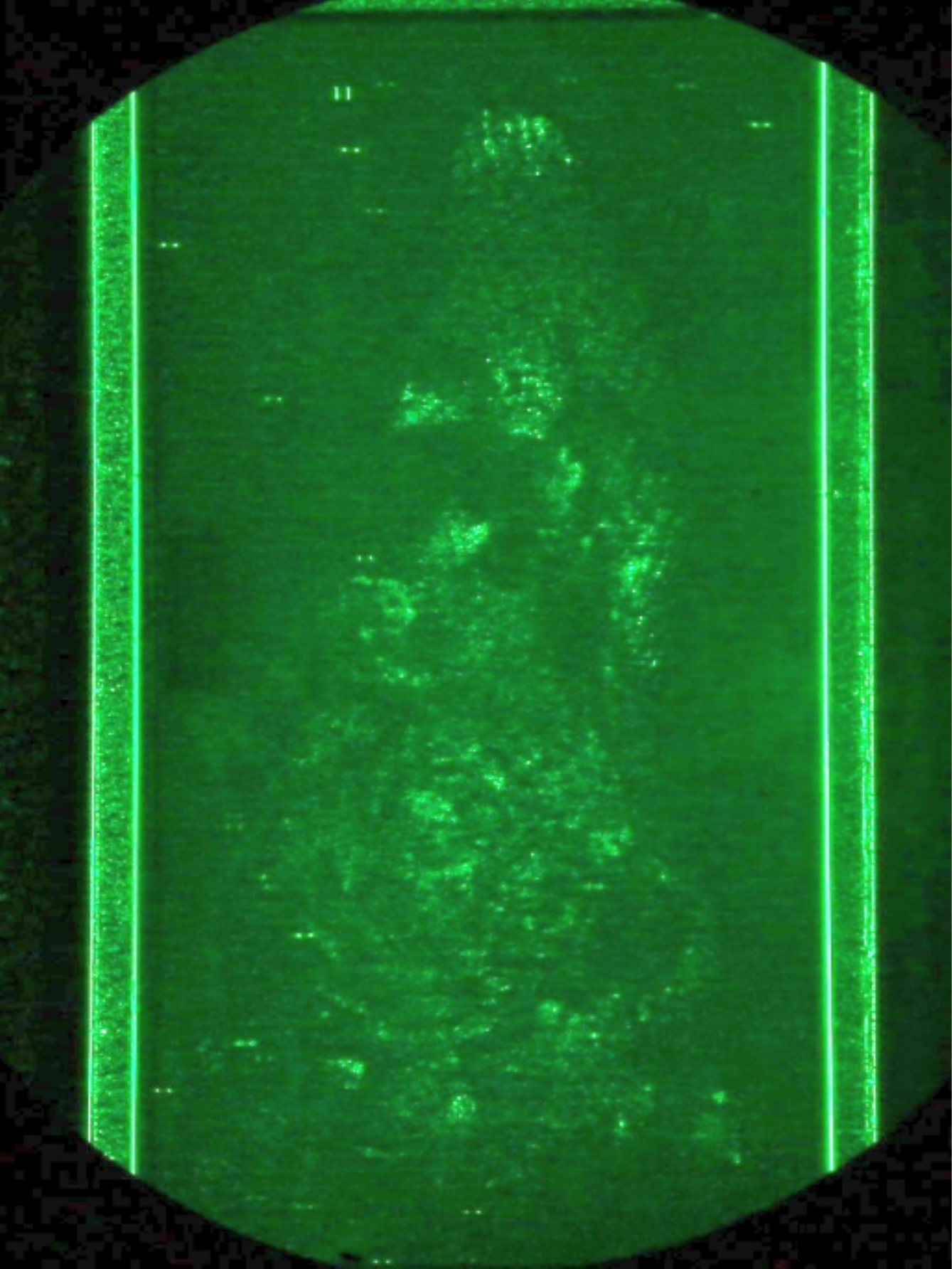}}\hfill
	    \subfloat[OR-Wolf]
	{\includegraphics[width=0.245\columnwidth]{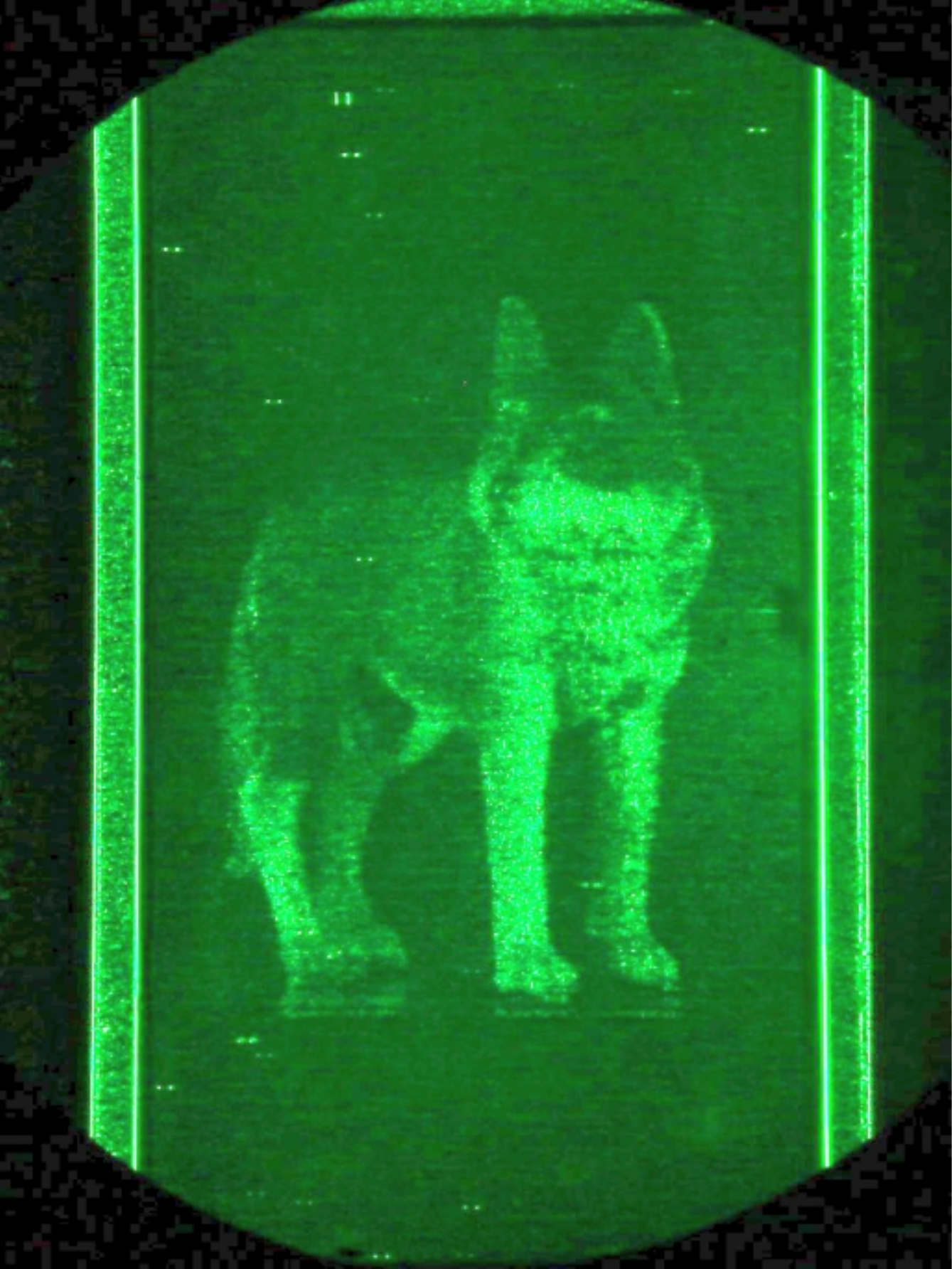}}

	\caption{Center views of the reference holograms optically reconstructed in the holographic display. The top row contains the 4 CGHs numerically generated from point-clouds and the bottom row shows the ORHs recorded from real objects.}
	\label{fig:Holos}
\end{figure}
The Fourier Holograms of the HoloDB database \cite{Ahar2019SuitabilityAO} are used for our experiments. The database consists of 8 reference holograms, 4 of which were captured from the macroscopic real objects and the other 4 were numerically generated from the point clouds. In the optical recording off-axis lensless Fourier holography was employed to obtain holograms of comparatively large objects with wide angular FoV while maximizing the used Space Bandwidth Product (SBP)~\cite{lohmann_spacebandwidth_1996,goodman_introduction_2004}. Compared to regular Fresnel holography, this method circumvent the common limitations on the sensor pixel pitch and resolution~\cite{LenslessSA,Stroke}. From the ORHs, on-axis holograms were obtained by bandpass filtering of positive frequencies components. The computer generated holograms were generated immediately as matching on-axis Fourier holograms. \figref{fig:Holos} shows the center view of the rendered reference holograms in their frontal focal distances.

\subsubsection{Compression specifications}\label{encodersdetails}
Each reference hologram later was separated into its algebraic components and quantized to 8 bits per pixel (bpp). Those quantized real and imaginary parts were encoded separately using the JPEG~2000~\cite{taubman_jpeg2000:_2002, schelkens_jpeg_2009}, intra 
H.265/HEVC~\cite{sullivan_overview_2012} and wave atom coding (WAC) ~\cite{birnbaum_wave_2019}. All three encoders compressed the reference holograms at bitrates of 0.25 bpp, 0.5 bpp, 0.75 bpp, and 1.5 bpp (bpp=bits per complex-valued pixel) creating a data set of 96 compressed holograms. 
We remark, that upon compression with JPEG~2000 aliasing artifacts were introduced in the reconstructions at the lower two bitrates. In these cases, fine scales of the $4$ level Mallat CDF $9/7$ wavelet decomposition were suppressed, leading to a downsampling of the hologram by a factor of $2$ per suppressed scale. This shrank the aliasing free cone~\cite{blinder_signal_2018} and caused the aliasing after reconstruction~\cite{Birnbaum2020_JPEG1}. The test subjects were instructed to ignore these artifacts during the subjective experiments where the MOS values were produced.

\subsubsection{Rating modalities}
Along with each hologram a total of 12 MOS scores are provided in HoloDB, accounting for renders from 2 focal distances and 2 perspectives (center view and right-corner view) which was displayed in 3 display setups (Holographic display, Light Field (LF) Display and regular 2D Monitor (2D)). The exceptions are the CG-Chess which was reconstructed in 3 focal distances (because of its deep volume of 310mm compared to recording distance of 491mm which includes multiple chess-pieces positioned in 3 rows) and the OR-Mermaid which was rendered in a single focal distance (due to its narrow thickness of only 5mm compared to the recording distance of 450mm).

\subsection{Quality assessment pipeline}
\label{sec:expSetup}
\begin{figure}
	\centering
	\includegraphics[width=0.99\columnwidth]{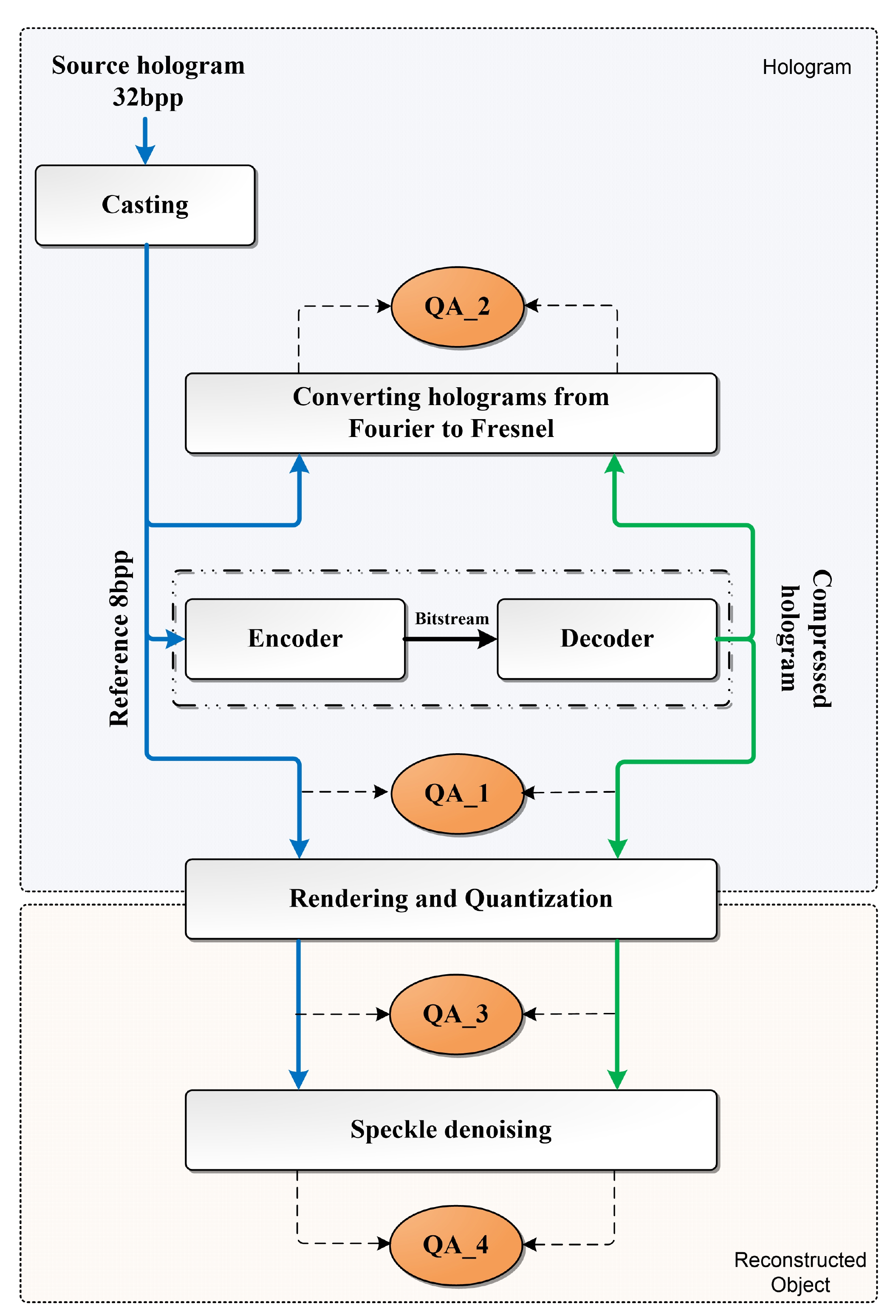} 
	\caption{A high-level scheme of the experimental pipeline showing where the predictions of the IQMs were compared with the MOS scores. The four Quality Assessment tracks in this paper are shown as QA\_1 to QA\_4 in the diagram.}
	\label{fig:pipeline}
	\vspace*{-1em}
\end{figure}

The typical procedure of visual quality assessment for a set of compressed digital images consists of simply comparing their perceptual appearance right after and before the compression step. However, in digital holography the quality assessment can be performed in more different points through the processing pipeline due to additional steps being involved for visualizing the 3D content in each hologram. Below, the implemented operations in our assessment pipeline and the assessment points are sequentially explained:\\
\begin{itemize}
    \item Quality assessment in hologram domain: 
    The real and imaginary parts of holograms are originally restored in floating point format. Though, the current implementation of the codecs like HEVC does not support such inputs and thus the algebraic components of holograms are quantized to 8bpp prior to the compression. Hereafter, the 8bpp hologram (8bpp for each of the real and imaginary channels) acts as the reference and undergoes the same operations as its compressed version. Choosing the 8bpp format is also related to the strict input requirements of some of the tested quality metrics as depicted in table~\ref{tab:IQMs}. 
    
    The reference is being encoded-decoded. The first measurement point(QA\_1) is here where the input of the encoder and the output of the decoder are given to the quality metrics to predict the visual quality of the holograms after reconstruction. These predictions are then compared w.r.t. the MOS.   
    
    To assess the impact of the hologram type on the performance of quality metrics, the compressed and reference holograms are converted to Fresnel type.The second measurement point(QA\_2) is here where similar quality evaluations as the QA\_1 are performed.
    \item Quality assessment after reconstructing the objects:
    Visualization of the holographic 3D contents, mainly requires two additional steps. First, the data values of the quantized hologram needs to be scaled back to the original data range of the source hologram. Second, each hologram should be back-propagated to reveal the captured 3D content.This process is repeated for both the reference and the compressed(decoded) holograms.
    
    After the center and corner views are rendered from both the hologram pair, those views are quantized to 8bpp again (to meet the input requirements of IQMs). Before quantization, data range clipping is performed to ensure few extreme values resulted from the speckle noise does not create an unnecessary expansion of dynamic range. Then again they are given to the quality metrics to predict their visual quality which is compared to their corresponding MOS(QA\_3).      
    
    Finally, to evaluate the impact of the speckle noise on the performance of the quality measures, an speckle denoising process is performed on the extracted views. Then in QA\_4 track, the same quality predictions and performance evaluations as the QA\_3 are conducted.
\end{itemize}
 \figref{fig:pipeline} shows an abstract scheme of the experimental pipeline and the four steps for which we conducted the quality evaluations in this paper. For the sake of clarity, we have omitted separate processing lines for the real and imaginary parts in the represented hologram domain.

\subsection{Conversion of Fourier to Fresnel holograms}
\label{subsec:Fourier2Fresnel}
Aside of the Fourier form of the considered holograms, which makes the most efficient use of the available space-bandwidth product, we also considered synthetically obtained Fresnel forms. Fourier holograms are marked by their wavefronts not being convergent and are thus under a planar reference wave reconstructed only in the conjugated plane of a lens. The wavefronts of Fresnel holograms converge under plane wave illumination without any additional lenses. Both forms are common and posses a substantially different space-frequency behaviour. Representing the same content as in a Fourier hologram $H_{Four}$ in a Fresnel hologram $H_{Fres}$ without aliasing, requires upsampling $US(\cdot)$ of $H_{Four}$, such that the effective pixel pitch is halved (sampled bandwidth doubled), and demodulated with a parabolic wavefront $K$ of mono-chromatic light of wavelength $\lambda$.
\begin{align}
    H_{Fres} = \underbrace{e^{\frac{\pi i}{\lambda z} (x^2 + y^2)}}_{=:K(z)}\cdot\text{US}(H_{Four})\label{eq:Fourier_to_Fresnel}
\end{align}
The distance $z$ in used in the kernel $K(z)$ is not the actual scene distance, as this would require upsampling by a large factor $m$. Instead, we compute $z$ such that objects are brought into focus upon direct observation as close to the hologram plane as possible without causing aliasing after upsampling. We will provide a general formula, but consider only $m=2$ as the smallest integer $m$. Only upsampling by an integer $m$ preserves the original samples and ensures invertibility. 

\begin{figure}
    \centering
    \subfloat[Fourier form\label{fig:Fourier}]{\includegraphics[width=0.49\columnwidth]{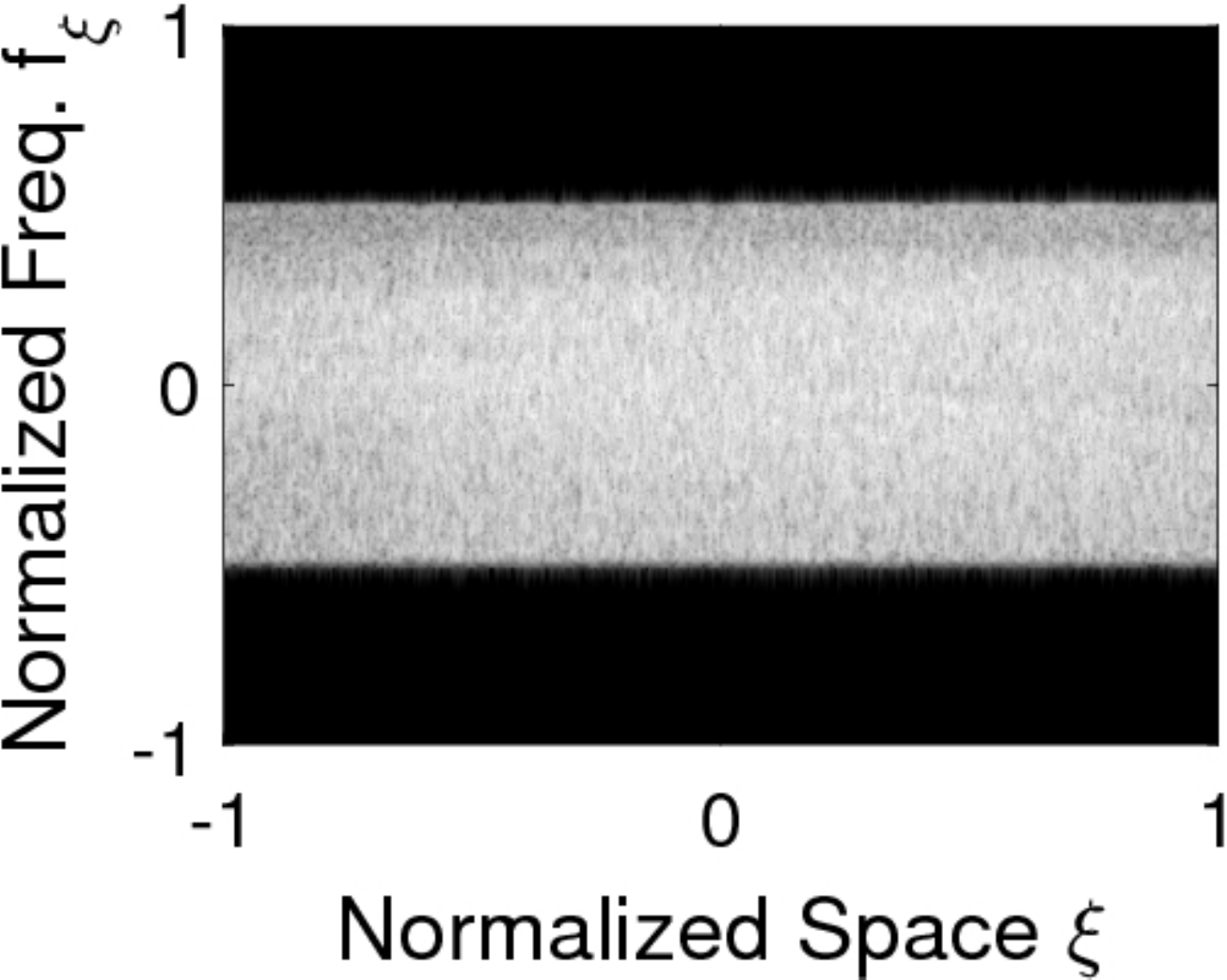}}\hspace{0.015\columnwidth}%
    \subfloat[Fourier form at half pixel pitch\label{fig:FourierSub}]{\includegraphics[width=0.49\columnwidth]{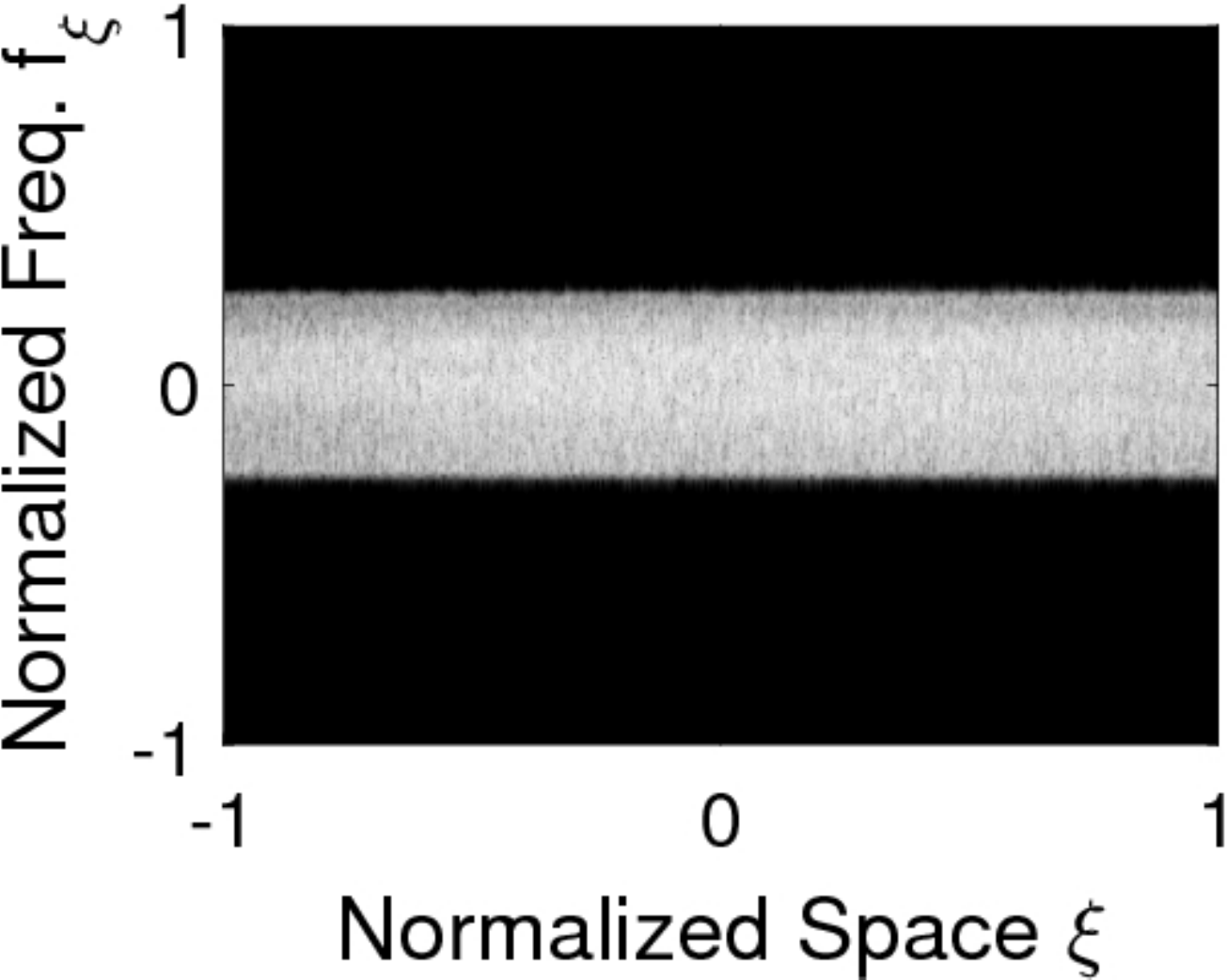}}
    
    \subfloat[Parabolic phase kernel $K$\label{fig:Kernel}]{\includegraphics[width=0.49\columnwidth]{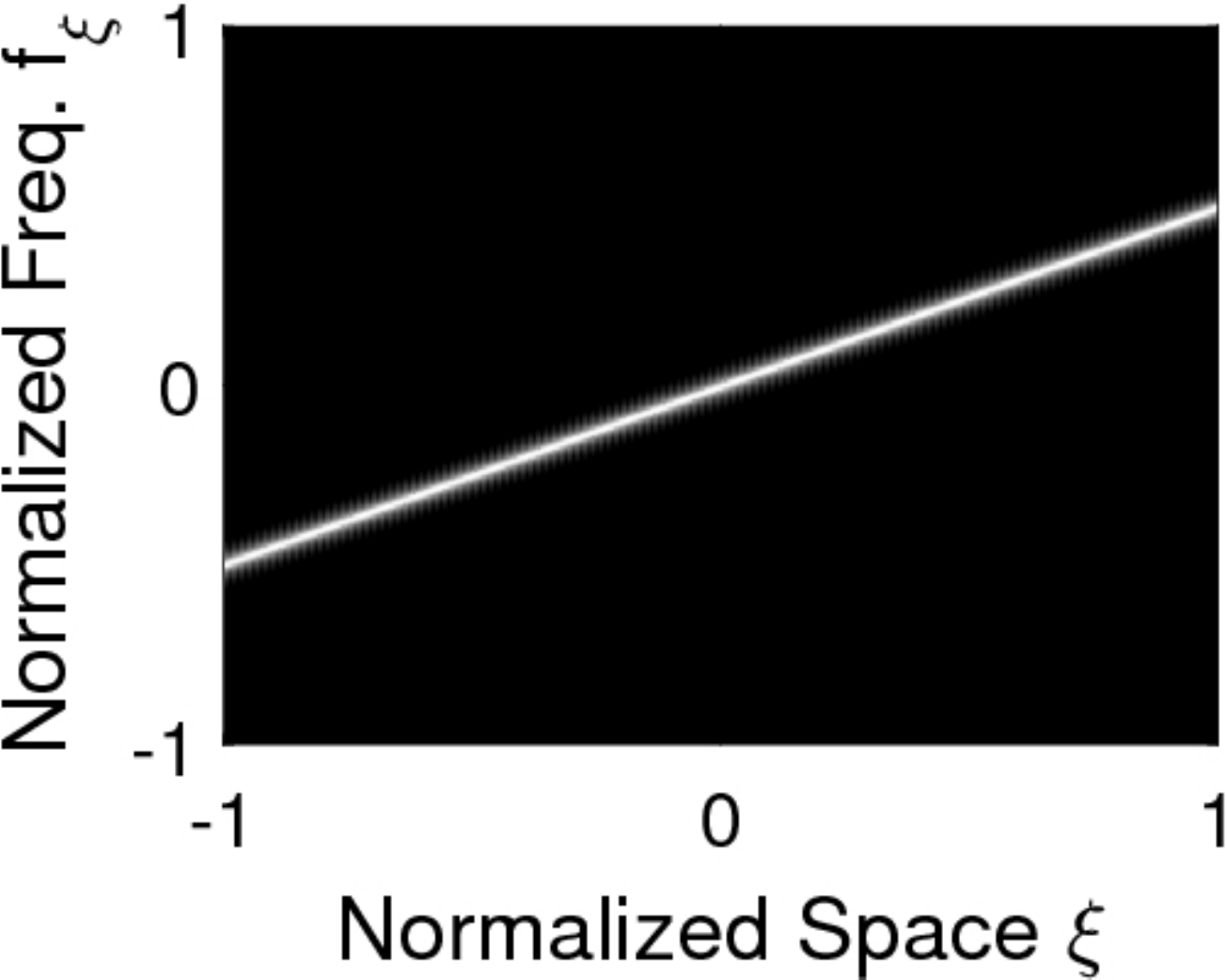}}\hspace{0.015\columnwidth}%
    \subfloat[Fresnel form\label{fig:Fresnel}]{\includegraphics[width=0.49\columnwidth]{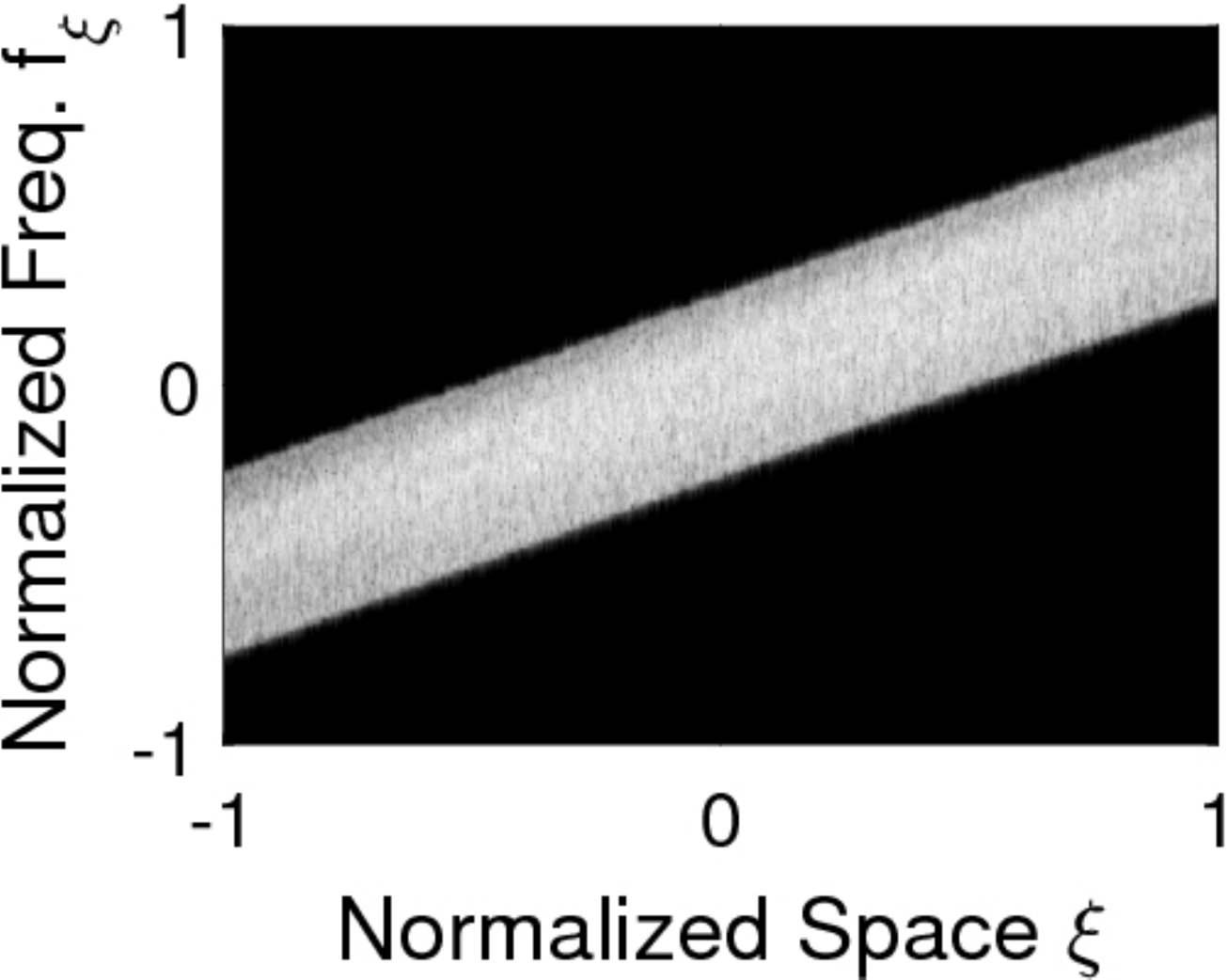}}
    \caption{Exemplary space-frequency footprint of 1D cross sections of the Fourier \protect\subref{fig:Fourier} and the Fresnel form \protect\subref{fig:Fresnel} of the OR-Squirrel hologram. The parabolic wavefront \protect\subref{fig:Kernel} focuses the subsampled Fourier hologram \protect\subref{fig:FourierSub} at a distance of $\sim$52.4cm in this case.}
    \label{fig:fourier_vs_fresnel}
\end{figure}

Given a Fourier hologram rectangular hologram recorded with a square pixel pitch $\hat{p}$ it is necessarily bandwidth-limited at $\pm(2\hat{p})^{-1}$ and it is sufficient to consider only the larger of the two hologram dimensions, i.e. $\hat{N}$~px. If the scene centre is in focus after a Fourier transform of the hologram, the main ridge of the phase space footprint will be aligned with the spatial axis, cf. \figref{fig:Fourier} and no additional 
offset of $z$ needs to be considered. Combining the ansatz, that after upsampling ($p=\hat{p}/m$, $N=m\hat{N}$) and demodulation at the edges (of the major dimension) of the Fresnel hologram $\abs{\xi}=Np/2$ the largest frequencies after $\abs{(2p)^{-1}}$ should be achieved; with the 
space-frequency law of Fresnel diffraction~\cite{Birnbaum2020TF}
\begin{align}
    f_\xi = \frac{\xi}{\lambda z}\;,
\end{align}
it is straightforward to show that $z$ is given as
\begin{align}
    z = \frac{\frac{m}{m-1}Np^2}{\lambda}\;.
\end{align}
%

Equation \eqref{eq:Fourier_to_Fresnel} is reversible without any loss and thus testing the reconstruction only from either representation is sufficient. The conversion may be understood as obtaining a Fresnel hologram from its compact space-bandwidth representation \cite{Kozacki:16}. Alternatively, the Fourier hologram can be interpreted as the interim hologram formed by a Fresnel hologram and the parabolic phase kernel $K$, which is used to facilitate numerical back-propagation of a hologram within the Fresnel approximation. The space-frequency  footprints of either form and the parabolic kernel are given in \figref{fig:fourier_vs_fresnel}. As we will see they will significantly influence the performance of the quality metrics.

\subsection{Quality metrics}
\label{sec:QMs}
For the current experiment, we use the mathematical measures which can directly evaluate the complex data, including Peak Signal-to-Noise Ratio (PSNR), Mean Squared Error (MSE) and its normalized form (NMSE), where the MSE is normalized by the Frobenius norm of the reference hologram. We also test our recently proposed IQM called Sparsness Significance Ranking Measure (SSRM) \cite{ahar2017sparse} which is native on the complex domain and showed a good compatibility for the quality evaluation of a limited set of CGHs\cite{ahar2018performance}. In its original form, SSRM operates solely in Fourier domain and predicts the similarity by comparing Fourier coefficients of reference and the impaired data. It calculates a separate quality score for the DC term which later is combined with the score of other coefficients. However, in some preliminary tests, we found out that in case of Fourier holograms, this may not necessarily result in a more robust performance. Consequently, we considered a version of this method which treats the DC term just like other Fourier coefficients. The results for testing this version are presented under the name of SSRMt. Since the compression of holograms was performed separately on the real and imaginary parts of the holograms, we can also separately evaluate those real and imaginary parts with the state-of-the-art IQMs ---which by default can operate on the real-valued data--- and then calculate the arithmetic mean to achieve one quality prediction score for each compressed hologram. The goal is to check whether they can relate their signal fidelity measurements on holograms, to the overall perceptual quality of the rendered scenes (represented by the provided MOS). The tested IQMs include: FSIM\cite{zhang2011fsim}, IWSSIM\cite{wang2010information}, MS-SSIM\cite{wang2003multiscale}, VIF\cite{sheikh2006image}, NLP\cite{laparra2016perceptual}, GMSD\cite{xue2013gradient}. The UQI\cite{wang2002universal} and SSIM\cite{wang2004image} have been utilized in the context of holographic data \emph{e.g} in \cite{wu2019bright,PANG2019105590} and \cite{ahar2015subjective}, consequently we added them also for the experiments.It should be noted that the machine learning based IQMs along with the ones which require the information from the chrominance channels of the color images were omitted from this evaluation. The former group  obviously require large training sets to be adapted for the case of holography while our holographic data set is too small for such purposes. The later group also require information from the color channels while our holograms are all monochromatic. Table~\ref{tab:IQMs}, summarizes the input requirements of the tested quality metrics and provides their parameter settings as it was used in this experiment. It also summarizes their main features and utilized underlying principle. For the experiments in QA\_1 and QA\_2 tracks where the input data are the complex-valued holograms, the real and imaginary parts of holograms were separately tested by all of the studied 13 methods. Although, for the methods 1 to 5 in the table~\ref{tab:IQMs} which can directly be deployed for the complex valued data, we tested them on the complex values. Their results are hereafter distinguished by the suffix ``\_C ''. Note that for the MSE, NMSE and the PSNR, calculation results averaged over real and imaginary parts in QA\_1 and QA\_2, are not reported due to being almost identical to those directly calculated on complex values.

\begin{table}
\caption{Information of the compared quality prediction algorithms. If necessary, the required dynamic range (DR) of input is identified.}
\resizebox{\columnwidth}{!}{
\begin{tabular}{|l|l|l|l|l|}
\hline
 & \textbf{IQM} & \textbf{\begin{tabular}[c]{@{}l@{}}Input \\Requirements\end{tabular}} & \textbf{Parameter settings} & \textbf{Underlying Principle} \\ \hline
1 & MSE & Real/Complex & N\textbackslash{}A & \begin{tabular}[c]{@{}l@{}}Pixel-based squared \\ difference measurement \\ with average pooling\end{tabular} \\ \hline
2 & NMSE & Real/Complex & \begin{tabular}[c]{@{}l@{}}Normalized by Frobenius\\ norm of reference data\end{tabular} & \begin{tabular}[c]{@{}l@{}}Normalized pixel-based\\ squared difference \\ measurement\end{tabular} \\ \hline
3 & PSNR & Real/Complex & \begin{tabular}[c]{@{}l@{}}Real Peak = 255,    \\ Complex Peak = 1\end{tabular} & \begin{tabular}[c]{@{}l@{}}Ratio between peak signal energy \\and pixel-based squared\\ difference measurement \\ on logarithmic scale\end{tabular} \\ \hline
4 & SSRM & Real/Complex & N\textbackslash{}A & \begin{tabular}[c]{@{}l@{}}Expresses correlation between \\ ranked amplitudes in frequency\\ domain based on their \\ sparseness significance\end{tabular} \\ \hline
5 & SSRMt & Real/Complex & N\textbackslash{}A & \begin{tabular}[c]{@{}l@{}}Adaptation of the SSRM where\\ the DC term is not separately\\ evaluated but is scored \\along with other coefficients\end{tabular} \\ \hline
6 & SSIM & Real & \begin{tabular}[c]{@{}l@{}}Dynamic Range(L)= 255\\      Exponents = {[}1,1,1{]} \\Regularization Constants: \\      C1 = (0.01*L)\textasciicircum{}2\\      C2 = (0.03*L)\textasciicircum{}2\\      C3 = C2/2\end{tabular} & \begin{tabular}[c]{@{}l@{}}Regularized weighted \\ comparison of local luminance\\ mean, variance (contrast) \\ and covariance (structure)\end{tabular} \\ \hline
7 & IWSSIM & Real & \begin{tabular}[c]{@{}l@{}}Dynamic Range(L)= 255\\     Exponents = {[}1,1,1{]}\\     Regularization Constants: \\     C1 = (0.01*L)\textasciicircum{}2\\     C2 = (0.03*L)\textasciicircum{}2\\     C3 = C2/2\end{tabular} & \begin{tabular}[c]{@{}l@{}}Calculates a scale variant weighted \\ SSIM-based local similarity \\ over each scale of a\\ Laplacian pyramid decomposition. \\ The scale weights are calculated \\ based on Gaussian scale mixture \\ model of natural images\end{tabular} \\ \hline
8 & MS-SSIM & Real & \begin{tabular}[c]{@{}l@{}}Dynamic Range(L)= 255\\      Regularization Constants: \\      C1 = (0.01*L)\textasciicircum{}2\\      C2 = (0.03*L)\textasciicircum{}2\\      C3 = C2/2\\      Number of scales = 5\end{tabular} & \begin{tabular}[c]{@{}l@{}}Decomposes the data to several\\ resolutions scales by iterative\\ low-pass filtering and downsampling \\ and compares the contrast and \\ similarity at each level based \\ on original SSIM. The last stage \\ involves luminance comparisoin.\\ A weighted multiplication yield \\ the final score.\end{tabular} \\ \hline
9 & UQI & Real & Block size = 8 & \begin{tabular}[c]{@{}l@{}}Special case of SSIM\\ without weight and\\ reqularization\end{tabular} \\ \hline
10 & GMSD & Real (DR 8:8bit) & \begin{tabular}[c]{@{}l@{}}Regularization Constant: \\      T = 170\end{tabular} & \begin{tabular}[c]{@{}l@{}}Calculates the standard deviation\\ of a similarity map calculated \\ based on pixel-wise \\ gradient magnitude comparision\end{tabular} \\ \hline
11 & FSIM & Real (DR 8:8bit) & \begin{tabular}[c]{@{}l@{}}Regularization Constants: \\      T1 = 0.85\\      T2 = 160\\      T3 = 200\end{tabular} & \begin{tabular}[c]{@{}l@{}}Weighted similarity, calculated\\ on image gradient magnitudes\\ and phase congruency\end{tabular} \\ \hline
12 & NLPD & Real & \begin{tabular}[c]{@{}l@{}}Laplacian Pyramid \\ \\ levels =  5\end{tabular} & \begin{tabular}[c]{@{}l@{}}RMSE of weighted \\ Laplacian pyramid \\ decomposition\end{tabular} \\ \hline
13 & VIFp & Real (DR 8:8bit) & N\textbackslash{}A & \begin{tabular}[c]{@{}l@{}}Mutual information is calculated\\  after modeling the image source \\ using Gaussian scale mixture \\ model on wavelet coefficients \\ extracted from steerable pyramid \\ decomposition and modeling the\\  distortion and HVS channels\end{tabular} \\ \hline
\end{tabular}}
\label{tab:IQMs}%
\end{table}
\subsection{Statistical analysis}
\label{sec:StatAnalysis}
In this experiment, we follow the guidelines of the Video Quality Experts Group(VQEG)~\cite{video2003final} for evaluating the predictive performance of the tested IQMs. In this regard, three evaluation criteria are considered namely: prediction monotonicity, accuracy, and consistency. 

The Spearman Rank-order Correlation Coefficient \linebreak(SROCC) and Kendall's tau Rank-Order Correlation \linebreak(KROC) are utilized to measure the strength of the IQMs in predicting the rank-ordering of the MOS. Next, we utilize the Pearson Correlation Coefficient (PCC) to measure the linear correlation between the MOS and the predicted scores. 

The score scale of several IQMs are not the same as the MOS and their score functions have a non-linear behaviour. To compensate, it is recommended in~\cite{video2003final} to fit a logistic function on the predicted scores with the constraint of being monotonic in the fitted interval. Afterwards, measuring the PCC and the Root Mean Squared Error(RMSE) between the MOS and the fitted scores will provide an estimation of prediction accuracy for the tested IQMs. In this experiment, we utilized a 4-parameter logistic function, as it is recently utilized in~\cite{bosse2018neural}, which automatically guaranties the monotone behaviour of the fit function. Hereafter, the PCC measured before and after the logistic regression is referred as PCC\_NoFit and PCC\_Fitted respectively. The recommended measure of prediction consistency is the outlier ratio. It is simply calculated as the ratio of the number of fitted predictions out of the $95\%$ confidence intervals of the MOS divided to the total number of MOS. 

Additionally, we perform a statistical significance test to determine whether the difference between performance of two quality metrics(here represented by the absolute difference between the fitted scores and the MOS) is statistically significant. A two-sided t-test is performed~\cite{hanhart2013benchmarking} where the null hypothesis is that there is no difference between performance of quality metrics, against the alternative hypothesis that the difference in performance is significant. The null hypothesis is rejected at a $5\%$ significance level. 

\section{Experimental results and analysis}
\label{sec:Results}

\subsection{Evaluation of quality metrics in hologram plane}
\label{sec:PreReconBenchmark}
In this section, we discuss the outcome of the first (QA\_1) test track referring to the evaluation of the performance of the studied IQAs in case of compressing Fourier holograms in the hologram plane, and a second (QA\_2) test track performing this test after converting the reference and decoded Fourier holograms to Fresnel holograms.
Following the procedure depicted in Fig.\ref{fig:pipeline}, for every complex-valued hologram, the same synthetic aperture used in \cite{Ahar2019SuitabilityAO} to extract the centre and right-corner views to display and obtain the MOS, was utilized to select exactly the same data from the original hologram and the corresponding compressed versions. Note, that the compression process was done on the real and imaginary parts of the complete holograms and then the synthetic apertures were applied, not vice versa. Thereafter, the algebraic components of those cropped compressed holograms were compared to their non-compressed unsigned 8bit version, \emph{i.e.} a cropped part of the input data to the encoder was compared to the same cropped part of the output of the decoder. The obtained results are compared to the MOS for the corresponding perspective averaged over the focal distances. In order to adhere to the page limits, only the overall statistics across all perspectives are presented, not the per view evaluations.

\subsubsection{QA\_1 - Evaluation on Fourier Holograms }
\label{sec:PreRecon8bit}
The overall results of our statistical analysis for this experimental track is presented in \tabref{tab:CorrsPreReconFourier}. The results are shown based on each set of MOS obtained from conducting the subjective test for holographic display (MOS\_OPT), the light field display (MOS\_LF) and the regular 2D display (MOS\_2D). 

A quick look at the table makes it evident that SSIM ranks best based on almost all evaluation criteria, independent of which MOS it was compared to. 
Seen the fact that the majority of the tested IQMs previously shown to be superior to the SSIM at least for the case of classical digital images, such a good performance of SSIM here was not expected. In digital imaging, SSIM is designed to measure the visual similarities between the structural information of the compared pair of data. However, when measured in hologram domain in particular, there is no specific structure in the data and the holograms in general mostly appear as noise. So, how SSIM is able to make such an accurate prediction by both correct ranking of the distorted holograms(high SROCC) and linear correlation of the predicted scores compared to the MOS (high PCC)? 

\begin{figure*}
\centering
	    \subfloat[HEVC 0.75~bpp]
	{\includegraphics[width=0.225\textwidth]{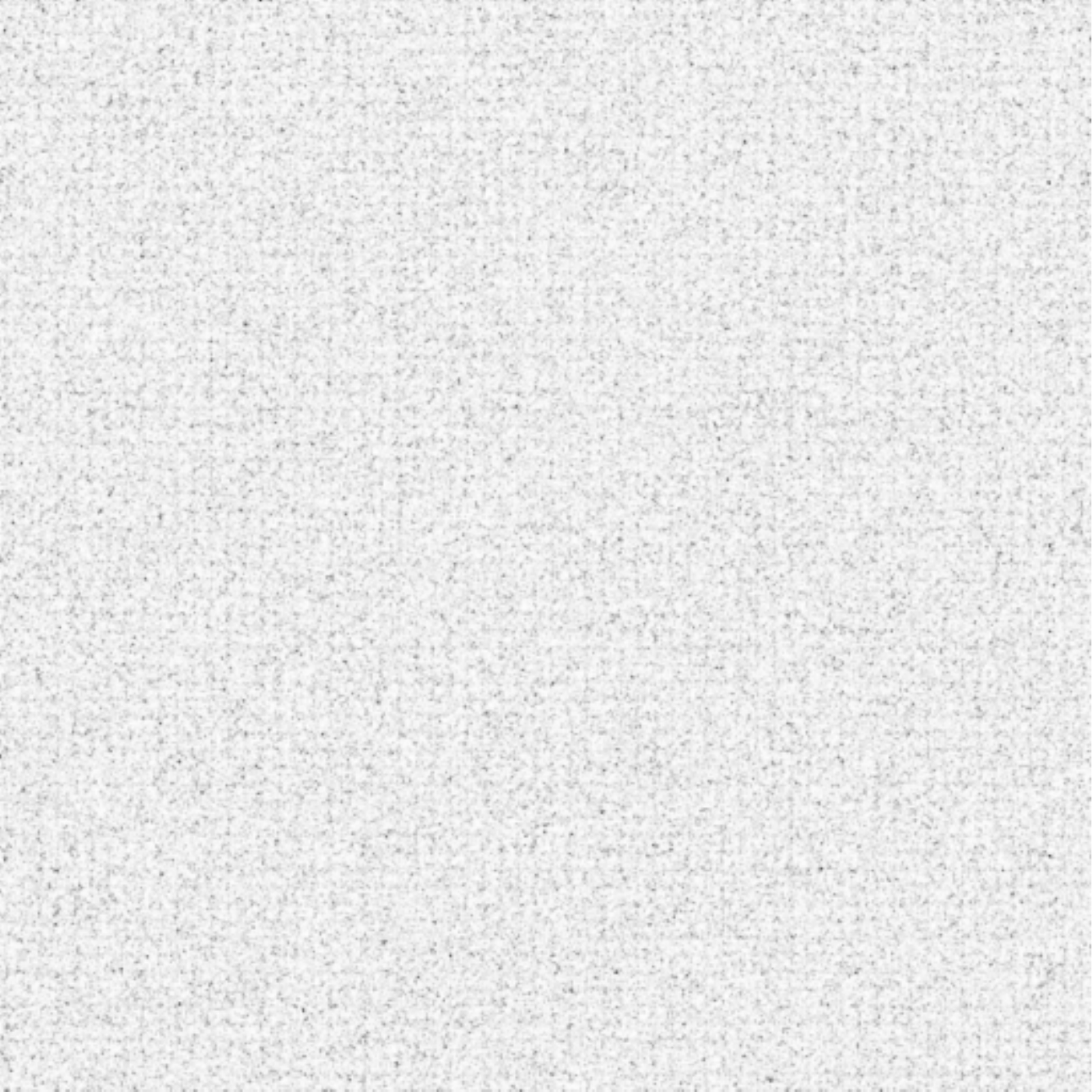}}\hspace*{0.025em}
	    \subfloat[HEVC 0.375~bpp]
	{\includegraphics[width=0.225\textwidth]{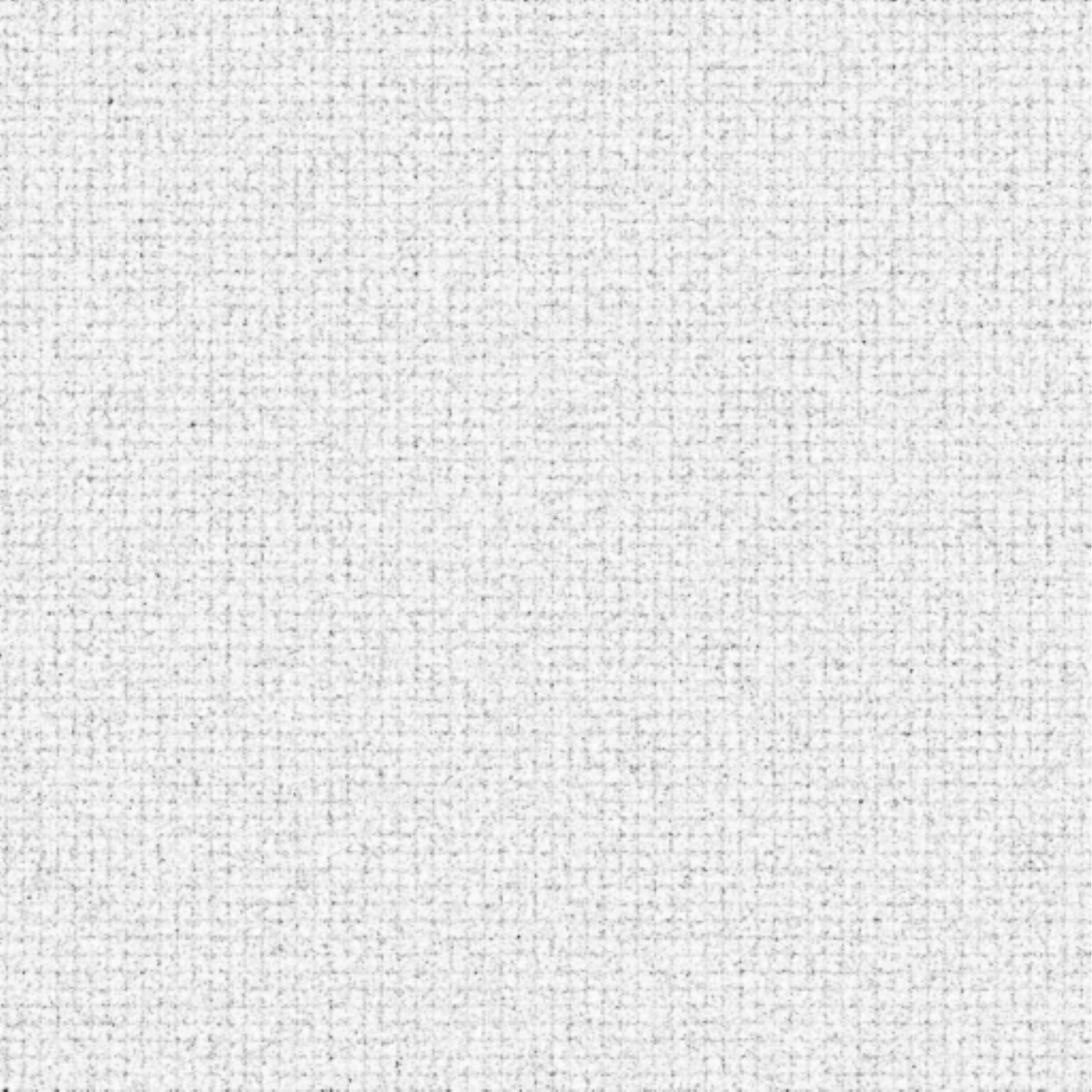}}\hspace*{0.025em}
	    \subfloat[HEVC 0.25~bpp]
	{\includegraphics[width=0.225\textwidth]{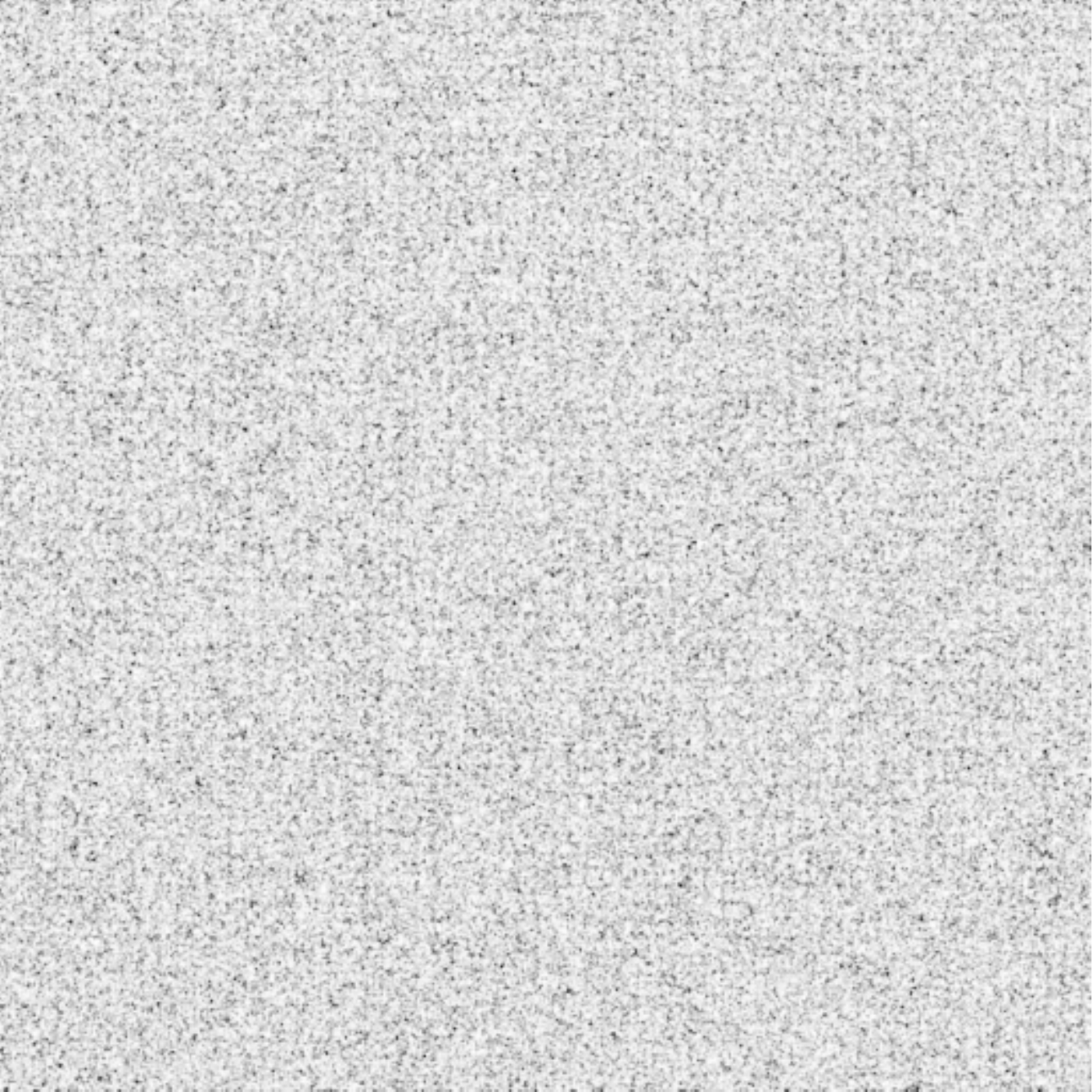}}\hspace*{0.025em}
	    \subfloat[HEVC 0.125~bpp]
	{\includegraphics[width=0.225\textwidth]{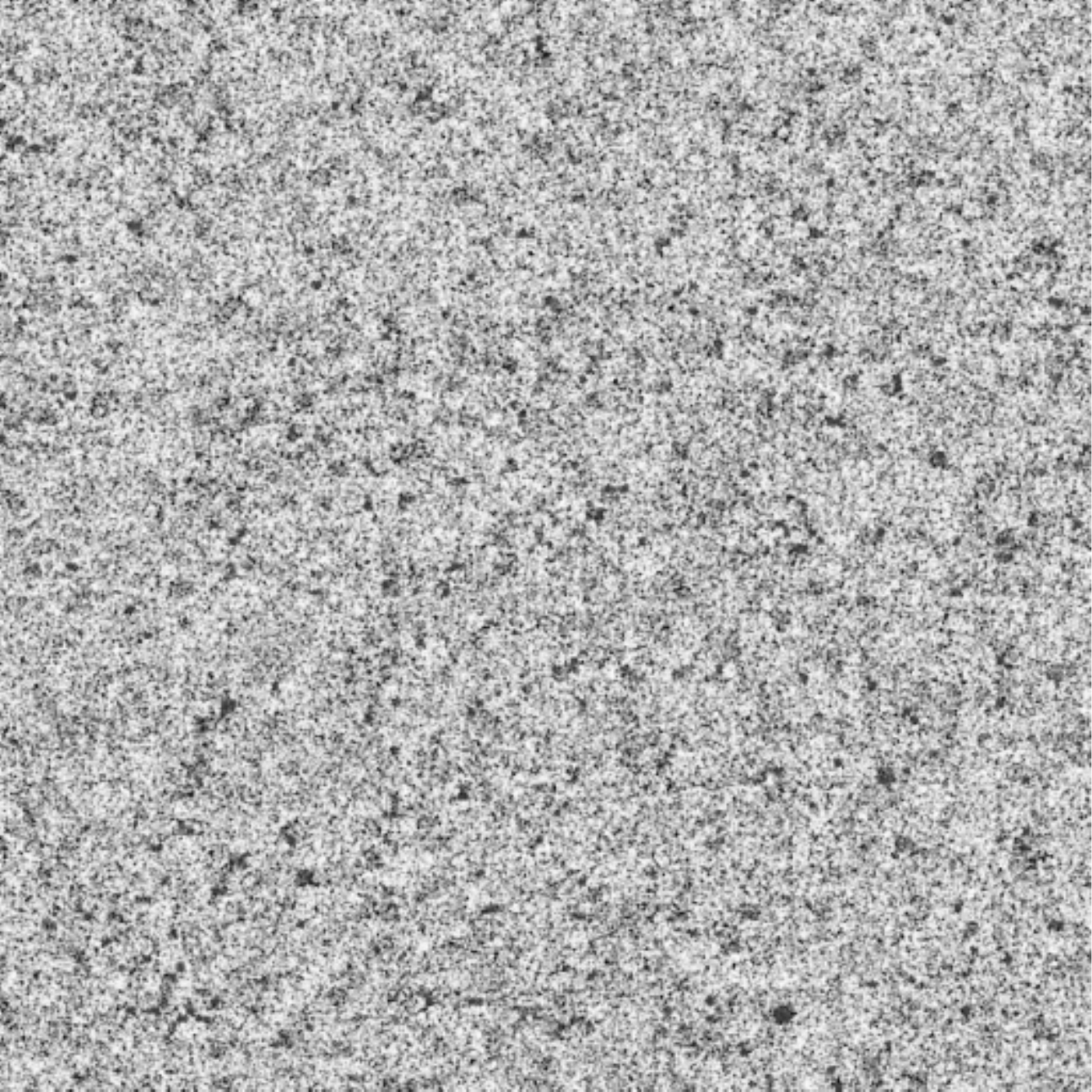}}
				
	\vspace*{-1em}
	    \subfloat[JP2K 0.75~bpp]
	{\includegraphics[width=0.225\textwidth]{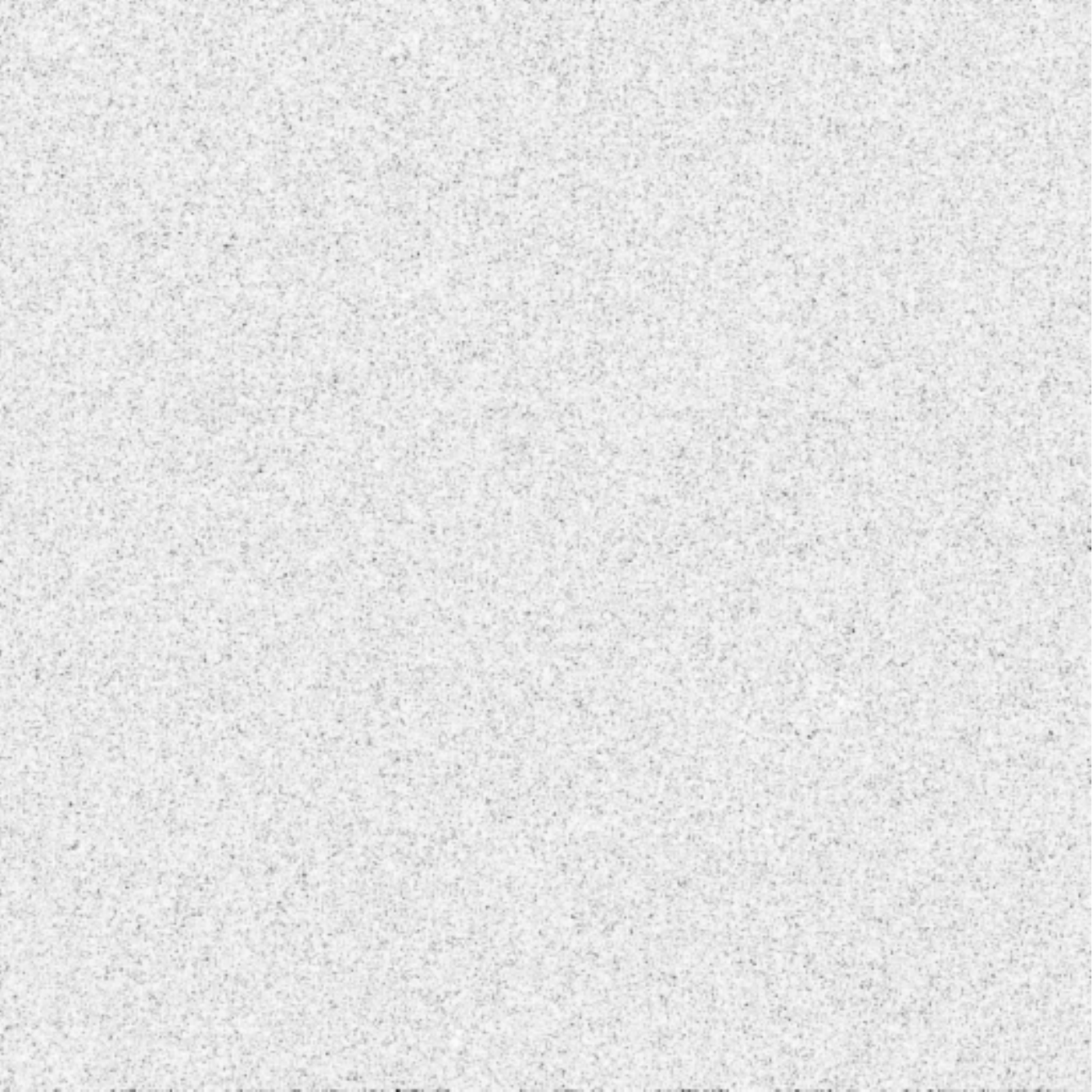}}\hspace*{0.025em}
	    \subfloat[JP2K 0.375~bpp]
	{\includegraphics[width=0.225\textwidth]{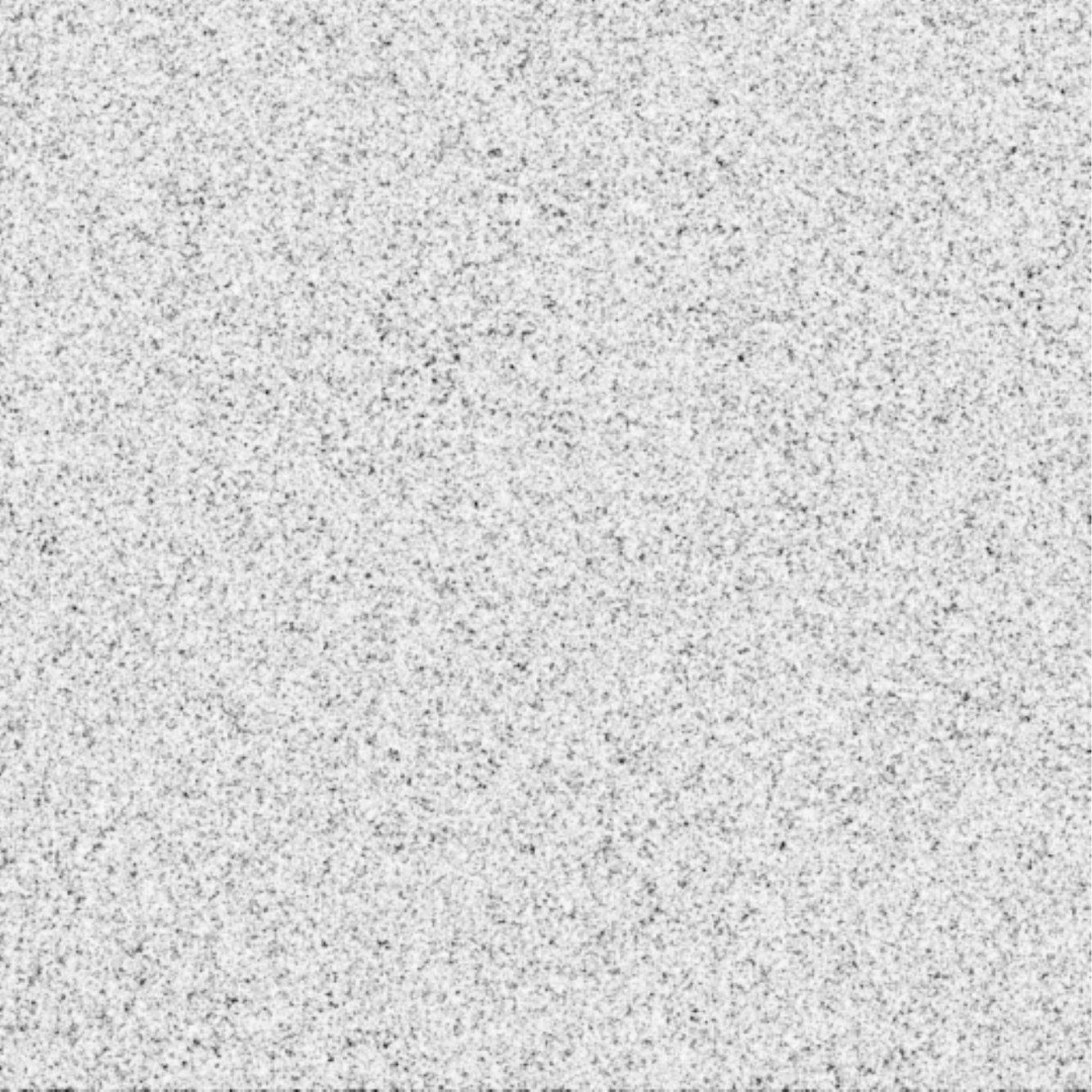}}\hspace*{0.025em}
	    \subfloat[JP2K 0.25~bpp]
	{\includegraphics[width=0.225\textwidth]{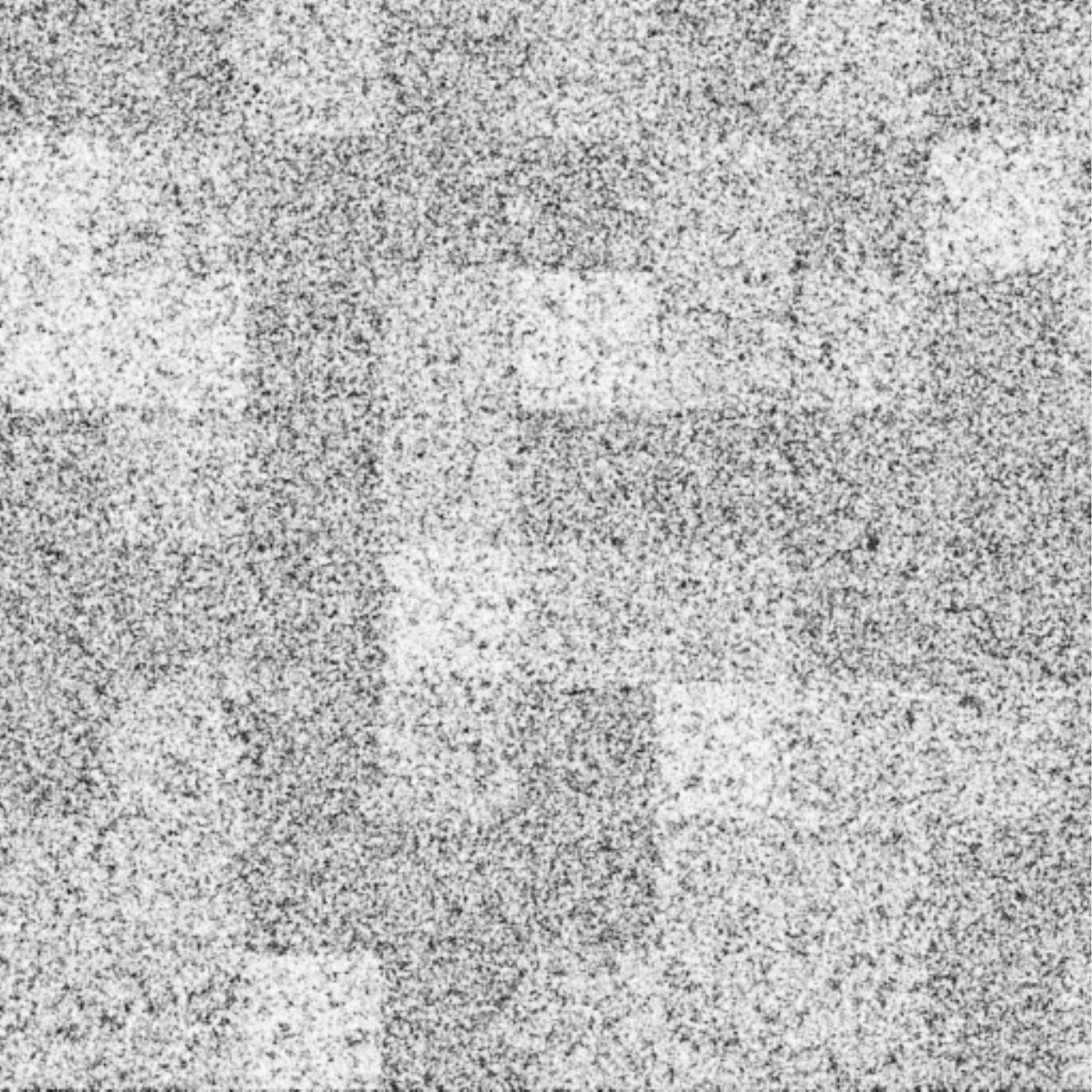}}\hspace*{0.025em}
	    \subfloat[JP2K 0.125~bpp]
	{\includegraphics[width=0.225\textwidth]{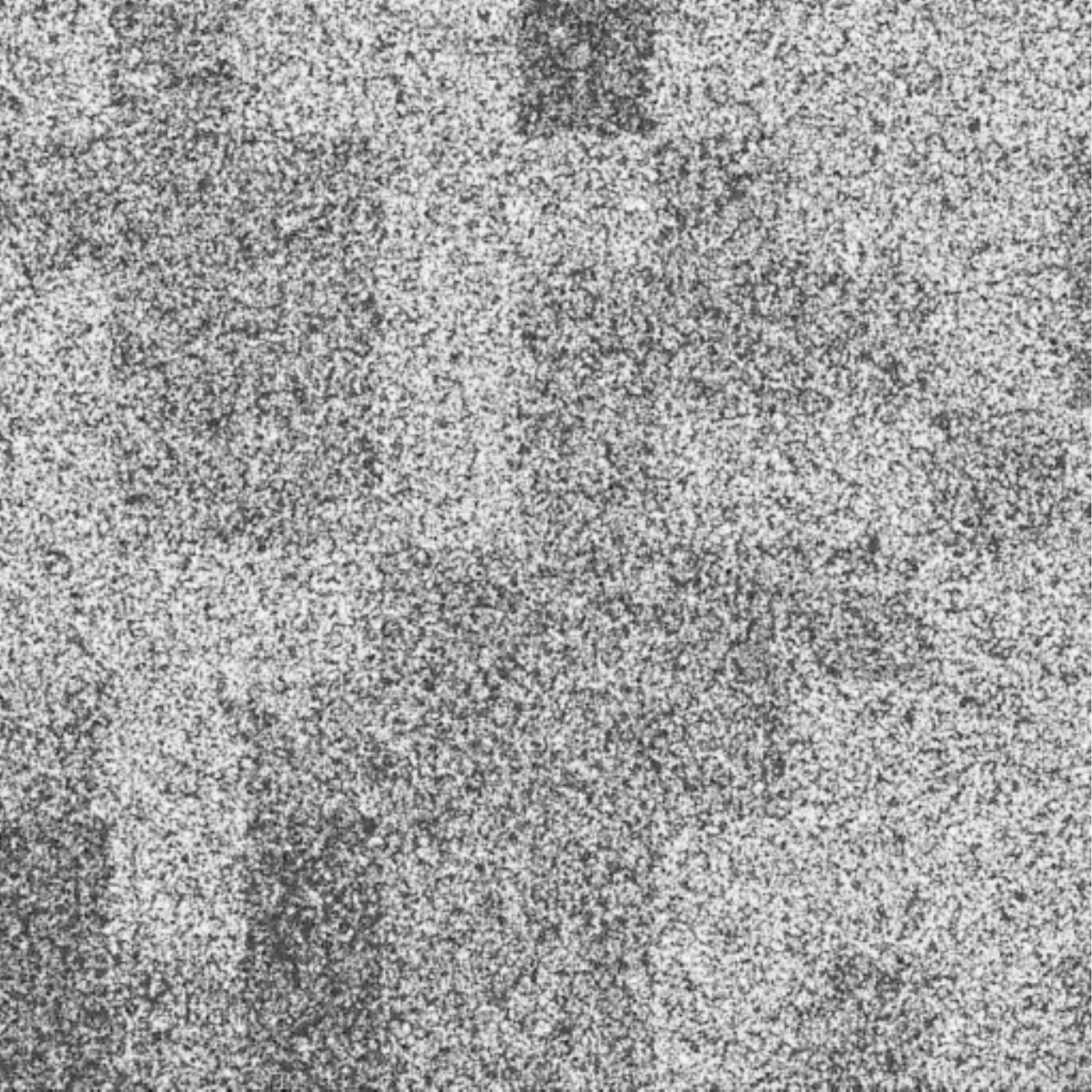}}
				
	\vspace*{-1em}
	    \subfloat[WAC 0.75~bpp]
	{\includegraphics[width=0.225\textwidth]{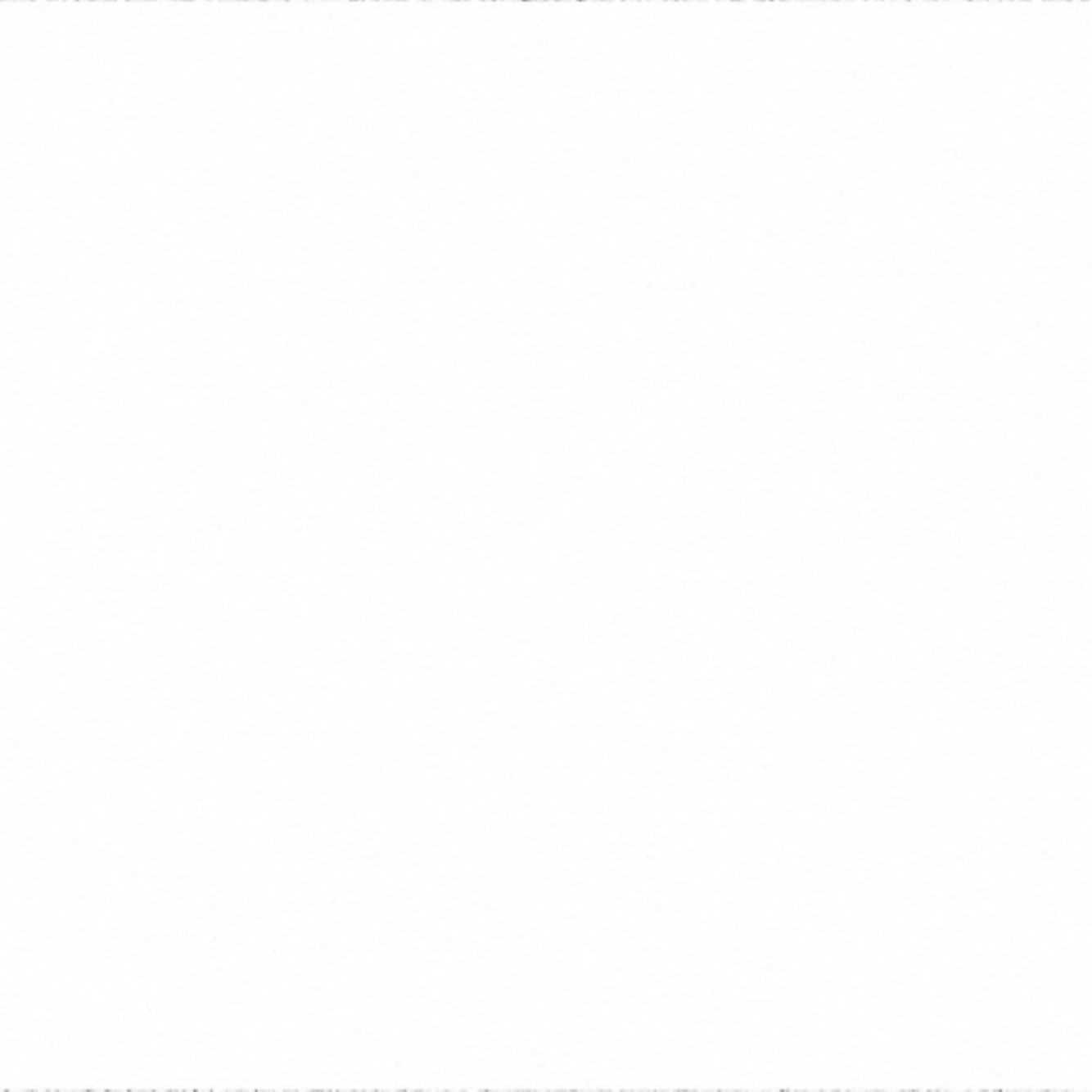}}\hspace*{0.025em}
	    \subfloat[WAC 0.375~bpp]
	{\includegraphics[width=0.225\textwidth]{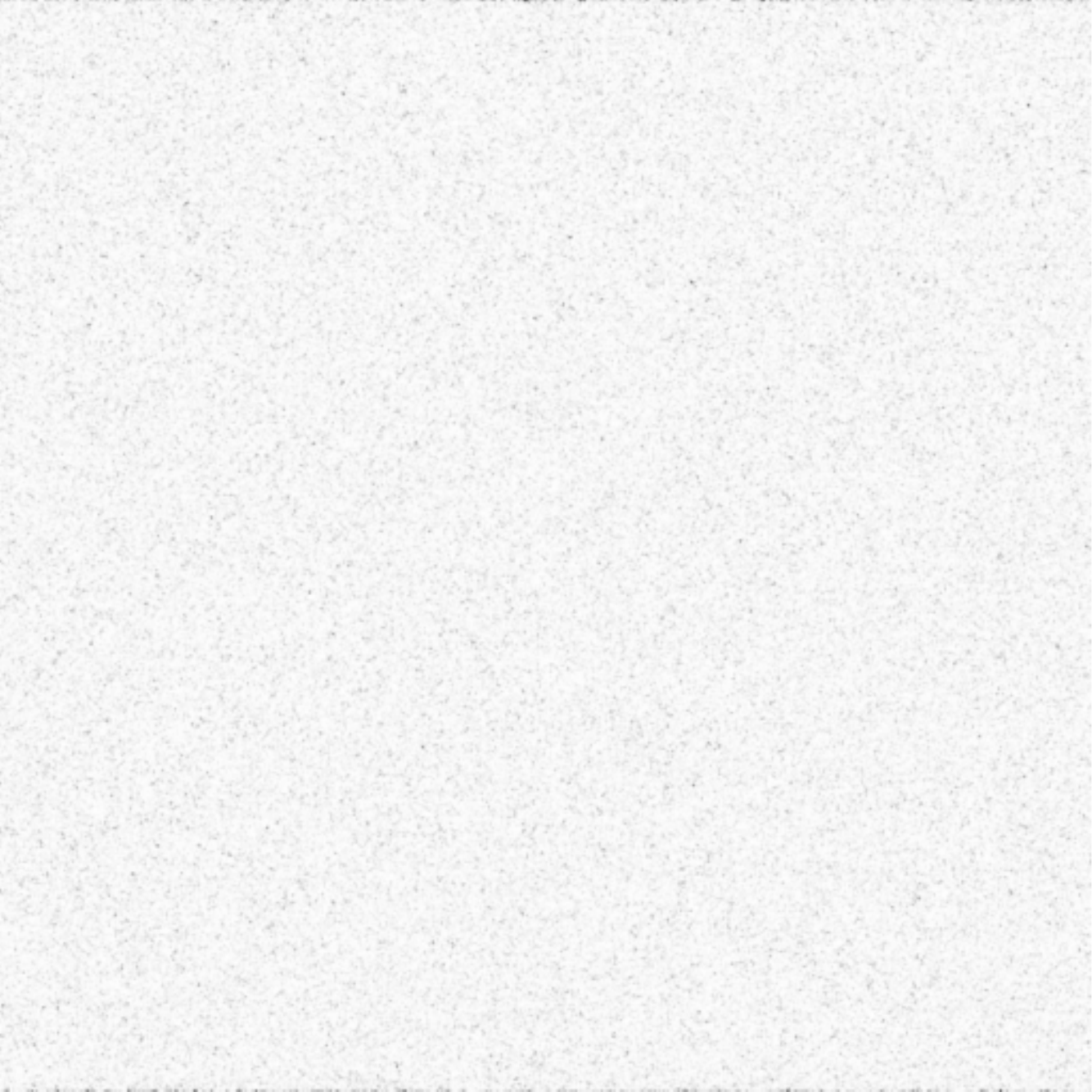}}\hspace*{0.025em}
	    \subfloat[WAC 0.25~bpp]
	{\includegraphics[width=0.225\textwidth]{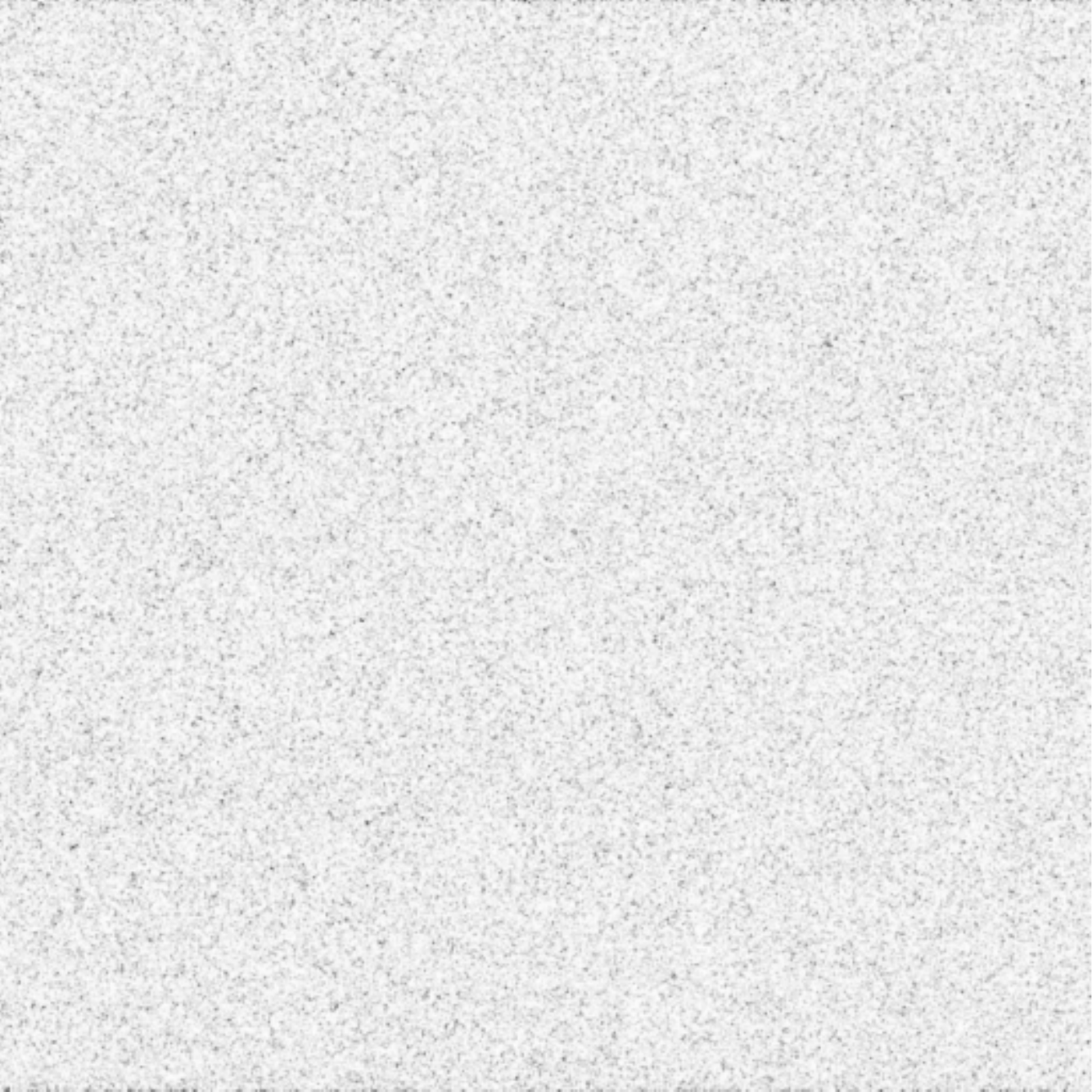}}\hspace*{0.025em}
	    \subfloat[WAC 0.125~bpp]
	{\includegraphics[width=0.225\textwidth]{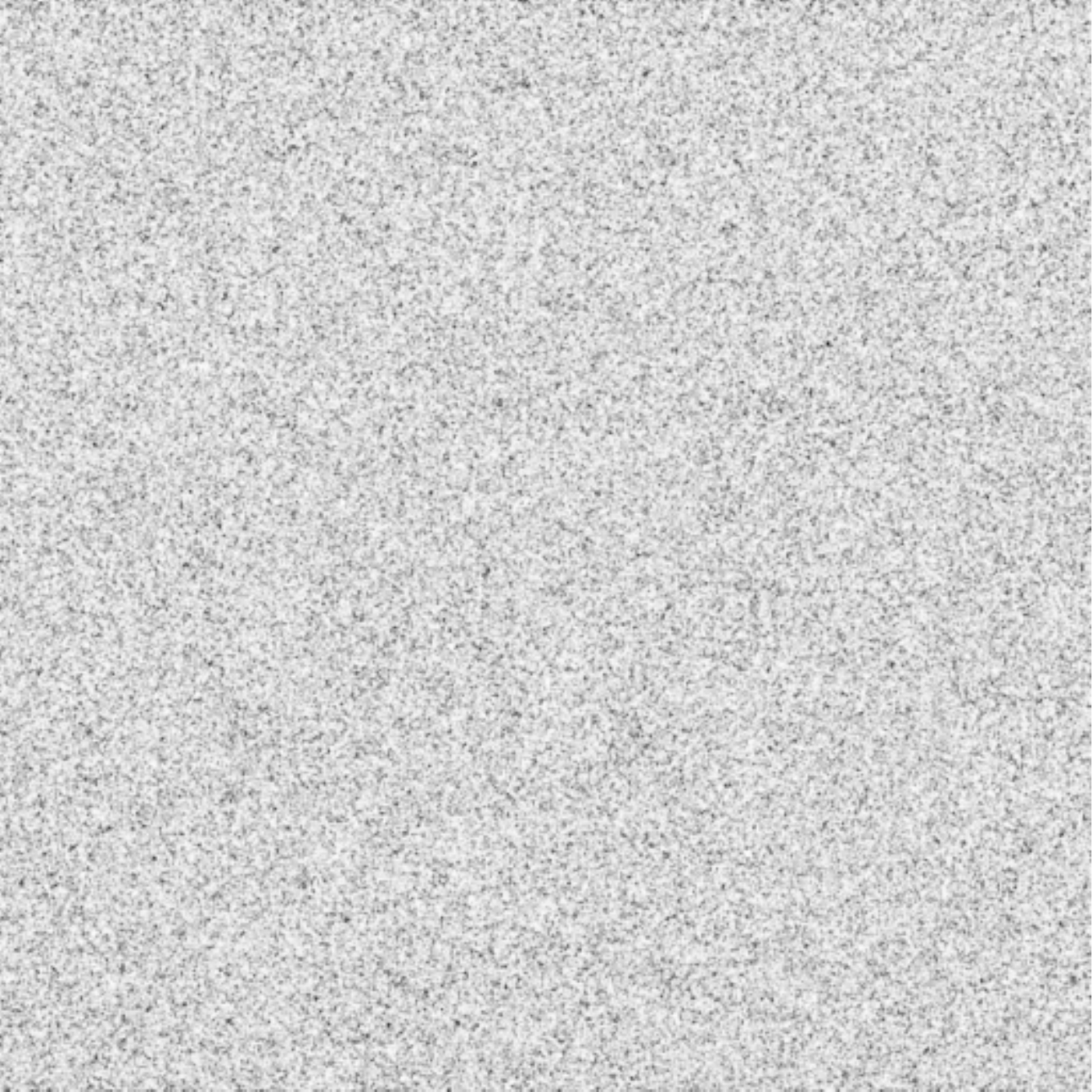}}

\caption{SSIM quality maps for the real part of the compressed CG-Ball. Bright and dark areas account for high and low similarity accordingly. It appears that all encoders, especially HEVC-Intra mode and JPEG~2000, create particular compression artifacts, which clearly intensify when operating at lower bit rates. This creates a distinctive gradient over quality maps, enabling the SSIM to correctly rank and predict their relative quality score.}
\label{fig:SSIMQmaps}%
\vspace*{-1em}
\end{figure*}

To answer this question, we took the CG-Ball hologram as an example and calculated the SSIM quality map for the real and imaginary parts of the compressed holograms separately. Fig.\ref{fig:SSIMQmaps} shows these quality maps. Due to resemblance of the quality maps for real and imaginary parts we only show the results for the real parts. The top row shows the results for the CG-Ball compressed with HEVC-Intra mode. While not directly visible in the compressed data itself, a certain grid artifact (related to the boundaries of the HEVC code blocks) appears in the SSIM quality map where the data is normalized based on the luminance and the standard deviation of pixel values (and compared to the non-compressed data using SSIM formulae). In the second row, quality maps for the hologram compressed with JPEG~2000 are depicted. Here, also heavy blocking artifacts show up in SSIM quality maps which otherwise are not clearly distinguishable in the compressed data, when observed directly. These artifacts at least partially could be related to what after the reconstruction appears as the aliasing artifacts(see section~\ref{encodersdetails}). In the third row, where quality maps for WAC are shown, no structured compression artifact are found. However, just like with the other two encoders, SSIM is able to correctly distinguish the compression levels by verifying those increases or decreases in the magnitude of similarity (\emph{i.e.} changes in brightness of quality maps observed in passing from one quality map to another. See the brightness of Fig.\ref{fig:SSIMQmaps}.(a) to (d) or Fig.\ref{fig:SSIMQmaps}.(i) to (l) and also note the Fig.\ref{fig:SSIMQmaps}.(i) compared to (e) and (a) ). Overall, it appears that by amplification of the compression artifacts which otherwise are wrapped-up and masked by the heavily noisy environment of the hologram (i.e.the interference pattern), SSIM is able to rather easily predict the relative quality score and rank of the decoded holograms without having any explicit knowledge about the actual appearance of the objects after the reconstruction.

Another observation based on the results of \tabref{tab:CorrsPreReconFourier} is that the multi-scale metrics are generally worse than single scale methods (such as SSIM, PSNR). This probably due to the fact that weights that assigned to the scores of each scale in multi-scale metrics. These weights are assigned based on characteristics of the human visual system (HVS) and psychophysical measurements of digital images and might not be relevant to complex holographic data before the reconstruction. 
However, it must be noted that even the predictions of the worst metrics in each category are not failing completely. For example, the UQI metric which seems to achieve the lowest correlation results, has an average SROCC of 0.8.

\begin{figure*}
\centering
	    \subfloat[ MOS-OPT\label{fig:StatSigPreReconFourier_OPT}]
	{\includegraphics[width=0.3\textwidth]{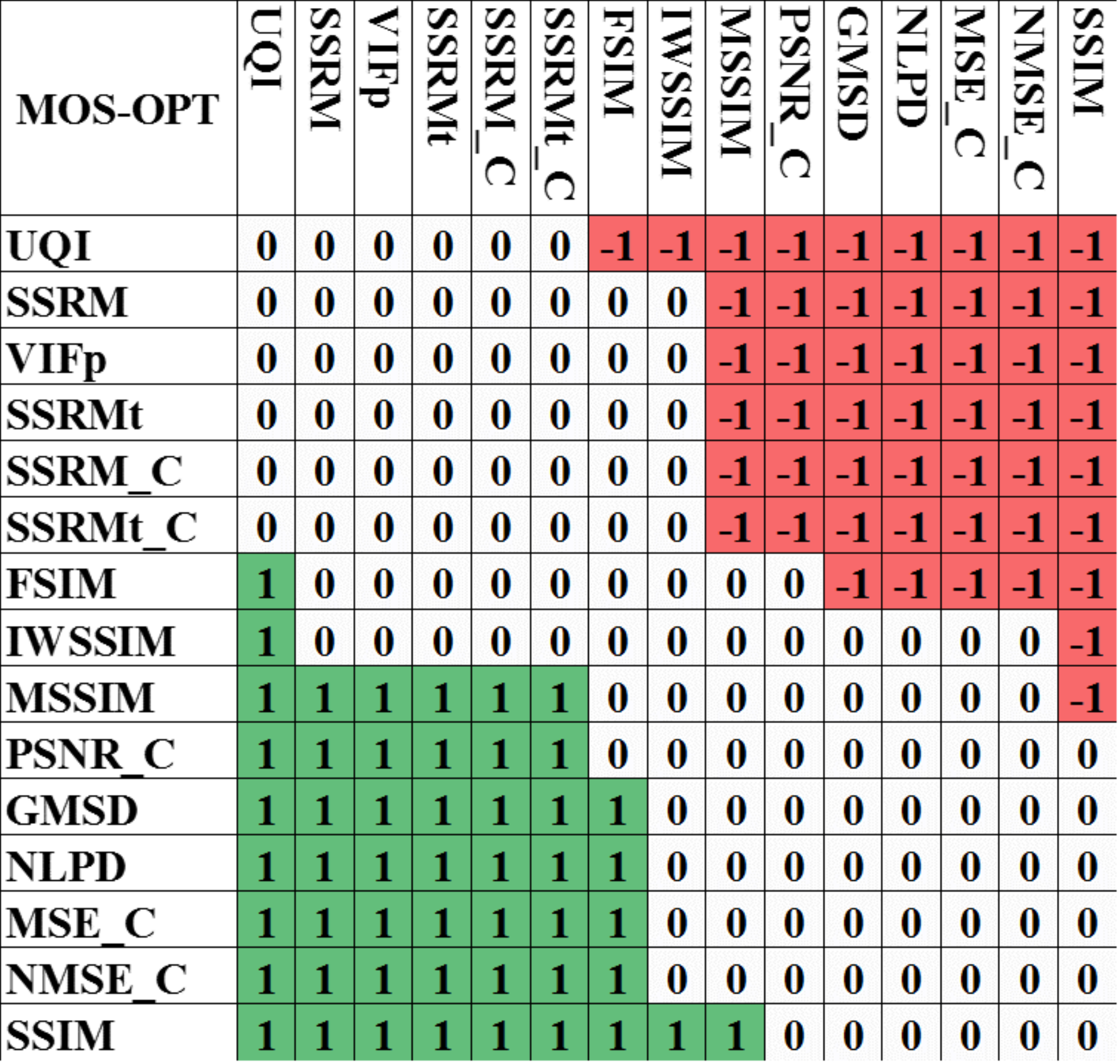}}\hspace*{0.75em}
	    \subfloat[ MOS-LF\label{fig:StatSigPreReconFourier_LF}]
	{\includegraphics[width=0.3\textwidth]{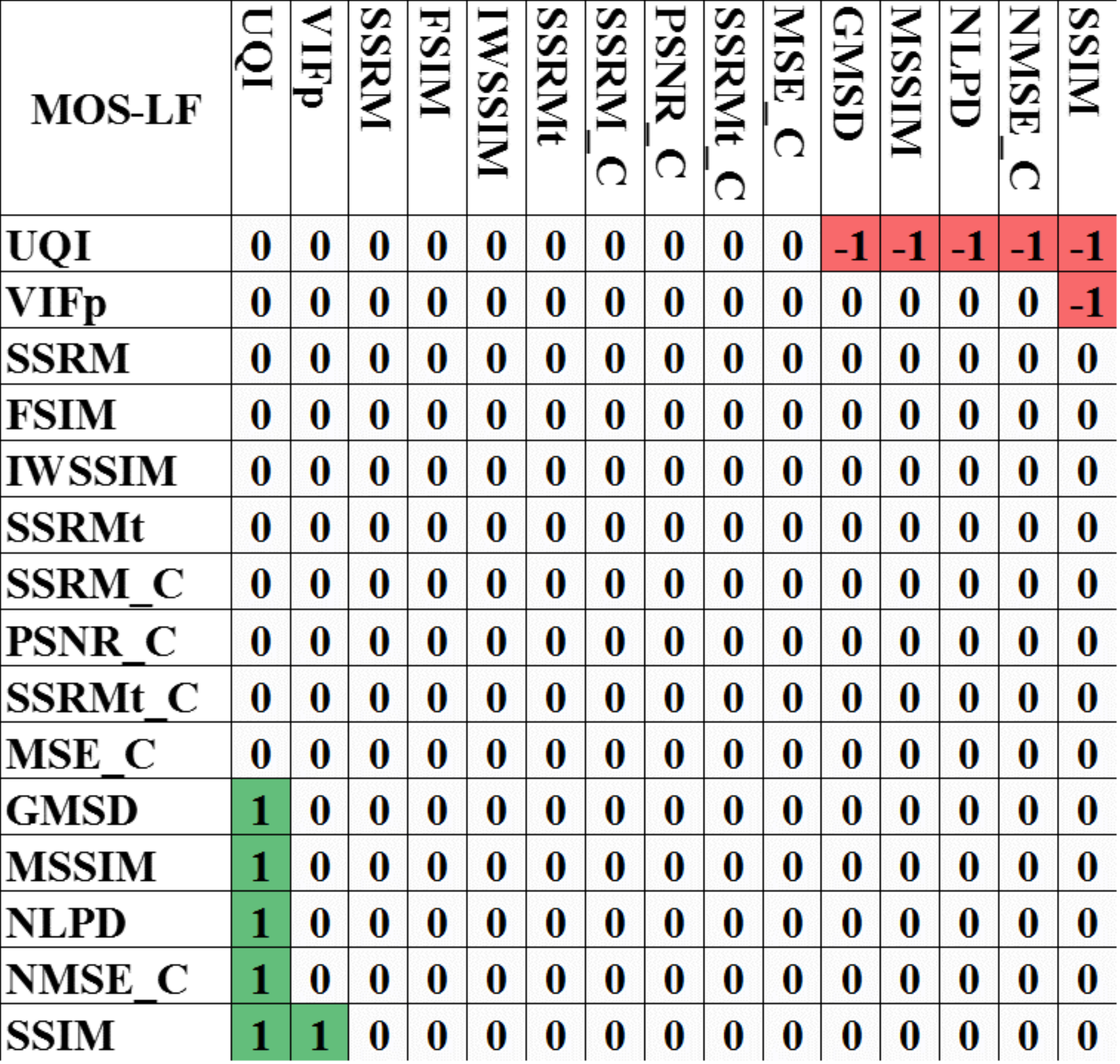}}\hspace*{0.75em}
	    \subfloat[ MOS-2D\label{fig:StatSigPreReconFourier_2D}]
	{\includegraphics[width=0.3\textwidth]{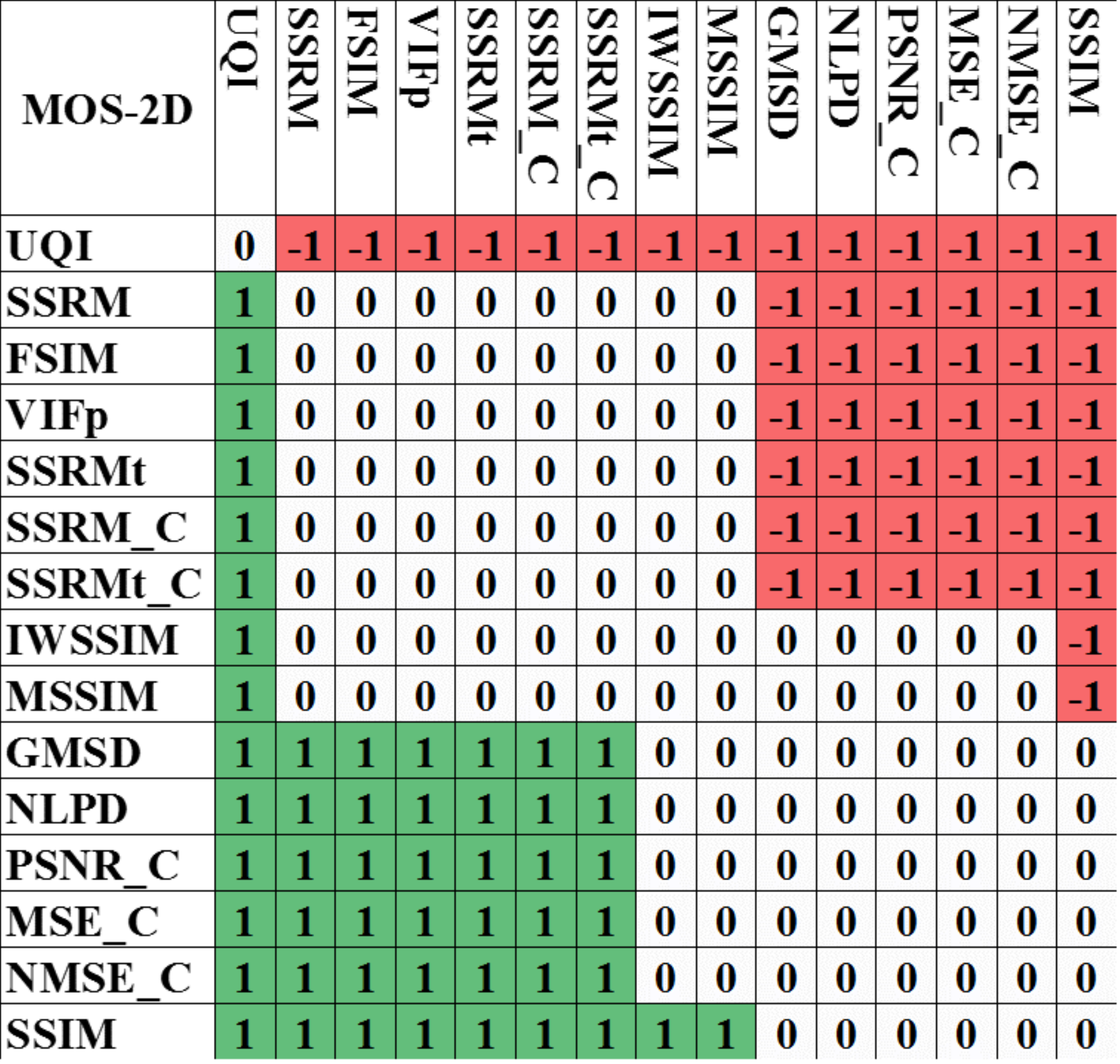}}
\caption{Statistical significance tables for the evaluation of quality metrics on Fourier holograms, rated after decoding. The statistics are separately shown based on the MOS scores obtained from optical holographic display (OPT) (a), the light field display (LF) (b) and the regular 2D display (2D) (c). The significance of the IQM performance in each row compared to the others is quantified by 1,0 and -1 representing significantly better performance, no significant difference and significantly worse performance respectively.}
\label{fig:StatSigPreReconFourier}
\vspace*{-1em}
\end{figure*}

\begin{table*}
\caption{Statistical evaluation of quality metrics for the Fourier holograms, rated after hologram decoding; thus before rendering. The statistics based on the MOS scores obtained from optical holographic display (OPT), the light field display (LF) and the regular 2D display (2D) and the IQM scores are separately shown. Quality metrics directly evaluated on the complex-valued data have "\_C" as postfix.}
\resizebox{\textwidth}{!}{%
\begin{tabular}{|l|l|l|l|l|l|l|l|l|l|l|l|l|l|l|l|l|}
\hline
\multicolumn{2}{|l|}{}                           & SSIM            & SSRM   & FSIM   & GMSD   & IWSSIM & MS-SSIM  & UQI    & VIFp   & NLPD           & SSRMt  & SSRM\_C & PSNR\_C         & SSRMt\_C & MSE\_C & NMSE\_C         \\ \hline
\multirow{6}{*}{\rot{\textbf{MOS-OPT}}} & SROCC  & \textbf{0.9069} & 0.8382 & 0.8425 & 0.8855 & 0.8604 & 0.8833 & 0.8032 & 0.8365 & 0.8876         & 0.8379 & 0.8366  & 0.8821          & 0.8364   & 0.8821 & 0.8907          \\ \cline{2-17} 
                                  & KRCC         & \textbf{0.7314} & 0.6334 & 0.6424 & 0.698  & 0.6648 & 0.6913 & 0.6036 & 0.6331 & 0.6966         & 0.6332 & 0.6329  & 0.6851          & 0.6329   & 0.6851 & 0.6967          \\ \cline{2-17} 
                                  & PCC\_NoFit   & \textbf{0.9088} & 0.8244 & 0.806  & 0.8708 & 0.8277 & 0.8593 & 0.7441 & 0.8138 & 0.8667         & 0.8245 & 0.8252  & 0.7389          & 0.8254   & 0.8731 & 0.8805          \\ \cline{2-17} 
                                  & PCC\_Fitted  & \textbf{0.9129} & 0.8305 & 0.8429 & 0.8839 & 0.8565 & 0.8808 & 0.757  & 0.8329 & 0.8837         & 0.8305 & 0.8297  & 0.878           & 0.8297   & 0.8779 & 0.8873          \\ \cline{2-17} 
                                  & RMSE         & \textbf{0.4589} & 0.6261 & 0.6049 & 0.5258 & 0.5803 & 0.5323 & 0.7346 & 0.622  & 0.5262         & 0.6262 & 0.6275  & 0.538           & 0.6275   & 0.5382 & 0.5184          \\ \cline{2-17} 
                                  & OutlierRatio & 0.6042          & 0.6771 & 0.6875 & 0.6458 & 0.6823 & 0.6354 & 0.7031 & 0.6719 & 0.651          & 0.6771 & 0.6979  & \textbf{0.5781} & 0.6979   & 0.5938 & 0.599           \\ \hline
\multirow{6}{*}{\rot{\textbf{MOS-LF}}}  & SROCC        & \textbf{0.9012} & 0.8716 & 0.8762 & 0.8965 & 0.8847 & 0.8966 & 0.8158 & 0.8588 & 0.8898         & 0.8714 & 0.8725  & 0.8812          & 0.8726   & 0.8812 & 0.8894          \\ \cline{2-17} 
                                  & KRCC         & \textbf{0.7254} & 0.6927 & 0.6991 & 0.7225 & 0.7031 & 0.7176 & 0.6289 & 0.673  & 0.7078         & 0.6923 & 0.6939  & 0.6916          & 0.6942   & 0.6916 & 0.7022          \\ \cline{2-17} 
                                  & PCC\_NoFit   & \textbf{0.8776} & 0.857  & 0.8151 & 0.8871 & 0.8187 & 0.8368 & 0.7286 & 0.8569 & 0.884          & 0.8572 & 0.8651  & 0.8028          & 0.8655   & 0.8633 & 0.8705          \\ \cline{2-17} 
                                  & PCC\_Fitted  & \textbf{0.9006} & 0.8718 & 0.8774 & 0.8929 & 0.8817 & 0.8961 & 0.8285 & 0.8588 & 0.8897         & 0.8718 & 0.8729  & 0.8814          & 0.8731   & 0.8819 & 0.8888          \\ \cline{2-17} 
                                  & RMSE         & \textbf{0.4531} & 0.5107 & 0.5002 & 0.4694 & 0.4919 & 0.4626 & 0.5838 & 0.5341 & 0.476          & 0.5107 & 0.5086  & 0.4925          & 0.5084   & 0.4915 & 0.4778          \\ \cline{2-17} 
                                  & OutlierRatio & \textbf{0.5729} & 0.599  & 0.6198 & 0.5833 & 0.5781 & 0.5781 & 0.6406 & 0.6146 & 0.5938         & 0.599  & 0.6042  & 0.6354          & 0.6042   & 0.6198 & 0.6302          \\ \hline
\multirow{6}{*}{\rot{\textbf{MOS-2D}}}  & SROCC        & \textbf{0.9001} & 0.8365 & 0.8386 & 0.8795 & 0.8608 & 0.8685 & 0.7901 & 0.8249 & 0.8738         & 0.8364 & 0.8365  & 0.8775          & 0.8365   & 0.8775 & 0.8846          \\ \cline{2-17} 
                                  & KRCC         & \textbf{0.7215} & 0.6424 & 0.6438 & 0.6892 & 0.6713 & 0.6703 & 0.5859 & 0.6218 & 0.6804         & 0.6419 & 0.6427  & 0.6864          & 0.6427   & 0.6864 & 0.6967          \\ \cline{2-17} 
                                  & PCC\_NoFit   & 0.8651          & 0.8229 & 0.7708 & 0.8767 & 0.7904 & 0.8026 & 0.7105 & 0.8346 & \textbf{0.878} & 0.8232 & 0.8316  & 0.8168          & 0.832    & 0.8561 & 0.8625          \\ \cline{2-17} 
                                  & PCC\_Fitted  & \textbf{0.9052} & 0.8454 & 0.8482 & 0.8859 & 0.8682 & 0.8771 & 0.7286 & 0.8366 & 0.8874         & 0.8454 & 0.8455  & 0.8894          & 0.8456   & 0.89   & 0.8938          \\ \cline{2-17} 
                                  & RMSE         & \textbf{0.4697} & 0.5903 & 0.5854 & 0.5126 & 0.5484 & 0.5309 & 0.757  & 0.6054 & 0.5094         & 0.5902 & 0.59    & 0.5052          & 0.5899   & 0.504  & 0.4957          \\ \cline{2-17} 
                                  & OutlierRatio & 0.6354          & 0.7552 & 0.7396 & 0.6667 & 0.6719 & 0.6875 & 0.7604 & 0.7396 & 0.6458         & 0.7552 & 0.7604  & 0.6198          & 0.7604   & 0.6302 & \textbf{0.6198} \\ \hline
\end{tabular}}
\label{tab:CorrsPreReconFourier}%
\end{table*}
To better understand the significance of SSIM performance in relation to the other metrics, we have provided 3 significance tables, one per set of MOSs in \figref{fig:StatSigPreReconFourier}. Here, it can be seen that the performance of the SSIM compared to the IQMs like MSE, NMSE, PSNR, GMSD, and NLPD is not significantly better, while compared to the other metrics it shows a clear superiority. Accounting for the MOS\_LF scores shown in \figref{fig:StatSigPreReconFourier}.(b), the results are even closer and most of the quality metrics are on par with respect to the criterion explained in section \ref{sec:StatAnalysis}. 
The reason for these results could be related to the nature of Fourier holography. The object-field wavefronts captured by a Fourier hologram are not convergent at any finite distance. Thus, we may interpret them as being convergent at a point at infinity. Propagation over infinite distances is described by the Fraunhofer propagation and essentially reduced to a Fourier transform. Henceforth, when analyzing Fourier holograms we essentially study the Fourier domain of the in-focus light field of an object. Thus, the metrics which try to relate their local (\emph{e.g} SSIM, GMSD) or global error measurements (\emph{e.g} PSNR, MSE) to the perceptual quality often perform better than metrics which rely on internal transformations. (\emph{e.g} SSRM has an internal Fourier transform which, when applied to a Fourier hologram, brings the data back to the object plane).          

\subsubsection{QA\_2 - Evaluation on Fresnel holograms}
\label{sec:PreReconFT2Fresnel}
The next test relates to the synthesized Fresnel holograms. The same analysis process as for the Fourier holograms in the previous section is deployed. The Fourier to Fresnel hologram conversion was detailed in section~\ref{subsec:Fourier2Fresnel}. While the space-frequency behaviour of the holograms changed drastically (sse Fig.~\ref{fig:fourier_vs_fresnel}, the data still represents wavefields with statistics very different from natural images. 

\begin{figure*}
\centering
	    \subfloat[MOS-OPT\label{fig:StatSigPreReconFresnel_OPT}]
	{\includegraphics[width=0.3\textwidth]{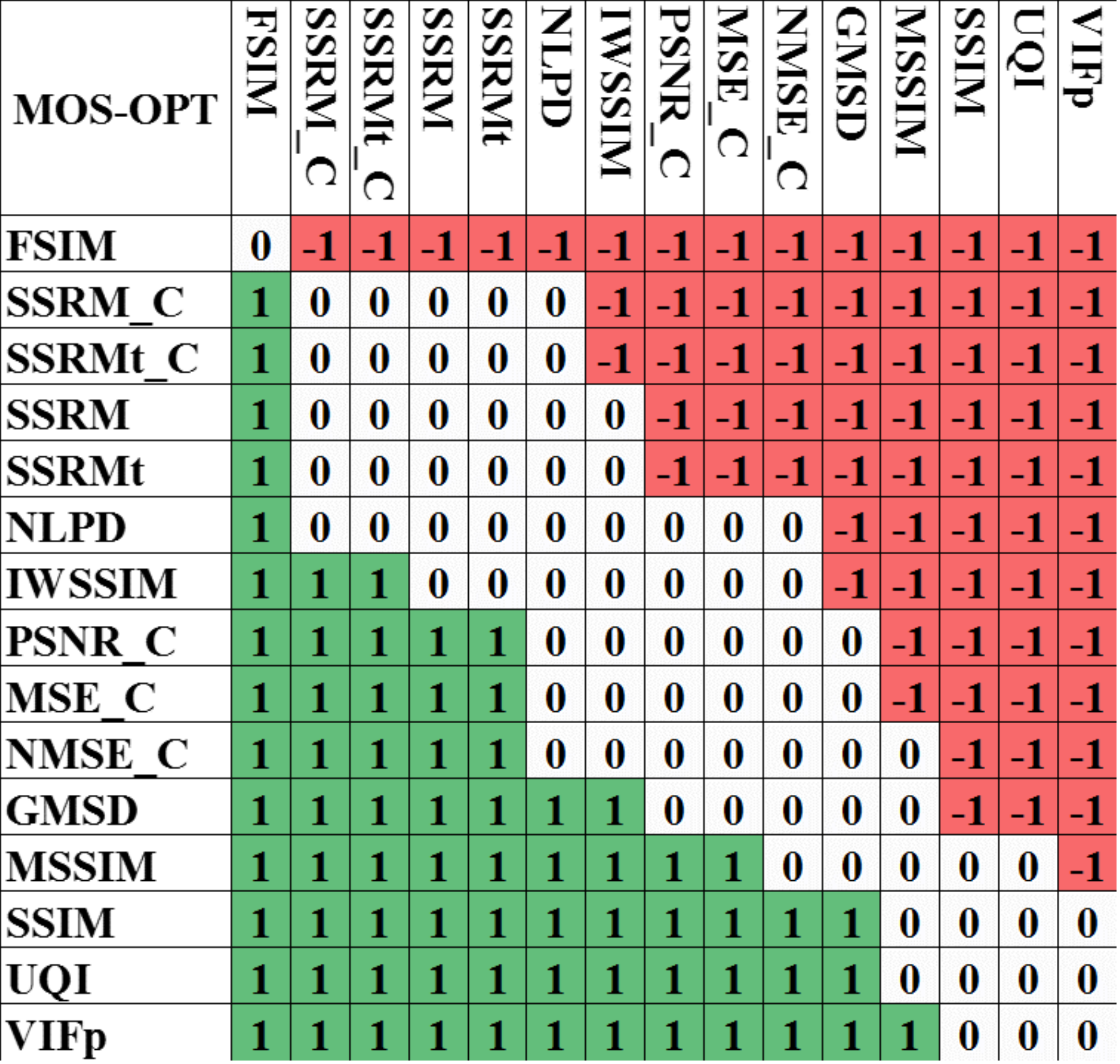}}\hspace*{0.75em}
	    \subfloat[MOS-LF\label{fig:StatSigPreReconFresnel_LF}]
	{\includegraphics[width=0.3\textwidth]{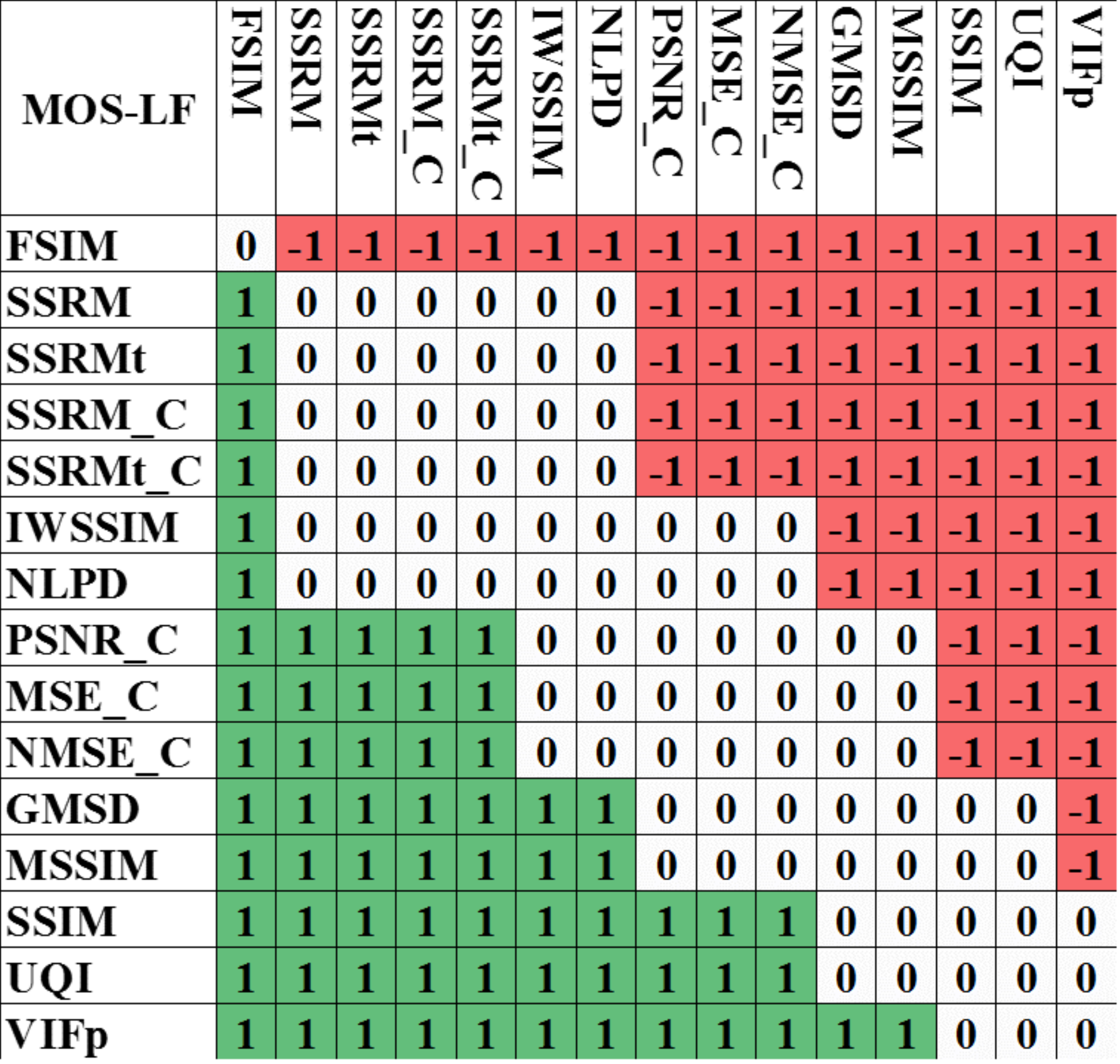}}\hspace*{0.75em}
	    \subfloat[MOS-2D\label{fig:StatSigPreReconFresnel_2D}]
	{\includegraphics[width=0.3\textwidth]{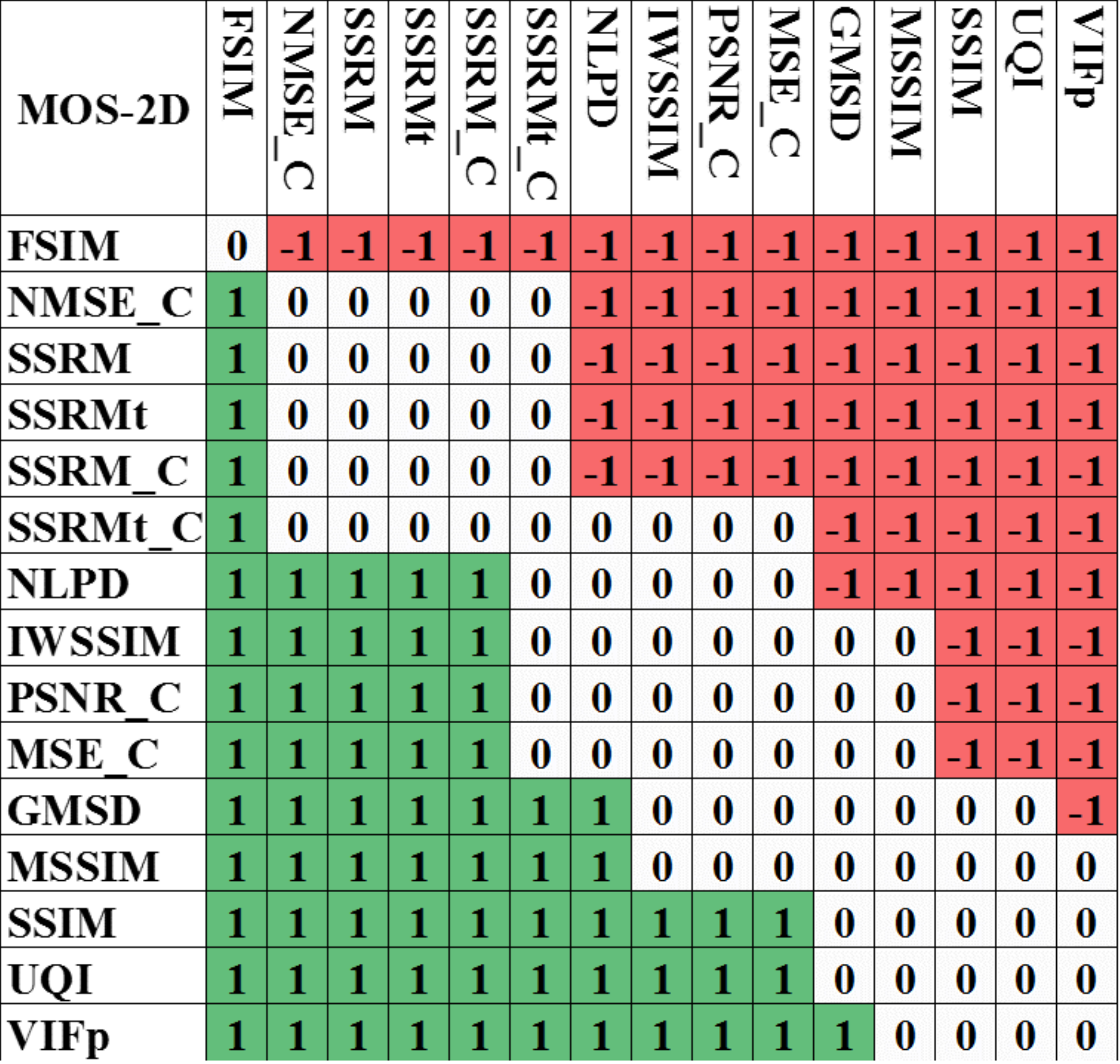}}
\caption{Statistical significance tables for the evaluation of quality metrics on the synthesized Fresnel holograms, rated after decoding. The statistics are separately shown based on the MOS scores obtained from optical holographic display (OPT) (a), the light field display (LF) (b) and the regular 2D display (2D) (c) and the IQM scores.}
\label{fig:StatSigPreReconFresnel}
\vspace*{-1em}
\end{figure*}

Interestingly, VIFp shows the best performance across all 3 MOS values in \tabref{tab:CorrsPreReconFresnel}. Although its rather lower PCC\_NoFit values indicate that its behaviour with respect to the MOS is less linear compared to its competitors like SSIM --- closely following behind the VIFp. Another point to be noticed is the rather larger gap between the top and worst performing metrics w.r.t each criterion. This gap becomes more evident in the significance tables of \figref{fig:StatSigPreReconFresnel}. Here, it is clear that the top 3 quality metrics namely VIFp, UQI and SSIM are rather similar while their predictions are significantly better than others.

\begin{table*}
\caption{Statistical evaluation of quality metrics for the synthesized Fresnel holograms, rated before hologram reconstruction. The statistics based on the MOS scores obtained from optical holographic display (OPT), the light field display (LF) and the regular 2D display (2D) and the IQM scores are separately shown. Quality metrics directly evaluated on the complex-valued data are prefixed by "\_C".}

\resizebox{\textwidth}{!}{
\begin{tabular}{|l|l|l|l|l|l|l|l|l|l|l|l|l|l|l|l|l|}
\hline
\multicolumn{2}{|l|}{}                  & SSIM            & SSRM   & FSIM   & GMSD   & IWSSIM & MS-SSIM  & UQI    & VIFp            & NLPD   & SSRMt  & SSRM\_C & PSNR\_C & SSRMt\_C & MSE\_C & NMSE\_C \\ \hline
\multirow{6}{*}{\rot{\textbf{MOS-OPT}}} & SROCC        & 0.8912          & 0.7076 & 0.5179 & 0.8463 & 0.791  & 0.8769 & 0.8922 & \textbf{0.9088} & 0.7632 & 0.7078 & 0.6928  & 0.8015  & 0.693    & 0.8015 & 0.8112  \\ \cline{2-17} 
                         & KRCC         & 0.7038          & 0.5142 & 0.3614 & 0.6563 & 0.5912 & 0.6802 & 0.7067 & \textbf{0.7343} & 0.5642 & 0.5143 & 0.5021  & 0.6047  & 0.5022   & 0.6047 & 0.6117  \\ \cline{2-17} 
                         & PCC\_NoFit   & \textbf{0.8912} & 0.6547 & 0.4607 & 0.8212 & 0.7596 & 0.854  & 0.8893 & 0.8045          & 0.7326 & 0.6548 & 0.6441  & 0.6954  & 0.6445   & 0.7699 & 0.783   \\ \cline{2-17} 
                         & PCC\_Fitted  & 0.8944          & 0.6876 & 0.4833 & 0.8304 & 0.7721 & 0.8716 & 0.8953 & \textbf{0.916}  & 0.7388 & 0.6878 & 0.6715  & 0.7878  & 0.672    & 0.7885 & 0.8036  \\ \cline{2-17} 
                         & RMSE         & 0.5027          & 0.8163 & 0.9841 & 0.6264 & 0.7144 & 0.5512 & 0.5007 & \textbf{0.4509} & 0.7576 & 0.816  & 0.833   & 0.6924  & 0.8325   & 0.6914 & 0.6691  \\ \cline{2-17} 
                         & OutlierRatio & 0.6042          & 0.7396 & 0.8073 & 0.6302 & 0.724  & 0.6198 & 0.599  & \textbf{0.5625} & 0.6719 & 0.7396 & 0.7344  & 0.6406  & 0.7344   & 0.651  & 0.651   \\ \hline
\multirow{6}{*}{\rot{\textbf{MOS-LF}}}  & SROCC        & 0.887           & 0.6973 & 0.5102 & 0.8399 & 0.7561 & 0.8687 & 0.8858 & \textbf{0.9067} & 0.7597 & 0.6968 & 0.6842  & 0.8148  & 0.6835   & 0.8148 & 0.8253  \\ \cline{2-17} 
                         & KRCC         & 0.7022          & 0.5138 & 0.3674 & 0.6598 & 0.5644 & 0.68   & 0.7004 & \textbf{0.733}  & 0.5714 & 0.5134 & 0.5055  & 0.6199  & 0.5045   & 0.6199 & 0.6297  \\ \cline{2-17} 
                         & PCC\_NoFit   & \textbf{0.8724} & 0.6494 & 0.4672 & 0.8348 & 0.7105 & 0.8196 & 0.8621 & 0.855           & 0.7501 & 0.6494 & 0.642   & 0.7675  & 0.642    & 0.7889 & 0.803   \\ \cline{2-17} 
                         & PCC\_Fitted  & 0.8859          & 0.7096 & 0.5034 & 0.8378 & 0.7594 & 0.8698 & 0.8849 & \textbf{0.9064} & 0.7636 & 0.7097 & 0.6945  & 0.8155  & 0.6946   & 0.8155 & 0.8278  \\ \cline{2-17} 
                         & RMSE         & 0.4837          & 0.7345 & 0.9008 & 0.5692 & 0.6782 & 0.5144 & 0.4855 & \textbf{0.4403} & 0.6732 & 0.7344 & 0.7501  & 0.6034  & 0.7499   & 0.6034 & 0.5849  \\ \cline{2-17} 
                         & OutlierRatio & 0.625           & 0.724  & 0.8385 & 0.6198 & 0.6563 & 0.6354 & 0.6354 & \textbf{0.5625} & 0.7292 & 0.724  & 0.7292  & 0.6771  & 0.7292   & 0.6719 & 0.6875  \\ \hline
\multirow{6}{*}{\rot{\textbf{MOS-2D}}}  & SROCC        & 0.8901          & 0.6928 & 0.5213 & 0.8413 & 0.7769 & 0.8639 & 0.8874 & \textbf{0.901}  & 0.7635 & 0.6923 & 0.6805  & 0.8227  & 0.6797   & 0.8227 & 0.8344  \\ \cline{2-17} 
                         & KRCC         & 0.7062          & 0.5017 & 0.3679 & 0.652  & 0.5757 & 0.6678 & 0.7021 & \textbf{0.7205} & 0.5711 & 0.5011 & 0.4923  & 0.6184  & 0.4913   & 0.6184 & 0.6315  \\ \cline{2-17} 
                         & PCC\_NoFit   & \textbf{0.8674} & 0.6494 & 0.4745 & 0.851  & 0.7308 & 0.8074 & 0.8541 & 0.8667          & 0.7687 & 0.6493 & 0.6434  & 0.7922  & 0.6431   & 0.7976 & 0.812   \\ \cline{2-17} 
                         & PCC\_Fitted  & 0.8953          & 0.7324 & 0.543  & 0.8599 & 0.8009 & 0.8814 & 0.8932 & \textbf{0.9063} & 0.7967 & 0.7323 & 0.7193  & 0.8408  & 0.7192   & 0.8408 & 0.8512  \\ \cline{2-17} 
                         & RMSE         & 0.4924          & 0.7525 & 0.928  & 0.564  & 0.6618 & 0.5221 & 0.497  & \textbf{0.4671} & 0.668  & 0.7525 & 0.7677  & 0.5982  & 0.7679   & 0.5982 & 0.58    \\ \cline{2-17} 
                         & OutlierRatio & 0.6563          & 0.7604 & 0.8073 & 0.6667 & 0.7188 & 0.6615 & 0.6927 & \textbf{0.6146} & 0.7448 & 0.7604 & 0.7604  & 0.6771  & 0.7604   & 0.6771 & 0.6875  \\ \hline
\end{tabular}}
\label{tab:CorrsPreReconFresnel}%
\end{table*}

\subsection{Evaluation on reconstructed holograms}
\label{sec:PostReconBenchmark}
In this section we provide the evaluation results of the experimental tracks QA\_3 and  QA\_4, \emph{i.e.} the evaluations after reconstruction of the Fourier holograms and after reconstruction plus speckle denoising. Here, we used the reconstructions of the centre and right-corner views, and at different focal distances as provided by the HoloDB. These reconstructions are generated with the same synthetic aperture as was used in previous experimental tracks, as well as in \cite{Ahar2019SuitabilityAO} to obtain the MOS. Only the absolute amplitudes of the reconstructed wavefield are examined by the IQMs as it is the principal part of the light sensed by our eye and upon which the MOS scores are based. The predictions and the MOS scores for all views and focal depths per rate point and codec are averaged to facilitate a direct comparison of IQM performances before and after reconstruction. Also, similar to the previous test tracks QA\_1 and QA\_2, only the overall statistics are presented.
\subsubsection{QA\_3 - Evaluation on reconstructed Fourier holograms}
\label{sec:PostReconSpeckle}
\tabref{tab:CorrsPostReconNoisy} provides the statistical results for the IQMs, measured immediately after the reconstruction process, with respect to the measured MOSs. It is interesting to note that at this point, the reconstructions are much more similar to natural images. However, their statistical properties in general do not exactly follow that of natural imagery. The reconstructions are contaminated for example by speckle noise and do contain out of focus objects as it was explained in section \ref{sec:Intro}. Nonetheless, the IQMs are expected to predict perceptual quality much better than prior to reconstruction since most of them are optimized for natural images.

\begin{figure*}
\centering
	    \subfloat[MOS-OPT\label{fig:StatSigPostReconNoisy_OPT}]
	{\includegraphics[width=0.3\textwidth]{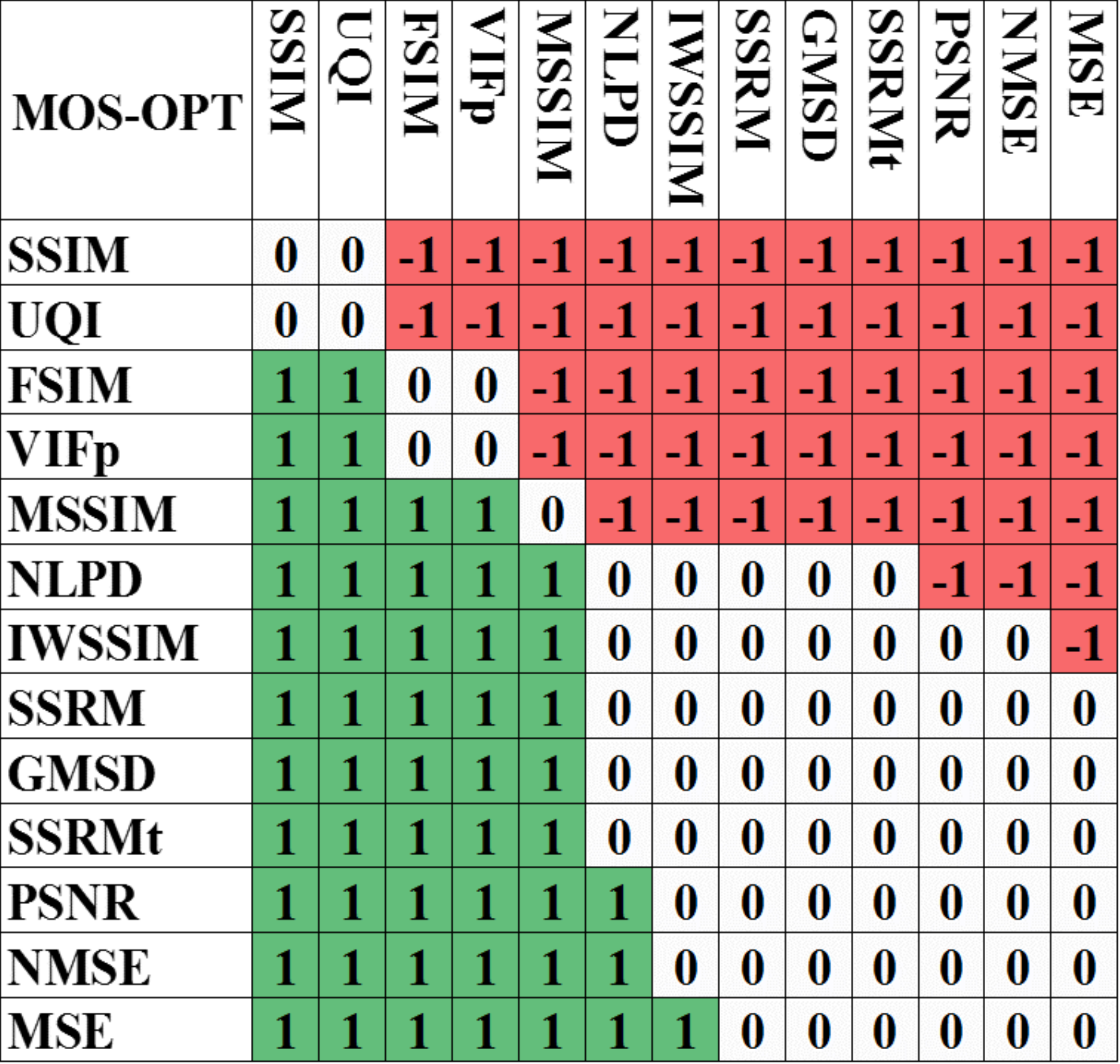}}\hspace*{0.75em}
	    \subfloat[MOS-LF\label{fig:StatSigPostReconNoisy_LF}]
	{\includegraphics[width=0.3\textwidth]{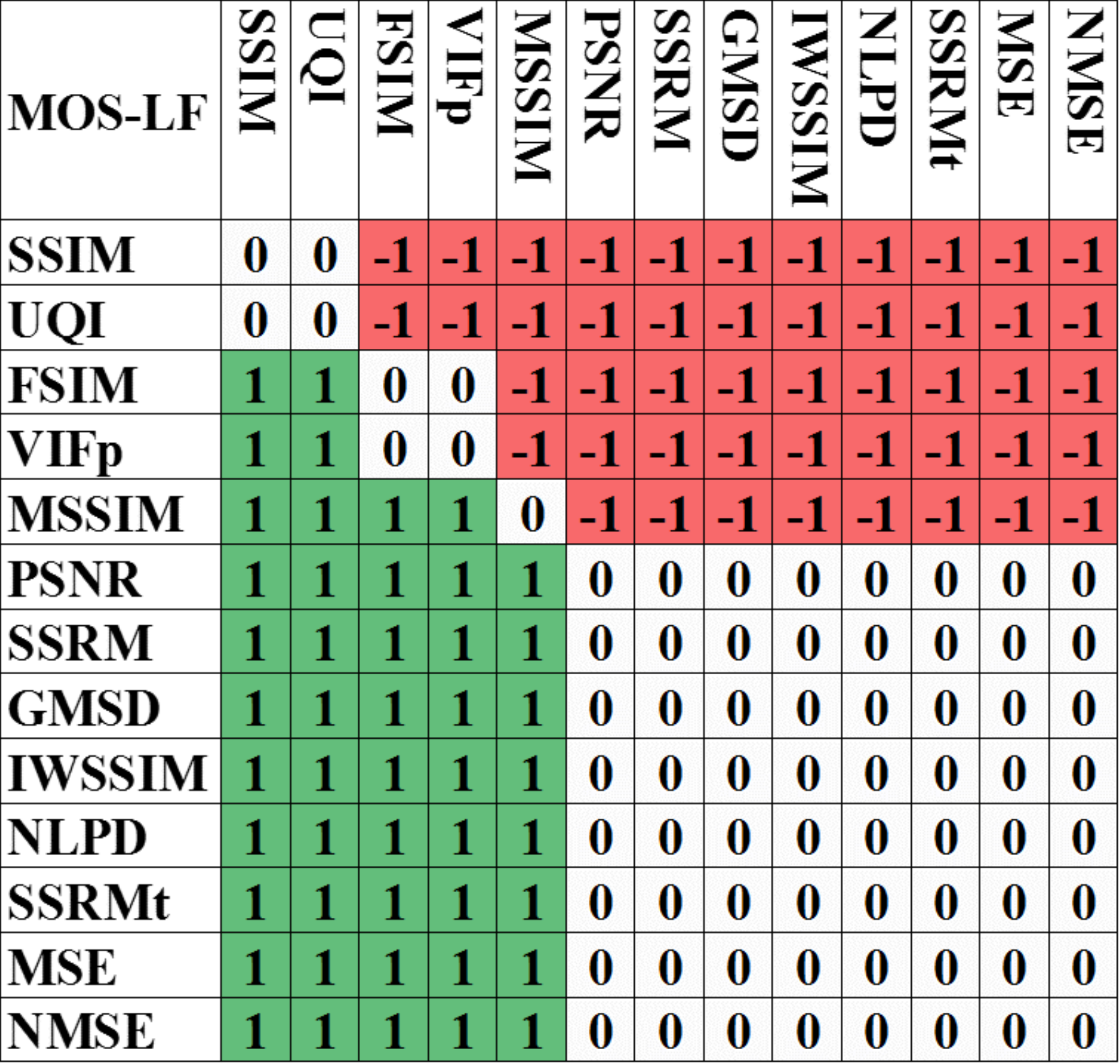}}\hspace*{0.75em}
	    \subfloat[MOS-2D\label{fig:StatSigPostReconNoisy_2D}]
	{\includegraphics[width=0.3\textwidth]{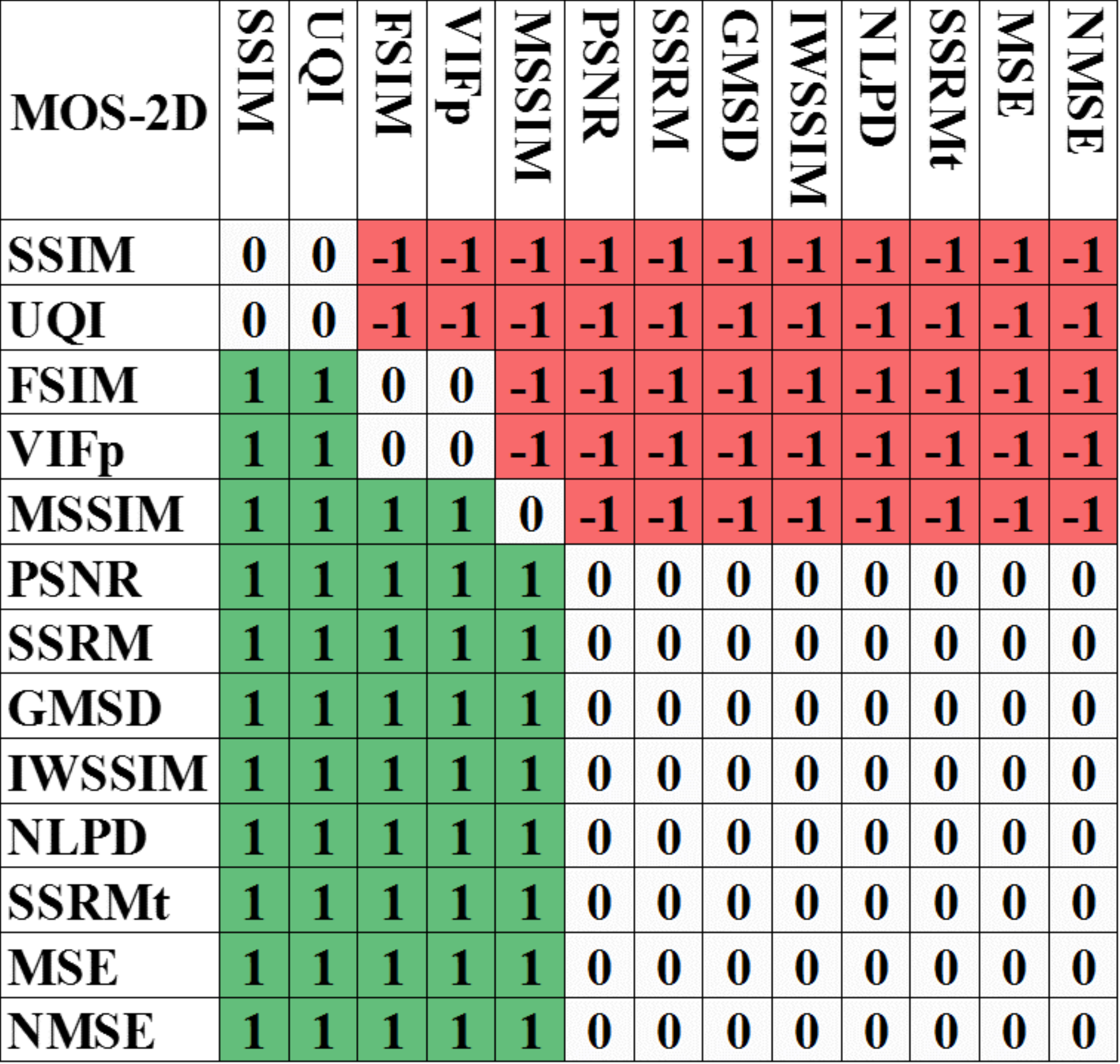}}
\caption{Statistical significance tables for the evaluation of quality metrics on the reconstructed Fourier holograms. The statistics are separately shown based on the MOS scores obtained from optical holographic display (OPT) (a), the light field display (LF) (b) and the regular 2D display (2D) (c) and the IQM scores. }
\label{fig:StatSigPostReconNoisy}
\vspace*{-1em}
\end{figure*}

For the majority of the tested IQMs we find a very competitive performance with the top ranks shared among the MSE family (MSE, NMSE, and PSNR) as well as the SSRM and SSRMt measures. However, the performances of the SSIM, UQI, and VIFp are this time however ranked worst --- being in strong contrast to their performance before rendering. The significance tables in \figref{fig:StatSigPostReconNoisy} clearly emphasize the gap between these three IQMs plus FSIM and the rest --- within which all candidates perform very similar.

\begin{table*}
\caption{Statistical evaluation of quality metrics for the reconstructed Fourier holograms. The statistics based on the MOS scores obtained from optical holographic display (OPT), the light field display (LF) and the regular 2D display (2D) and the IQM scores are separately shown.}

\resizebox{\textwidth}{!}{%
\begin{tabular}{|l|l|l|l|l|l|l|l|l|l|l|l|l|l|l|}
\hline
\multicolumn{2}{|l|}{}                           & PSNR            & SSIM   & SSRM            & FSIM   & GMSD   & IWSSIM & MS-SSIM  & UQI    & VIFp   & NLPD   & SSRMt           & MSE             & NMSE            \\ \hline
\multirow{6}{*}{\rot{\textbf{MOS-OPT}}} & SROCC        & \textbf{0.9319} & 0.5164 & 0.9218          & 0.7166 & 0.9185 & 0.9124 & 0.8394 & 0.4897 & 0.7682 & 0.9077 & \textbf{0.9319} & \textbf{0.9347} & 0.9306          \\ \cline{2-15} 
                                  & KRCC         & \textbf{0.7727} & 0.3648 & 0.7608          & 0.5322 & 0.7433 & 0.7343 & 0.6437 & 0.3468 & 0.5626 & 0.7267 & \textbf{0.7723} & \textbf{0.7771} & 0.7688          \\ \cline{2-15} 
                                  & PCC\_NoFit   & 0.8045          & 0.4847 & 0.9106          & 0.7036 & 0.897  & 0.9006 & 0.8256 & 0.5047 & 0.7002 & 0.882  & \textbf{0.9284} & 0.9267          & 0.9179          \\ \cline{2-15} 
                                  & PCC\_Fitted  & 0.9352          & 0.5044 & 0.9265          & 0.7161 & 0.9168 & 0.914  & 0.8389 & 0.5126 & 0.7628 & 0.9076 & 0.9294          & \textbf{0.9374} & 0.9341          \\ \cline{2-15} 
                                  & RMSE         & 0.3981          & 0.9707 & 0.4229          & 0.7847 & 0.4488 & 0.456  & 0.6118 & 0.9652 & 0.7268 & 0.4719 & 0.4149          & \textbf{0.3916} & 0.4012          \\ \cline{2-15} 
                                  & OutlierRatio & 0.5156          & 0.875  & \textbf{0.474}  & 0.6875 & 0.5521 & 0.6094 & 0.6458 & 0.8333 & 0.7083 & 0.5677 & 0.5208          & 0.5417          & 0.5365          \\ \hline
\multirow{6}{*}{\rot{\textbf{MOS-LF}}}  & SROCC        & 0.9343          & 0.5201 & 0.9333          & 0.758  & 0.9293 & 0.9175 & 0.8452 & 0.4383 & 0.7729 & 0.9278 & 0.9223          & \textbf{0.9378} & \textbf{0.9379} \\ \cline{2-15} 
                                  & KRCC         & 0.7824          & 0.3782 & 0.7801          & 0.5672 & 0.7787 & 0.7475 & 0.6742 & 0.3004 & 0.5725 & 0.7676 & 0.7673          & \textbf{0.7878} & \textbf{0.7878} \\ \cline{2-15} 
                                  & PCC\_NoFit   & 0.8609          & 0.5042 & \textbf{0.9293} & 0.7751 & 0.9249 & 0.915  & 0.847  & 0.4528 & 0.7476 & 0.9229 & 0.9131          & 0.9058          & 0.9031          \\ \cline{2-15} 
                                  & PCC\_Fitted  & 0.9309          & 0.5266 & 0.9302          & 0.778  & 0.9291 & 0.9166 & 0.8511 & 0.4614 & 0.7753 & 0.9289 & 0.9168          & 0.9338          & \textbf{0.9361} \\ \cline{2-15} 
                                  & RMSE         & 0.3808          & 0.8862 & 0.3827          & 0.655  & 0.3856 & 0.4169 & 0.5473 & 0.9249 & 0.6584 & 0.3861 & 0.4162          & 0.3731          & \textbf{0.3668} \\ \cline{2-15} 
                                  & OutlierRatio & 0.5469          & 0.9167 & \textbf{0.5104} & 0.7448 & 0.526  & 0.5781 & 0.7292 & 0.7969 & 0.6979 & 0.5573 & \textbf{0.5104} & 0.5417          & 0.5417          \\ \hline
\multirow{6}{*}{\rot{\textbf{MOS-2D}}}  & SROCC        & 0.9312          & 0.5496 & 0.9284          & 0.8154 & 0.9276 & 0.9204 & 0.8655 & 0.4161 & 0.8034 & 0.922  & 0.9178          & \textbf{0.9344} & 0.9314          \\ \cline{2-15} 
                                  & KRCC         & 0.7737          & 0.3923 & 0.773           & 0.6315 & 0.776  & 0.7523 & 0.6996 & 0.2855 & 0.6015 & 0.7569 & 0.7601          & \textbf{0.7804} & 0.7771          \\ \cline{2-15} 
                                  & PCC\_NoFit   & 0.875           & 0.5455 & \textbf{0.9304} & 0.7988 & 0.9299 & 0.9166 & 0.8611 & 0.443  & 0.7782 & 0.923  & 0.9104          & 0.8891          & 0.8827          \\ \cline{2-15} 
                                  & PCC\_Fitted  & 0.9334          & 0.5617 & 0.9312          & 0.8162 & 0.9341 & 0.9226 & 0.8697 & 0.4435 & 0.8069 & 0.9265 & 0.9275          & \textbf{0.9357} & 0.9345          \\ \cline{2-15} 
                                  & RMSE         & 0.3966          & 0.9143 & 0.4028          & 0.6385 & 0.3945 & 0.4262 & 0.5454 & 0.9905 & 0.6528 & 0.4158 & 0.4132          & \textbf{0.3898} & 0.3933          \\ \cline{2-15} 
                                  & OutlierRatio & 0.5417          & 0.875  & \textbf{0.4688} & 0.6667 & 0.474  & 0.5573 & 0.6406 & 0.8333 & 0.7344 & 0.5469 & \textbf{0.4688} & 0.5156          & 0.526           \\ \hline
\end{tabular}}
\label{tab:CorrsPostReconNoisy}%
\end{table*}

\subsubsection{QA\_4 - Evaluation after speckle denoising of reconstructed Fourier holograms}
\label{sec:PostReconNoSpeckle}
\begin{figure*}
\centering
	    \subfloat[ MOS-OPT]
	{\includegraphics[width=0.3\textwidth]{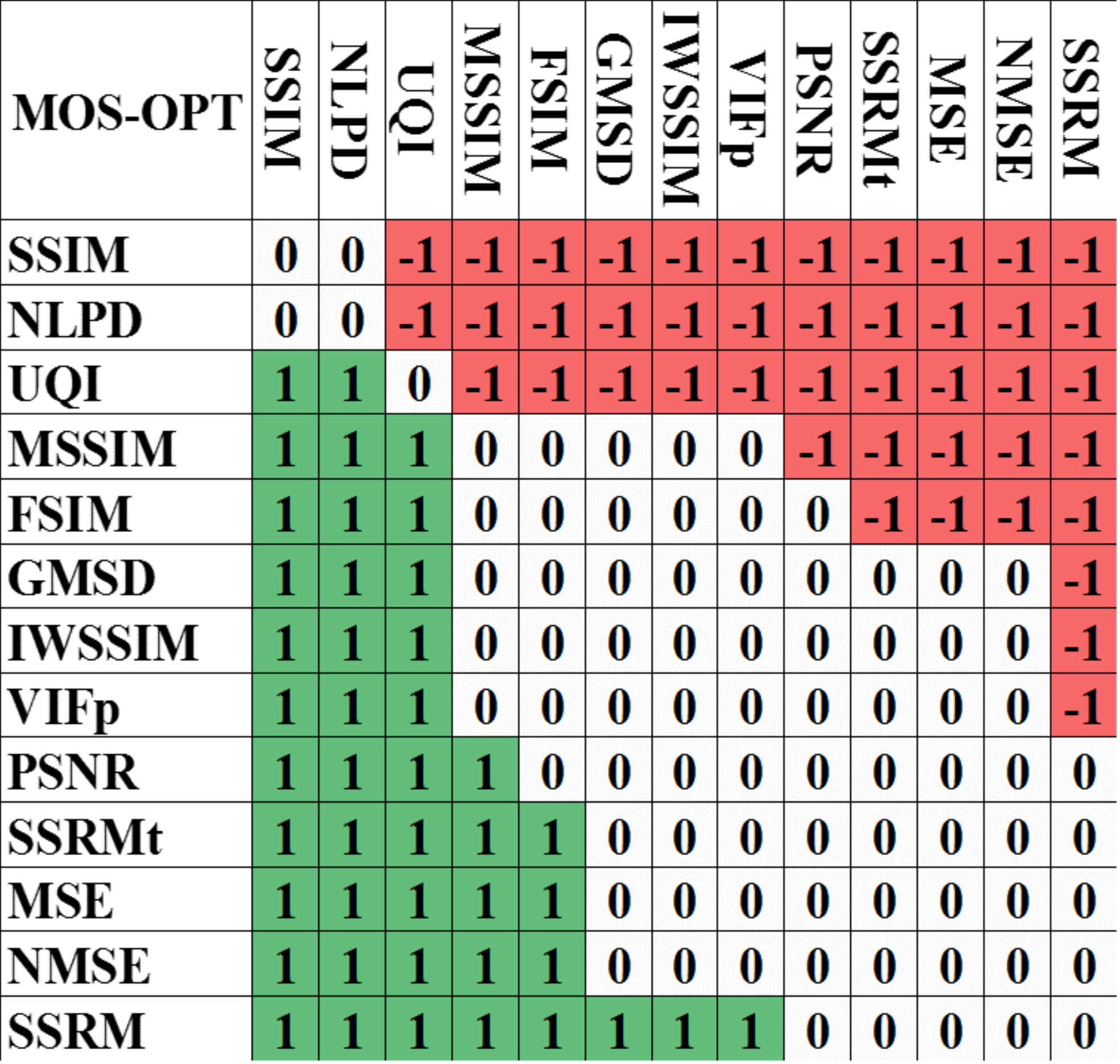}}\hspace*{0.75em}
	    \subfloat[ MOS-LF]
	{\includegraphics[width=0.3\textwidth]{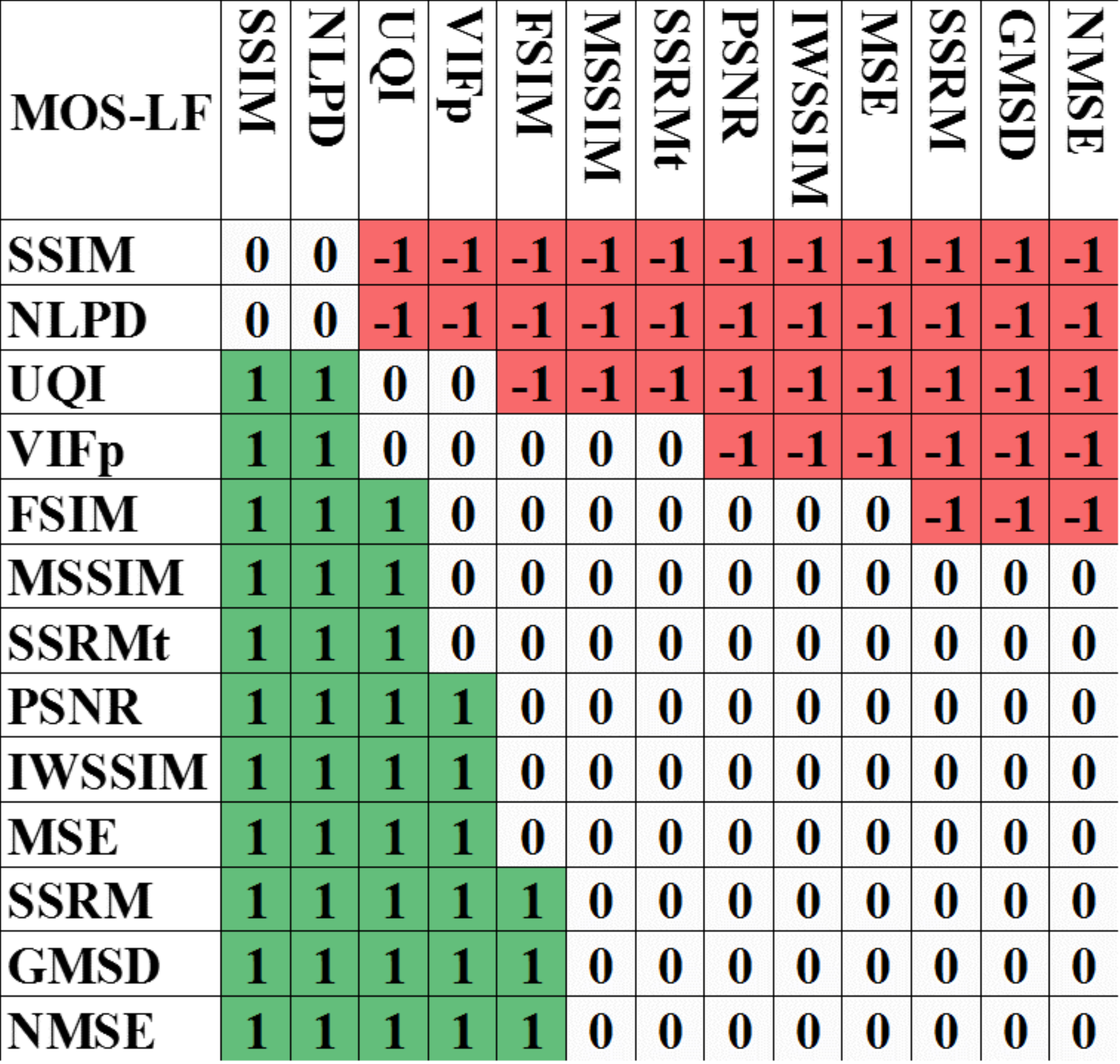}}\hspace*{0.75em}
	    \subfloat[ MOS-2D]
	{\includegraphics[width=0.3\textwidth]{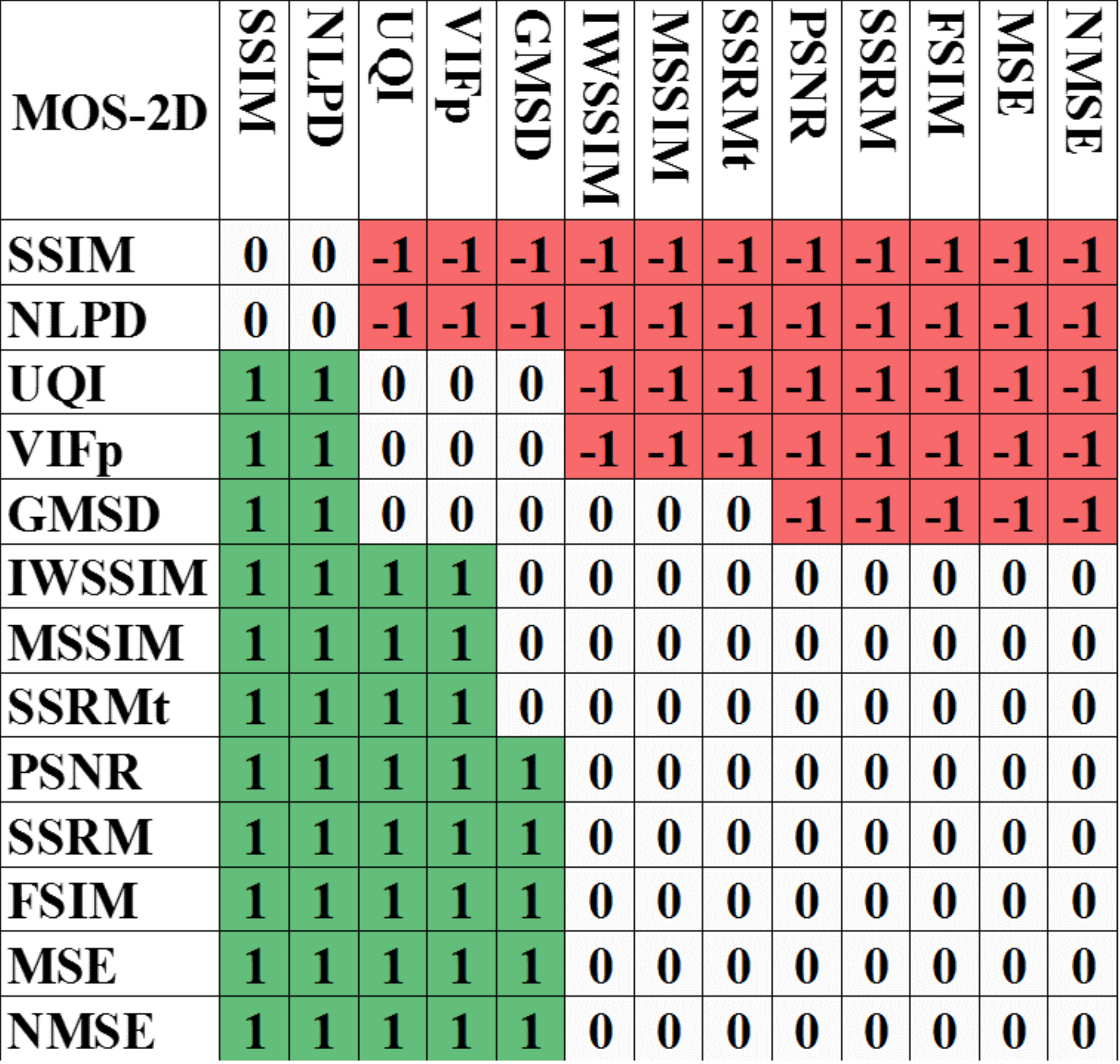}}
\caption{Statistical significance tables for the evaluation of quality metrics on the reconstructed Fourier holograms after speckle denoising. The statistics based on the MOS scores obtained from optical holographic display (OPT) (a), the light field display (LF) (b) and the regular 2D display (2D) (c) and the IQM scores, are separately shown. }
\label{fig:StatSigPostReconNoSpeckle}
\vspace*{-1em}
\end{figure*}

\begin{figure*}
\centering
	{\includegraphics[width=0.25\textwidth]{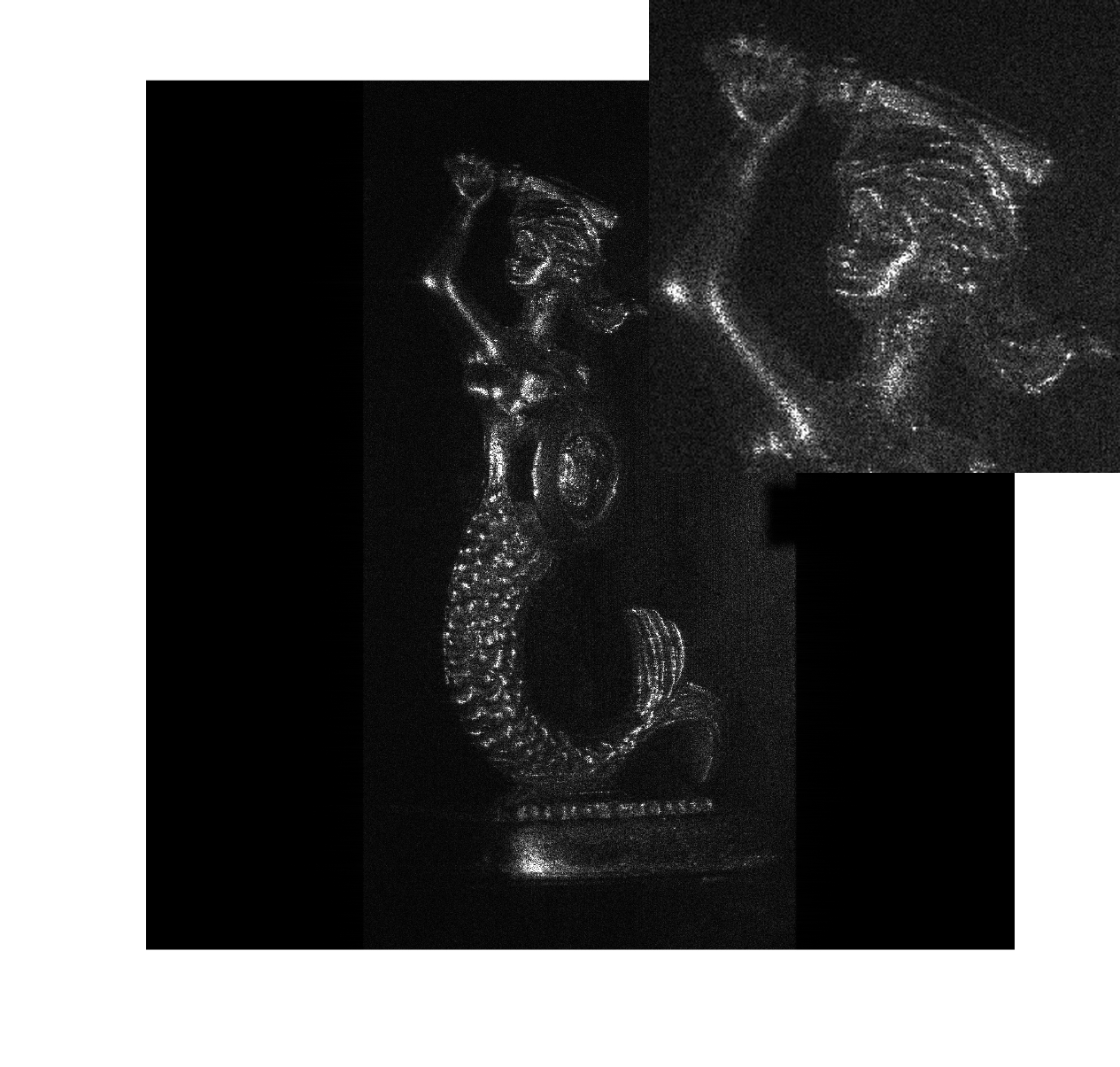}}\hfill
	{\includegraphics[width=0.25\textwidth]{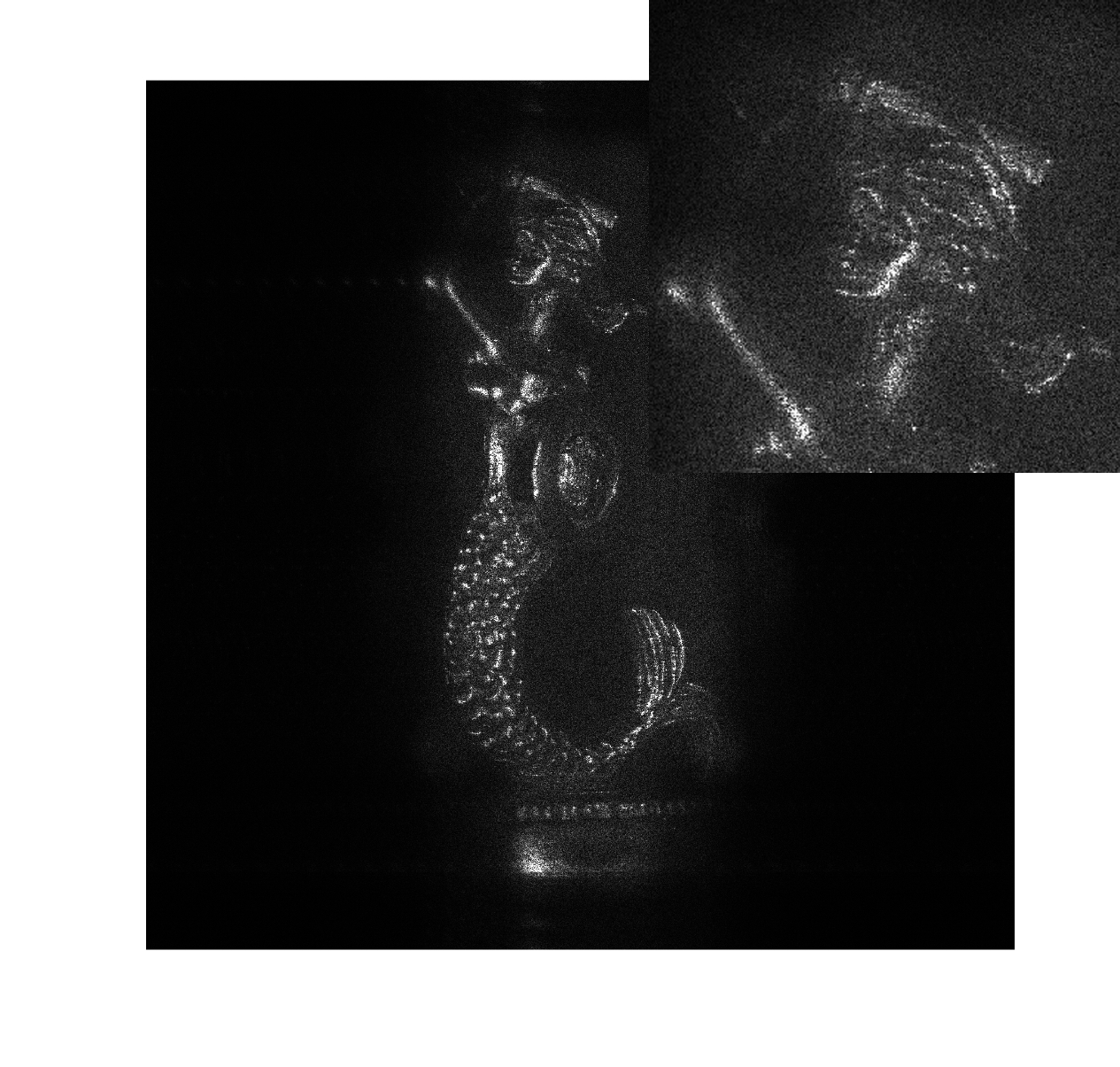}}\hfill
	{\includegraphics[width=0.25\textwidth]{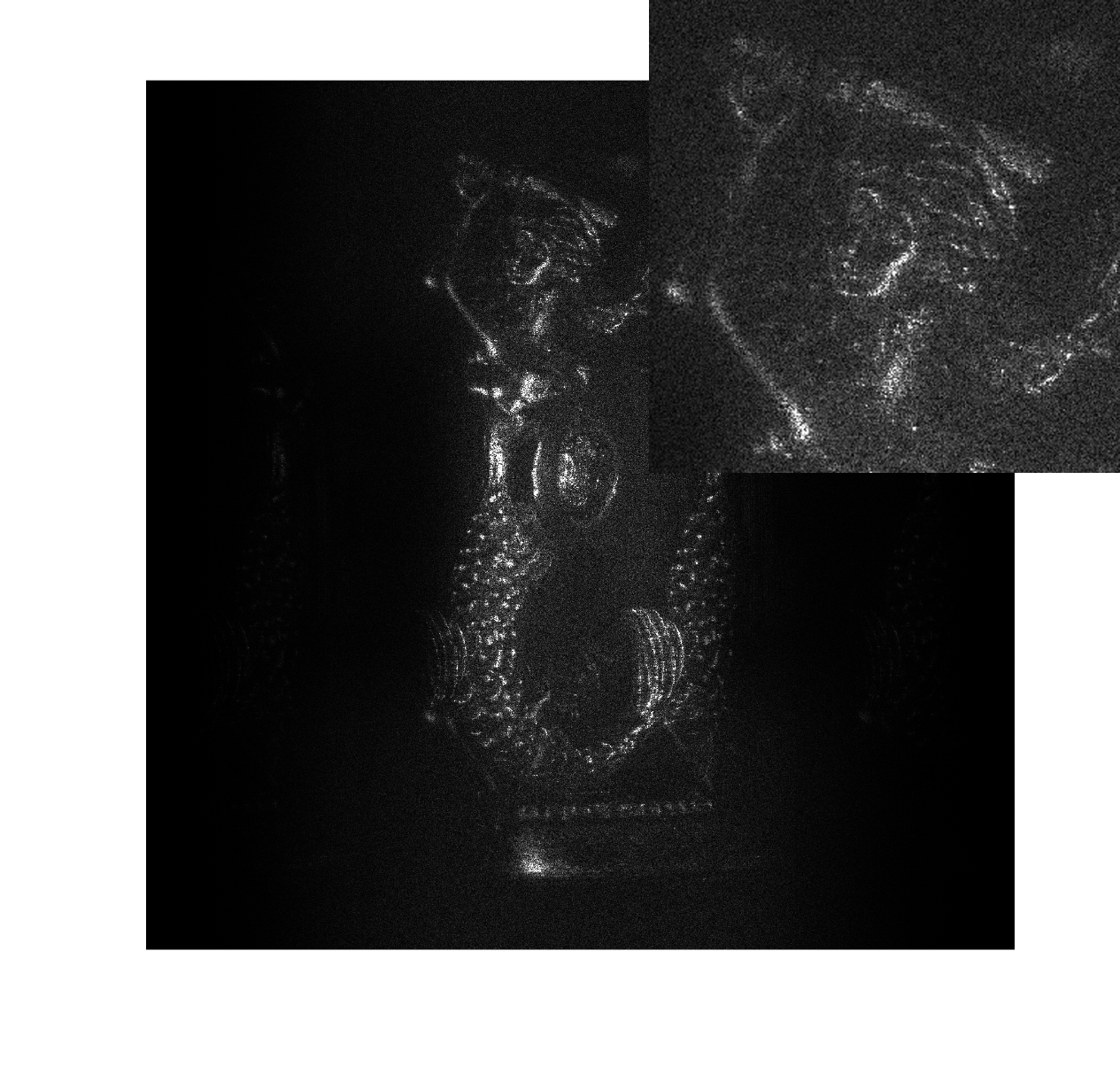}}\hfill
	{\includegraphics[width=0.25\textwidth]{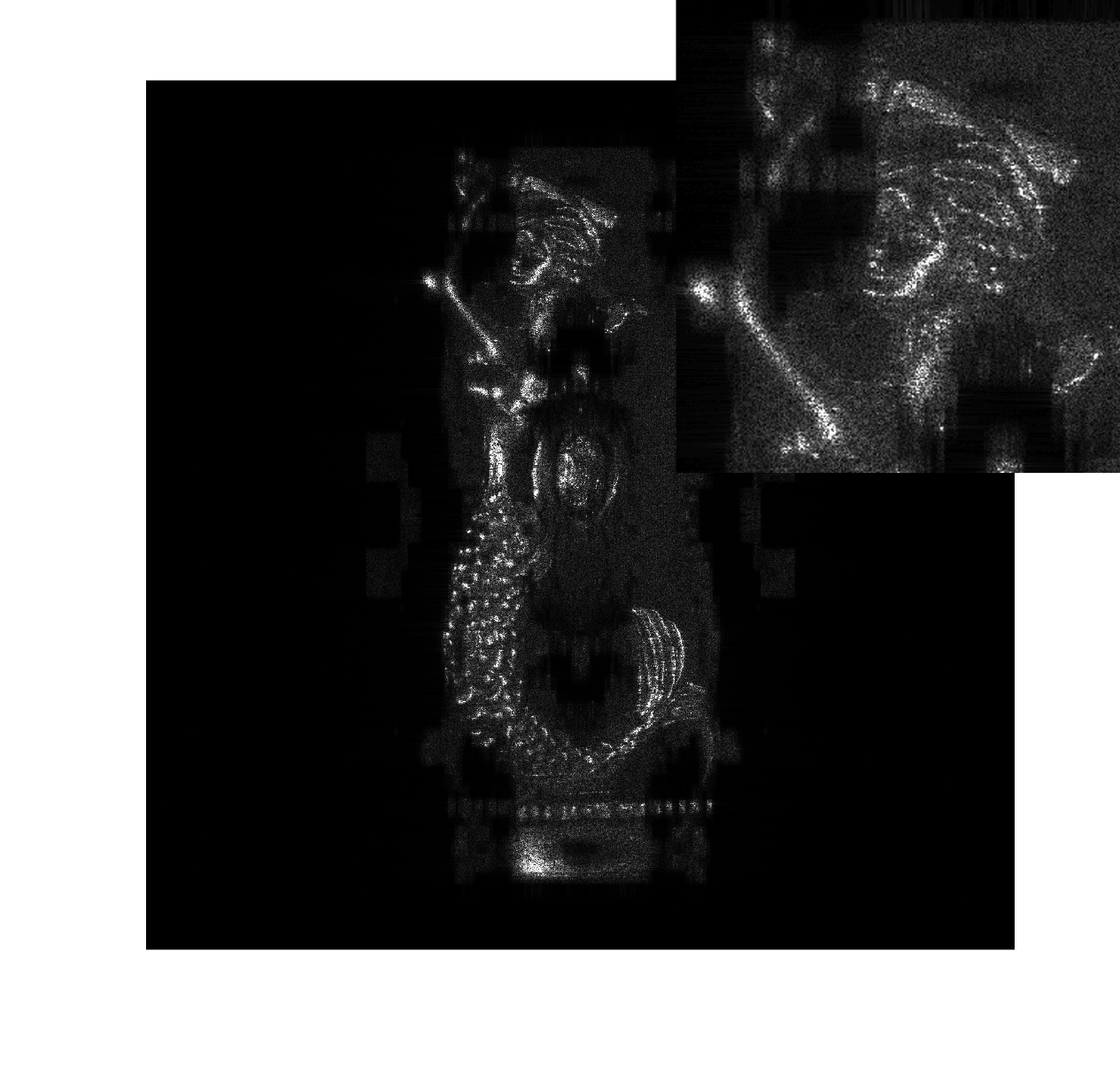}}
				
	\vspace*{-1em}
	    \subfloat[Ref 8~bpp]
	{\includegraphics[width=0.25\textwidth]{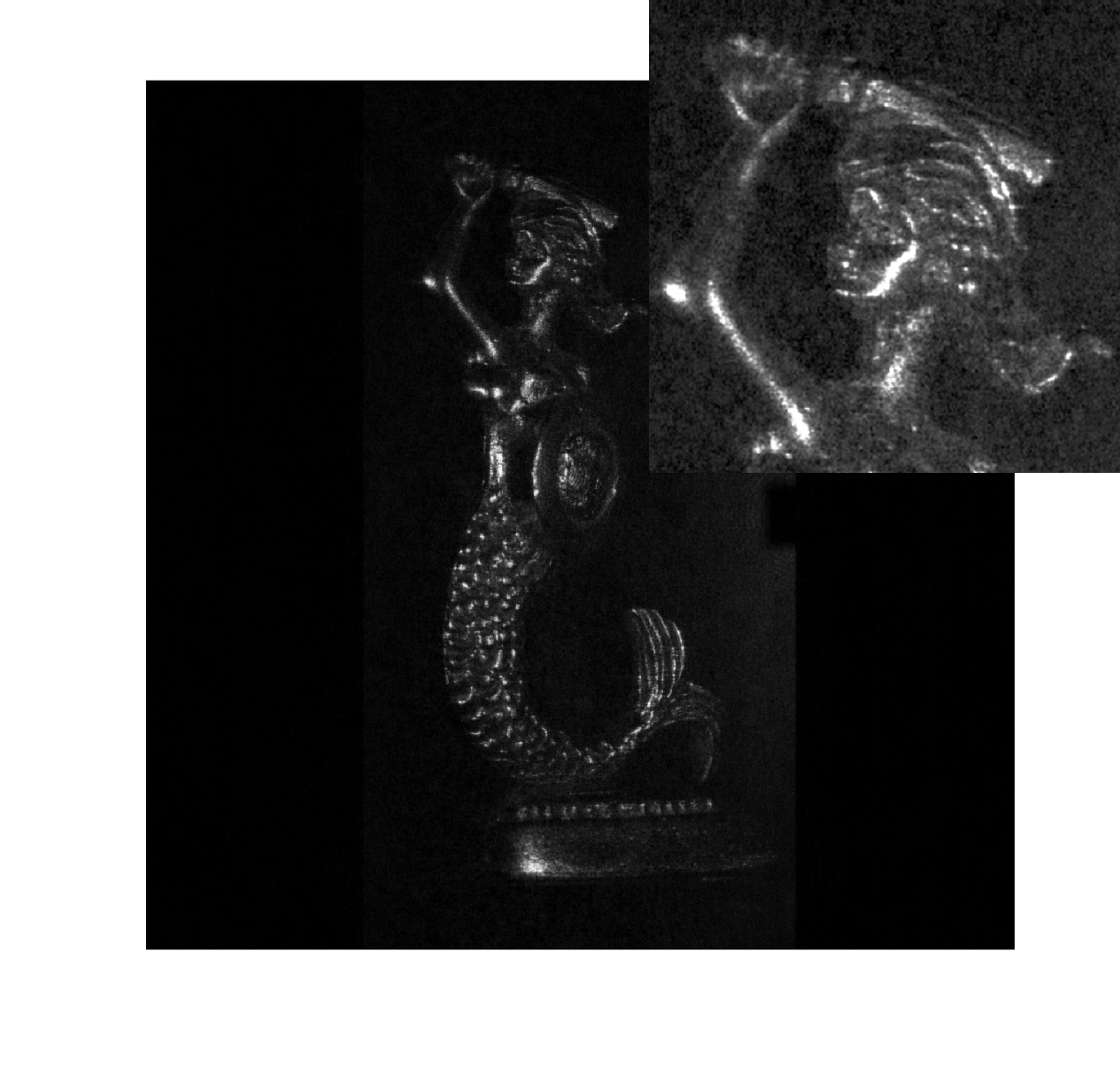}}\hfill
	    \subfloat[HEVC 0.75~bpp]
	{\includegraphics[width=0.25\textwidth]{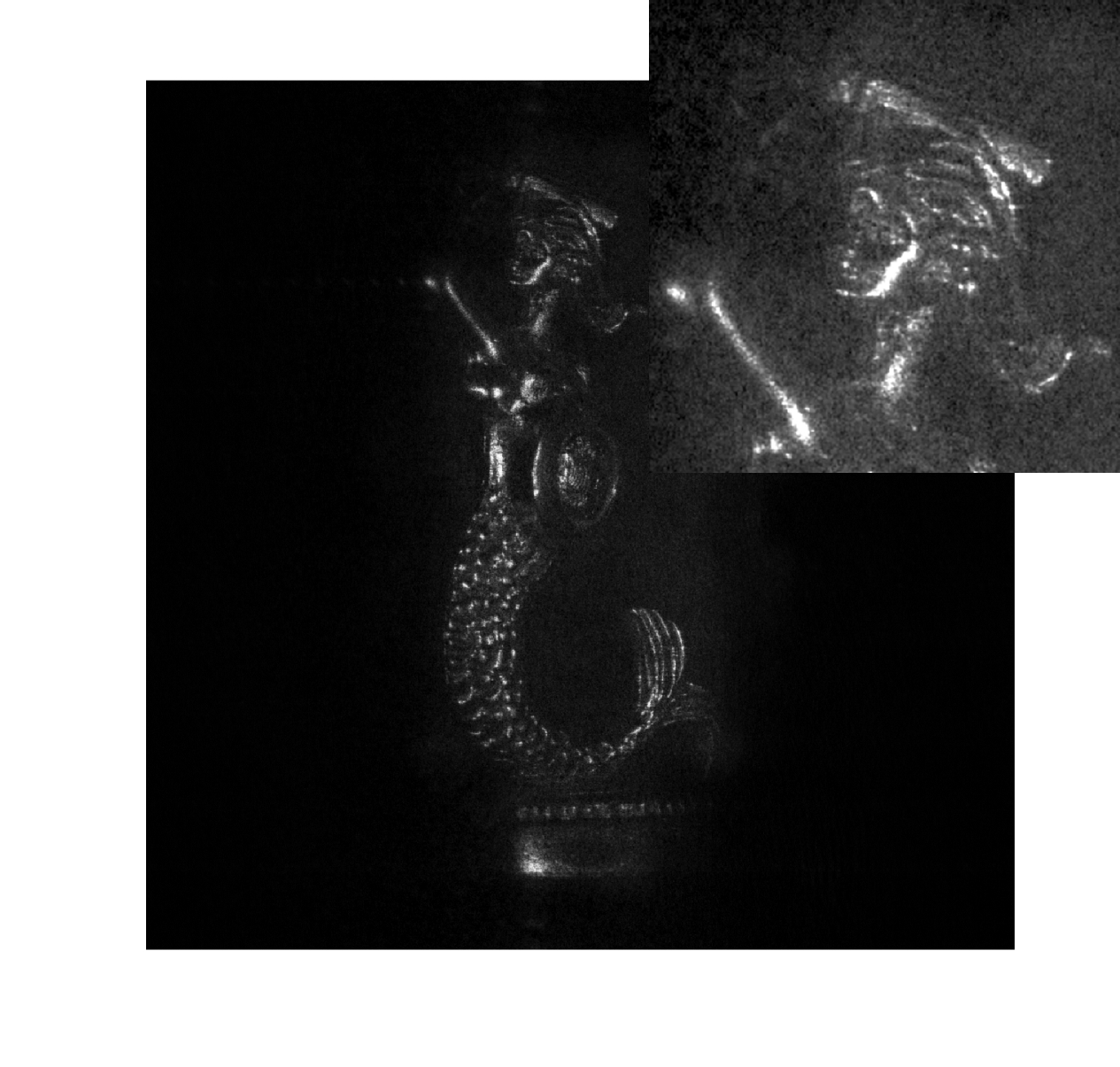}}\hfill
	    \subfloat[JP2K 0.75~bpp]
	{\includegraphics[width=0.25\textwidth]{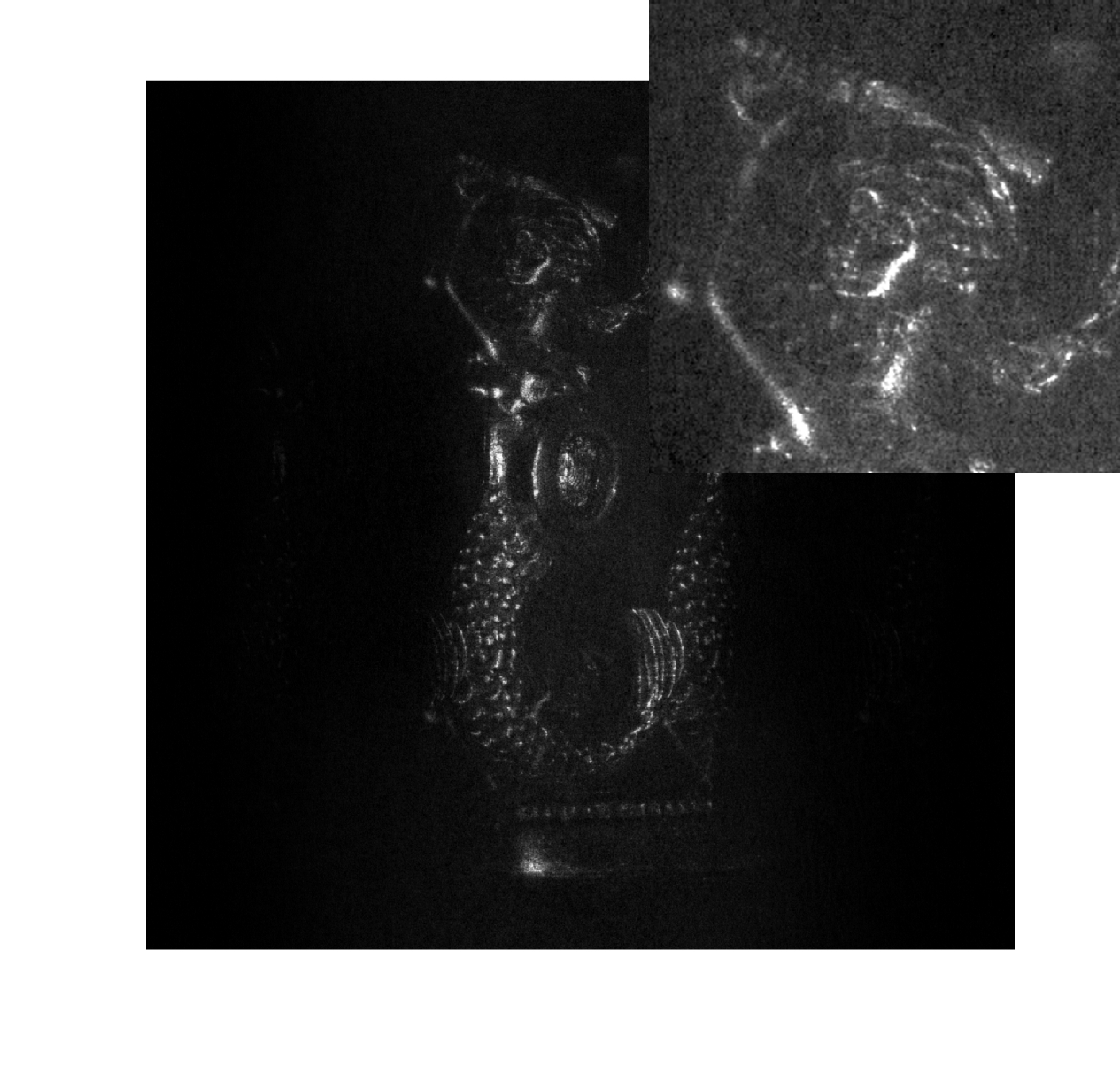}}\hfill
	    \subfloat[WAC 0.75~bpp]
	{\includegraphics[width=0.25\textwidth]{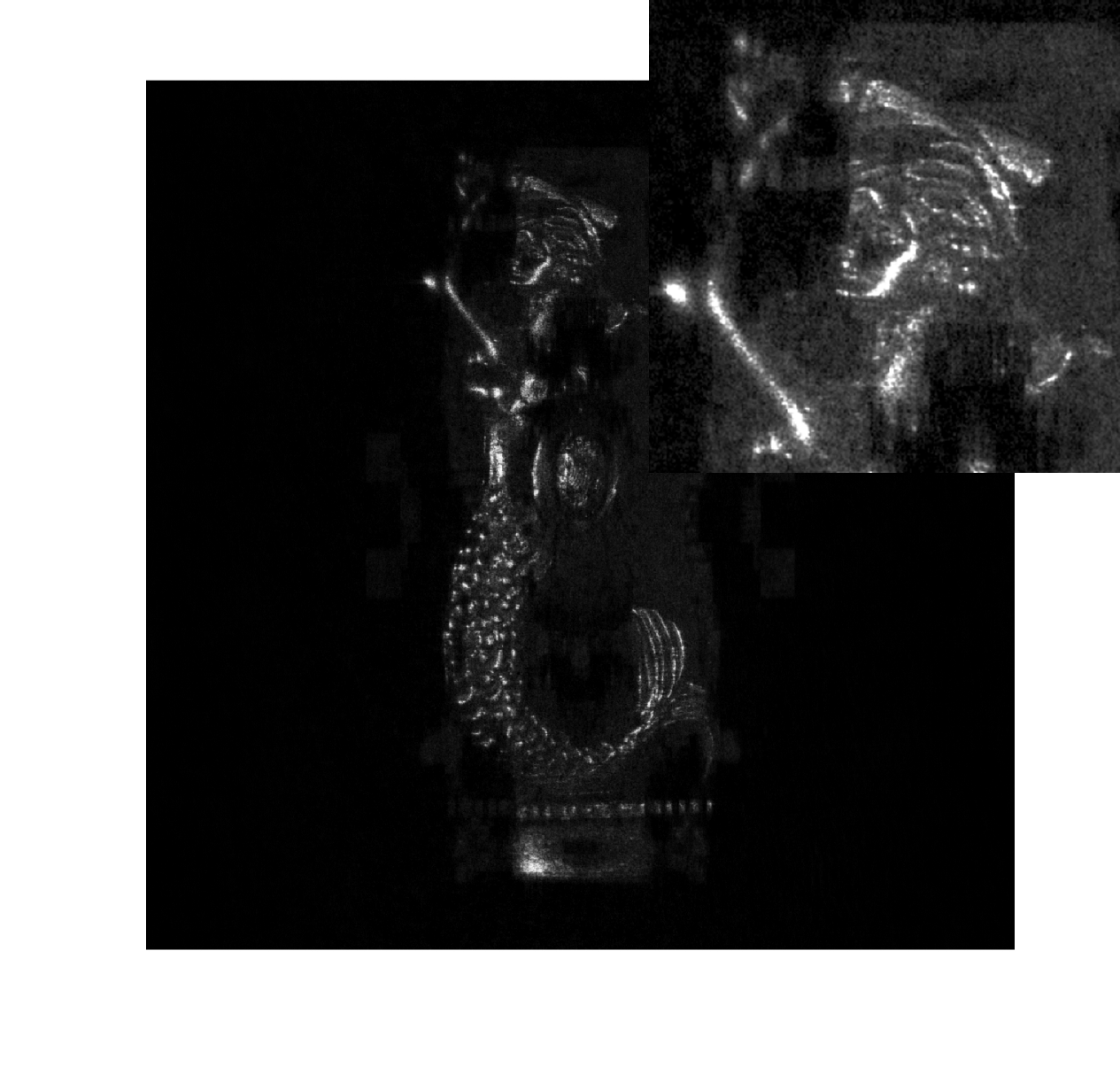}}
	
\caption{Exemplary numerical reconstruction of OR-Mermaid from center view. First row: Reference, encoded with HEVC, JP2K and WAC at 0.75~bpp, before speckle denoising(QA\_3). Second row: same as first row, after speckle denoising(QA\_4) }
\label{fig:SpeckleEx}%
\vspace*{-1em}
\end{figure*}

In this test track we try to address the concern regarding potential effect of the speckle noise on the IQM performances. To evaluate, we denoised the reconstructions via a two-dimensional Wigner filter. The same parameters for denoising the reconstructions of the reference and the compressed holograms were used. \figref{fig:SpeckleEx} demonstrates and exemplary case before and after the denoising. \tabref{tab:CorrsPostReconNoSpeckle} demonstrates the evaluation results for the IQMs tested on the denoised reconstructions. With the denoised reconstructions being much more similar to natural images, the Fourier transform-based SSRM dominates once more the top spots w.r.t most of the evaluation criteria; closely followed by its twin version the SSRMt, and the MSE family (NMSE, MSE, and PSNR). When compared to the results of \tabref{tab:CorrsPostReconNoisy}, we notice that on one side the top performing IQMs before denoising do not experience a significant improvement or degradation, while on the other side almost all of the remaining IQMs clearly step up their game, closing the gap. Solely the NLPD can be considered an outlier as it significantly drops in performance after the denoising step. One reason for this can be the sensitivity of the NLPD to the modifications on the image gradient which is exactly what a Wigner filter would flatten out and hence potentially directly impacting the prediction performance of NLPD. Although, the exact reason for such behaviour is not known to us. The statistical significance tables in \figref{fig:StatSigPostReconNoSpeckle} also emphasize our observations.

\begin{table*}
\caption{Statistical evaluation of quality metrics for the reconstructed Fourier holograms after speckle noise removal. The statistics based on the MOS scores obtained from optical holographic display (OPT), the light field display (LF) and the regular 2D display (2D) and the IQM scores are separately shown.}
\resizebox{\textwidth}{!}{%
\begin{tabular}{|l|l|l|l|l|l|l|l|l|l|l|l|l|l|l|}
\hline
\multicolumn{2}{|l|}{}                           & PSNR   & SSIM   & SSRM            & FSIM            & GMSD   & IWSSIM & MS-SSIM  & UQI    & VIFp   & NLPD   & SSRMt           & MSE             & NMSE            \\ \hline
\multirow{6}{*}{\rot{\textbf{MOS-OPT}}} & SROCC        & 0.919  & 0.675  & \textbf{0.9272} & 0.9011          & 0.9022 & 0.9056 & 0.8867 & 0.8486 & 0.8926 & 0.7516 & 0.9261          & 0.919           & 0.9217          \\ \cline{2-15} 
                                  & KRCC         & 0.7464 & 0.4974 & \textbf{0.7661} & 0.7144          & 0.7181 & 0.7256 & 0.6999 & 0.6472 & 0.7144 & 0.5551 & \textbf{0.7618} & 0.7457          & 0.7473          \\ \cline{2-15} 
                                  & PCC\_NoFit   & 0.8308 & 0.6702 & \textbf{0.9162} & 0.8819          & 0.8797 & 0.8861 & 0.8589 & 0.7434 & 0.7443 & 0.7228 & \textbf{0.921}  & 0.8852          & 0.8845          \\ \cline{2-15} 
                                  & PCC\_Fitted  & 0.9208 & 0.6719 & \textbf{0.93}   & 0.9035          & 0.9006 & 0.9032 & 0.8812 & 0.8369 & 0.8932 & 0.7343 & 0.9223          & 0.9222          & 0.9238          \\ \cline{2-15} 
                                  & RMSE         & 0.4383 & 0.8326 & \textbf{0.4133} & 0.4817          & 0.4887 & 0.4824 & 0.5314 & 0.6152 & 0.5054 & 0.7631 & 0.4343          & 0.4347          & 0.4303          \\ \cline{2-15} 
                                  & OutlierRatio & 0.5469 & 0.7396 & \textbf{0.5156} & 0.6563          & 0.599  & 0.5521 & 0.5833 & 0.651  & 0.6198 & 0.75   & \textbf{0.5156} & 0.5417          & 0.5417          \\ \hline
\multirow{6}{*}{\rot{\textbf{MOS-LF}}}  & SROCC        & 0.9271 & 0.7145 & \textbf{0.9282} & 0.9017          & 0.9211 & 0.9136 & 0.901  & 0.8647 & 0.885  & 0.7235 & 0.9069          & \textbf{0.9283} & \textbf{0.9313} \\ \cline{2-15} 
                                  & KRCC         & 0.7681 & 0.5413 & \textbf{0.7713} & 0.7217          & 0.7629 & 0.7452 & 0.7348 & 0.6781 & 0.6992 & 0.532  & 0.7458          & 0.7699          & \textbf{0.7718} \\ \cline{2-15} 
                                  & PCC\_NoFit   & 0.8801 & 0.7315 & \textbf{0.9241} & 0.8993          & 0.9102 & 0.9094 & 0.8908 & 0.8107 & 0.8031 & 0.7109 & 0.8983          & 0.8748          & 0.8729          \\ \cline{2-15} 
                                  & PCC\_Fitted  & 0.9261 & 0.7338 & 0.9243          & 0.907           & 0.9224 & 0.9145 & 0.9021 & 0.8613 & 0.8813 & 0.7427 & 0.9018          & 0.927           & \textbf{0.9296} \\ \cline{2-15} 
                                  & RMSE         & 0.3934 & 0.7083 & 0.3978          & 0.4391          & 0.4027 & 0.4218 & 0.4499 & 0.5297 & 0.4927 & 0.6981 & 0.4504          & 0.391           & \textbf{0.3842} \\ \cline{2-15} 
                                  & OutlierRatio & 0.6198 & 0.849  & \textbf{0.5365} & 0.6771          & 0.4896 & 0.5521 & 0.6302 & 0.6406 & 0.6458 & 0.7135 & 0.5677          & 0.6094          & 0.5781          \\ \hline
\multirow{6}{*}{\rot{\textbf{MOS-2D}}}  & SROCC        & 0.9306 & 0.7532 & 0.9262          & \textbf{0.9312} & 0.9014 & 0.9225 & 0.9107 & 0.8788 & 0.8924 & 0.722  & 0.9033          & \textbf{0.9312} & 0.9307          \\ \cline{2-15} 
                                  & KRCC         & 0.776  & 0.5743 & 0.7688          & 0.7733          & 0.7325 & 0.7597 & 0.749  & 0.6967 & 0.7044 & 0.5325 & 0.7365          & \textbf{0.7786} & 0.7771          \\ \cline{2-15} 
                                  & PCC\_NoFit   & 0.8997 & 0.7703 & \textbf{0.9283} & 0.918           & 0.8896 & 0.9141 & 0.8918 & 0.8464 & 0.8268 & 0.7286 & 0.8989          & 0.8552          & 0.8506          \\ \cline{2-15} 
                                  & PCC\_Fitted  & 0.9354 & 0.7704 & 0.9302          & \textbf{0.9395} & 0.9106 & 0.9277 & 0.9157 & 0.8911 & 0.8952 & 0.7831 & 0.9151          & 0.9353          & 0.9354          \\ \cline{2-15} 
                                  & RMSE         & 0.3909 & 0.7046 & 0.4057          & \textbf{0.3786} & 0.4567 & 0.4125 & 0.4442 & 0.5015 & 0.4925 & 0.6873 & 0.4456          & 0.3911          & 0.3908          \\ \cline{2-15} 
                                  & OutlierRatio & 0.4896 & 0.7292 & \textbf{0.4427} & 0.5156          & 0.6042 & 0.5    & 0.5625 & 0.6302 & 0.6823 & 0.724  & 0.5052          & 0.4688          & 0.474           \\ \hline
\end{tabular}}
\label{tab:CorrsPostReconNoSpeckle}%
\end{table*}

\subsection{Global analysis}
\label{sec:InterPipelineComps}
\begin{figure*}
\centering
	    \subfloat[$FSIM$\label{fig:Overall4s_FSIM}]
	{\includegraphics[width=0.25\textwidth]{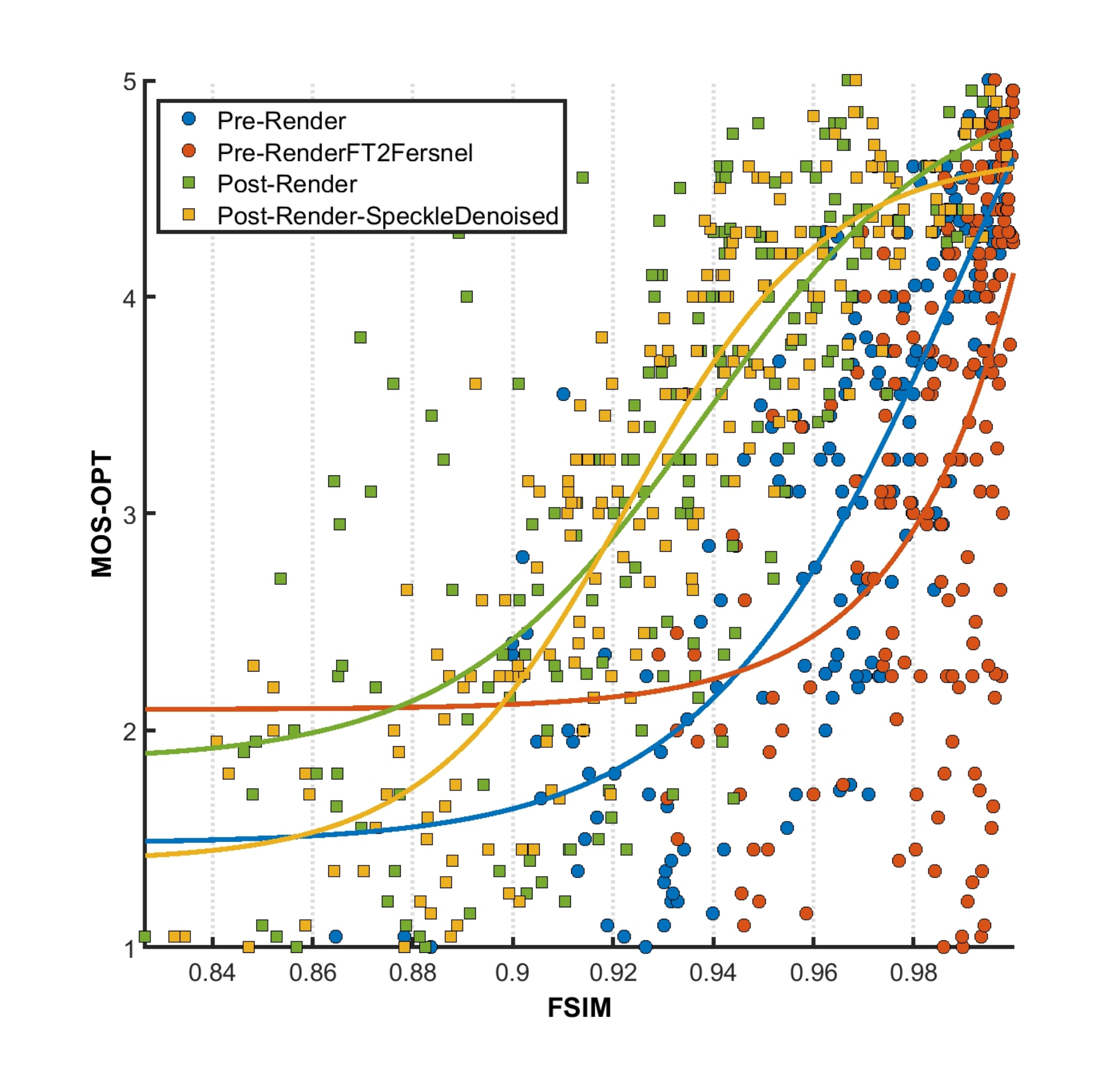}}
	    \subfloat[$GMSD$\label{fig:Overall4s_GMSD}]
	{\includegraphics[width=0.25\textwidth]{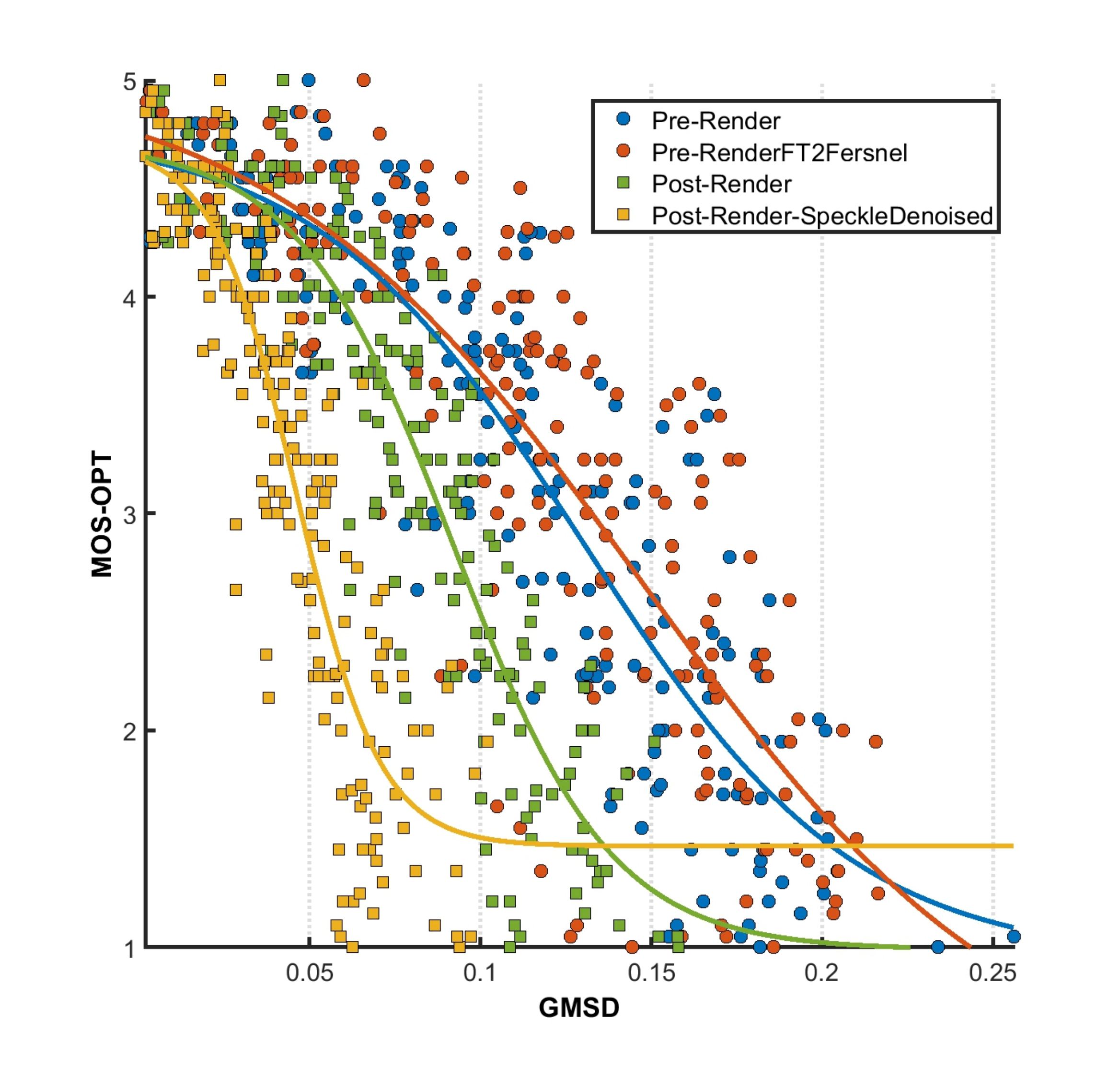}}
	    \subfloat[$IWSSIM$\label{fig:Overall4s_IWSSIM}]
	{\includegraphics[width=0.25\textwidth]{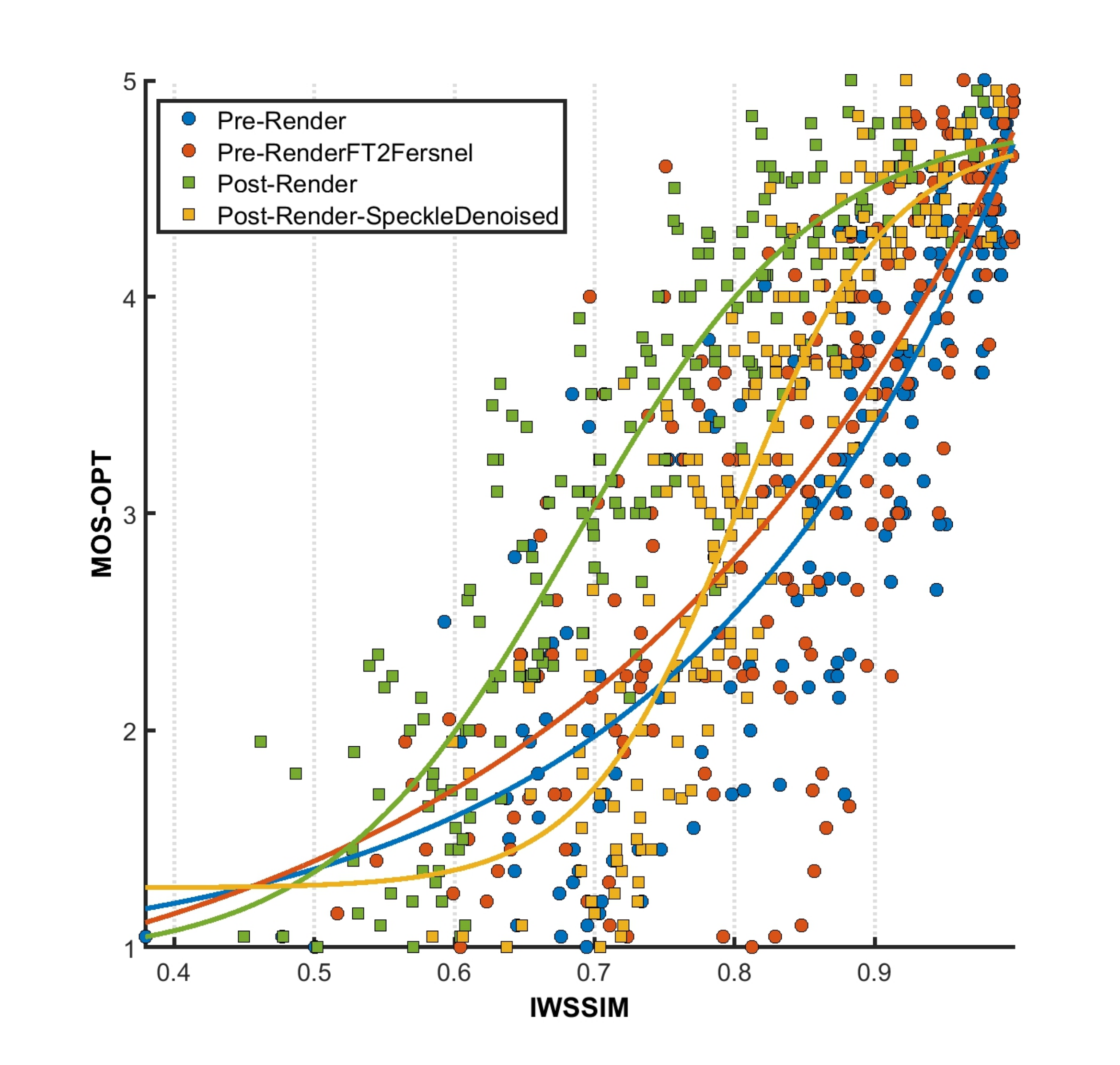}}
	    \subfloat[$MSE$\label{fig:Overall4s_MSE}]
	{\includegraphics[width=0.25\textwidth]{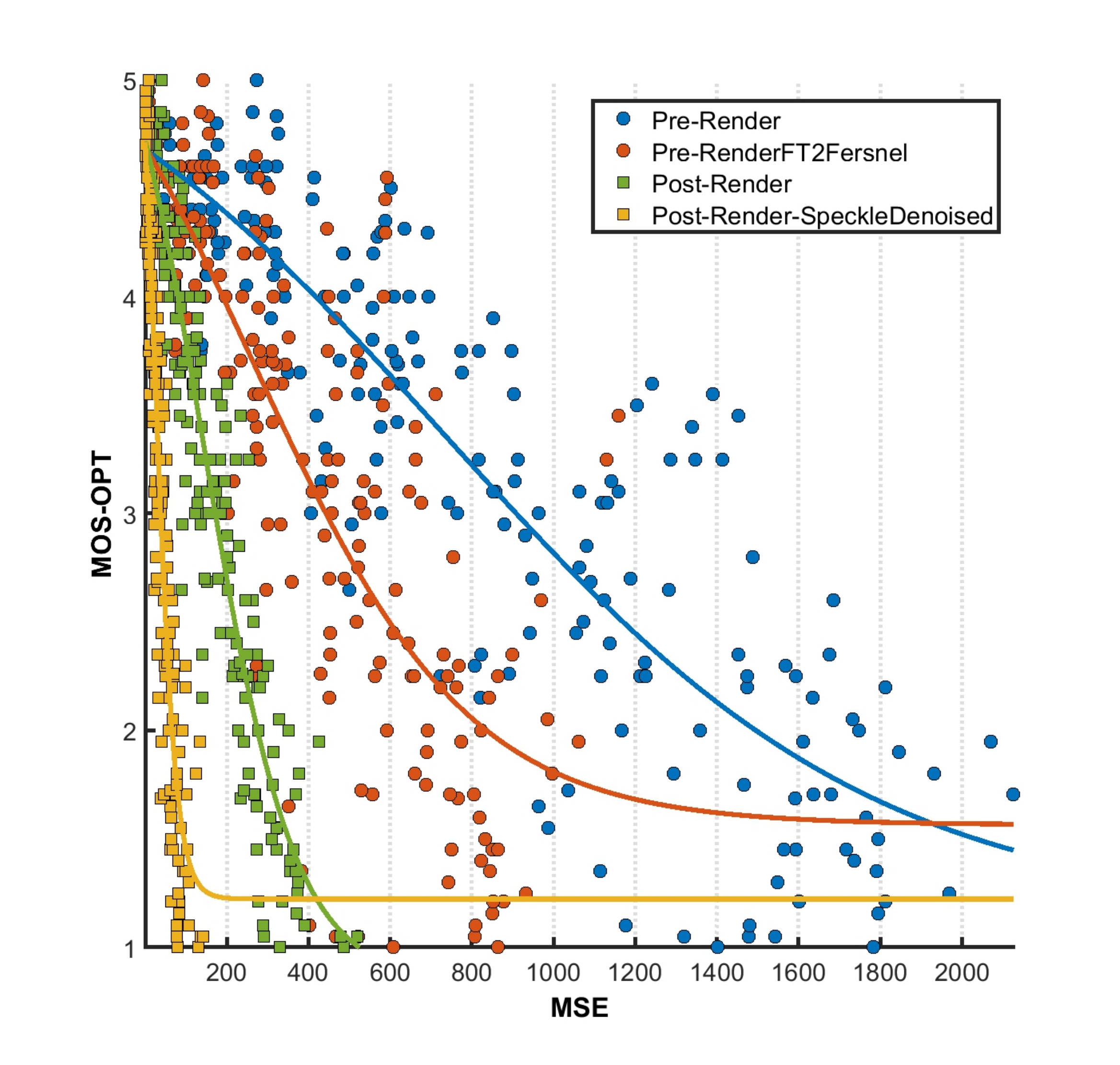}}
				
	\vspace*{-1em}
	    \subfloat[$MS-SSIM$\label{fig:Overall4s_MS-SSIM}]
	{\includegraphics[width=0.25\textwidth]{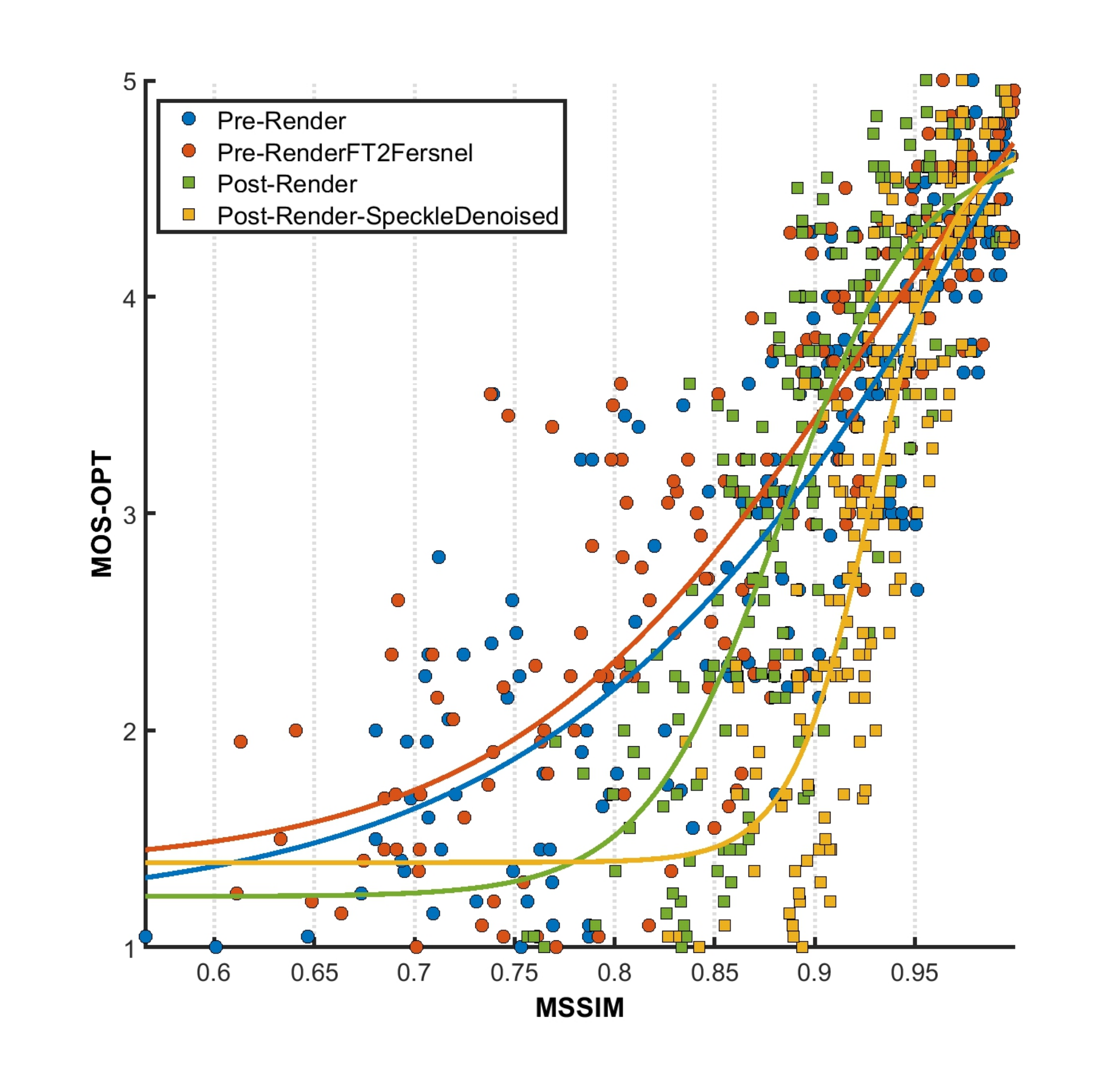}}
	    \subfloat[$NLPD$\label{fig:Overall4s_NLPD}]
	{\includegraphics[width=0.25\textwidth]{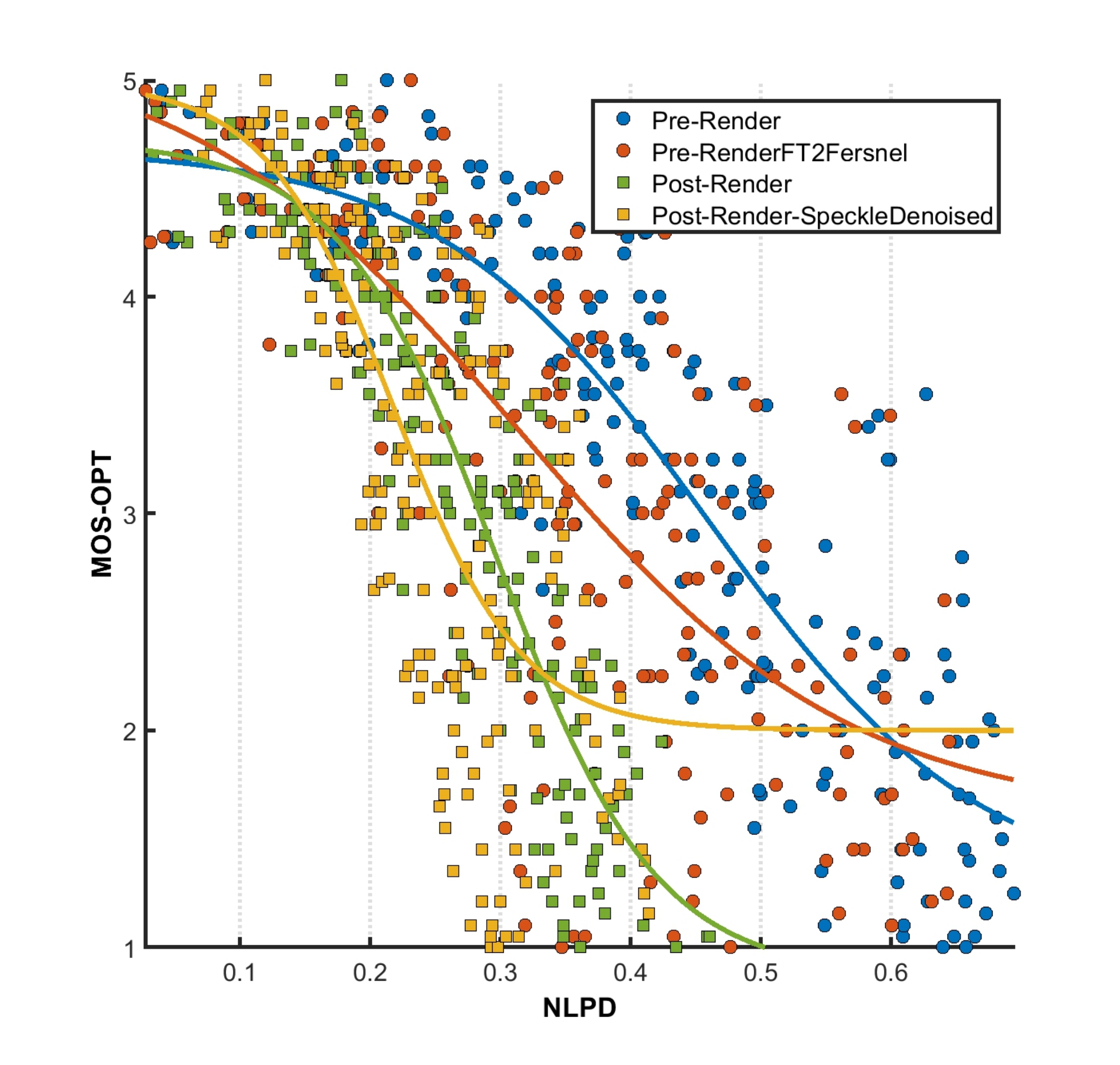}}
	    \subfloat[$NMSE$\label{fig:Overall4s_NMSE}]
	{\includegraphics[width=0.25\textwidth]{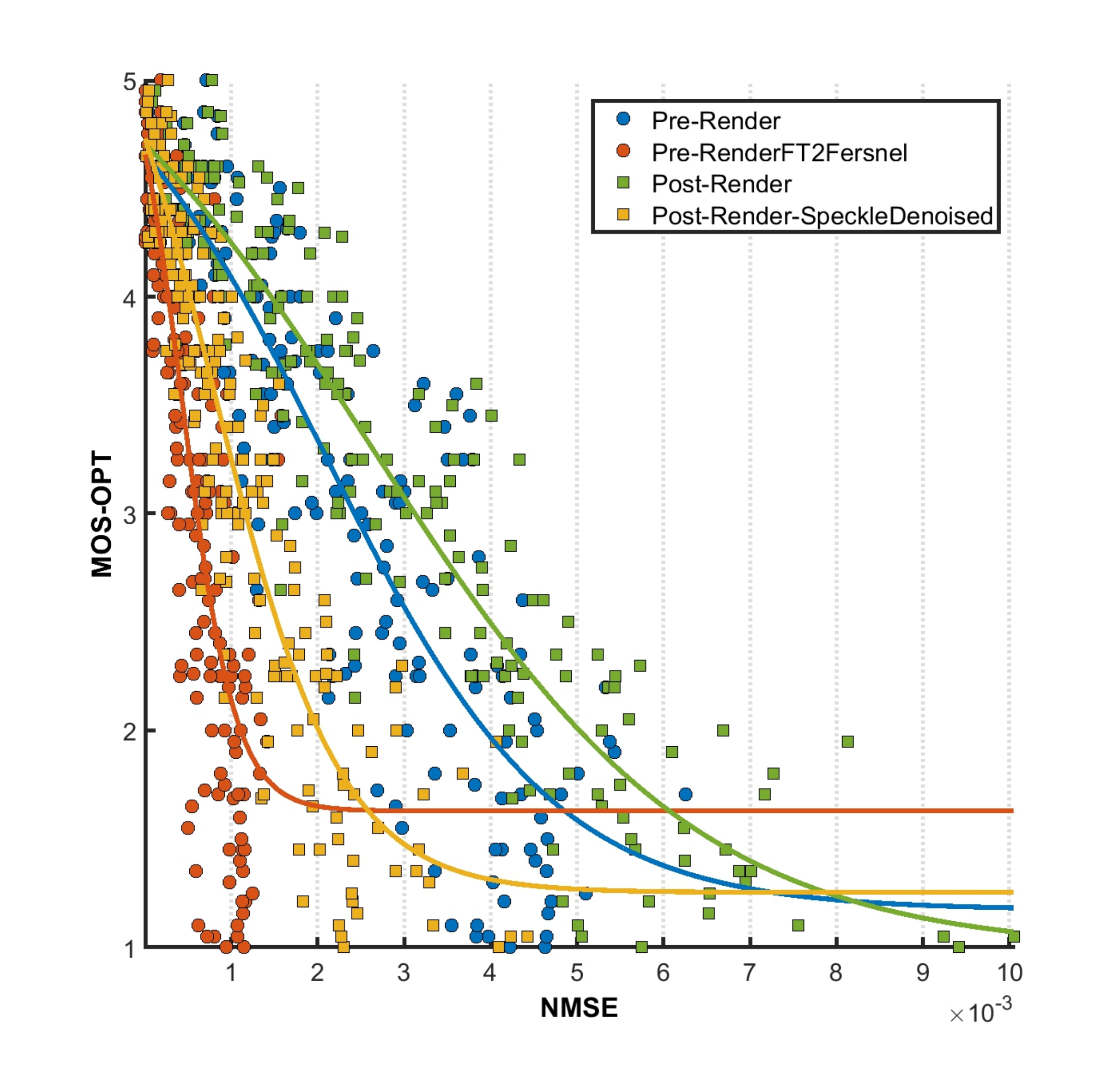}}
	    \subfloat[$PSNR$\label{fig:Overall4s_PSNR}]
	{\includegraphics[width=0.25\textwidth]{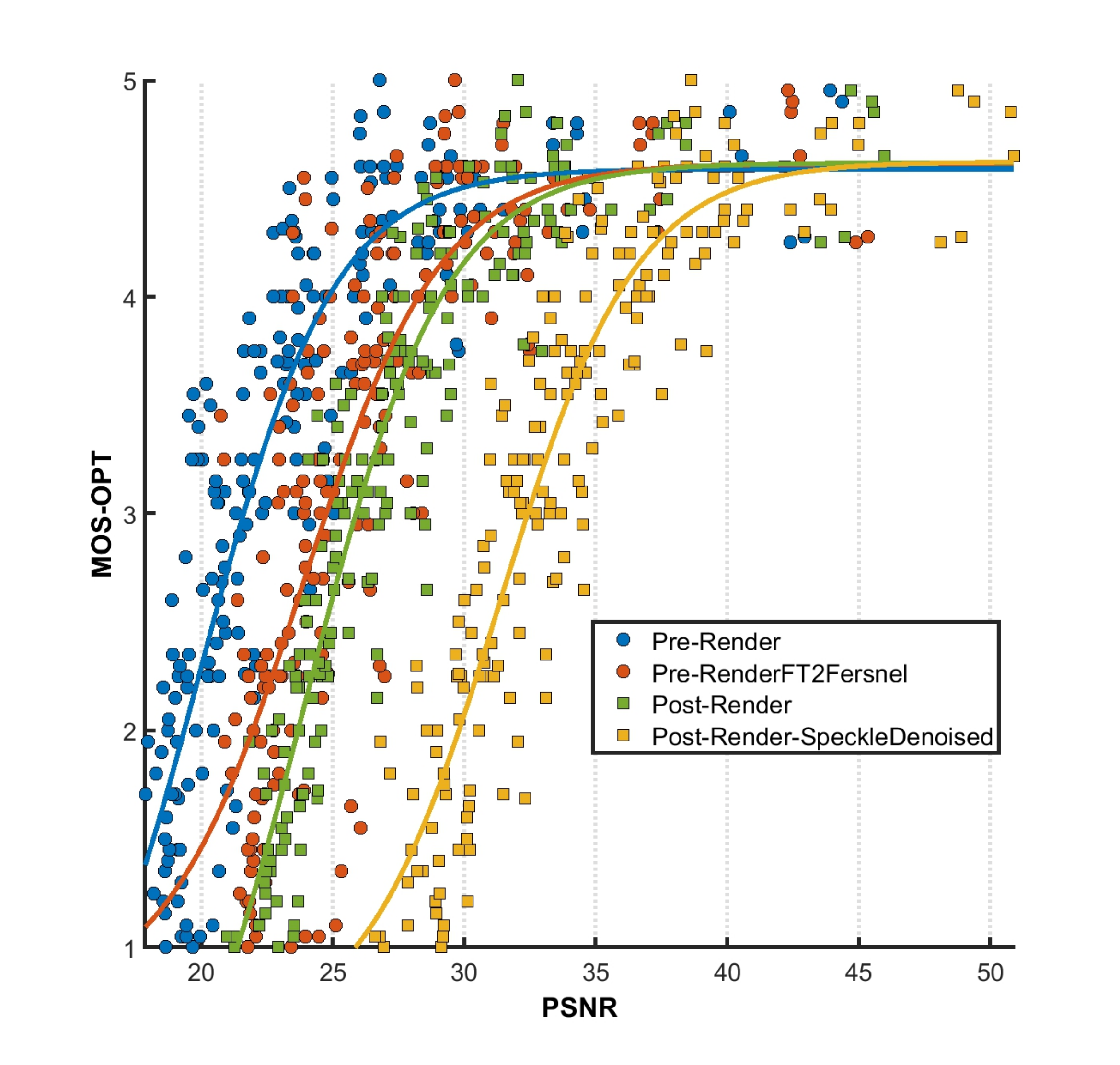}}
				
	\vspace*{-1em}
	    \subfloat[$SSIM$\label{fig:Overall4s_SSIM}]
	{\includegraphics[width=0.25\textwidth]{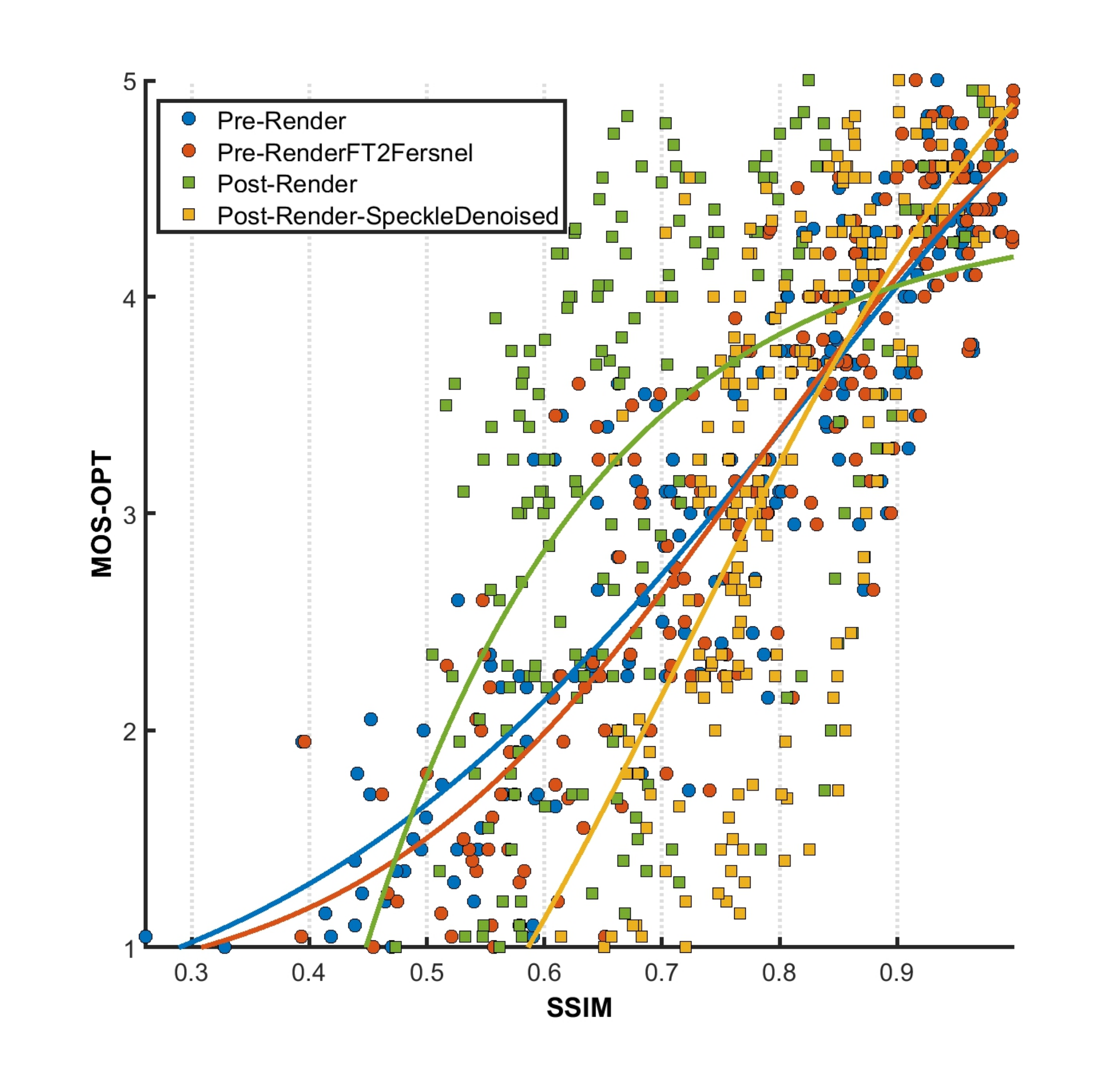}}
	    \subfloat[$SSRM$\label{fig:Overall4s_SSRM}]
	{\includegraphics[width=0.25\textwidth]{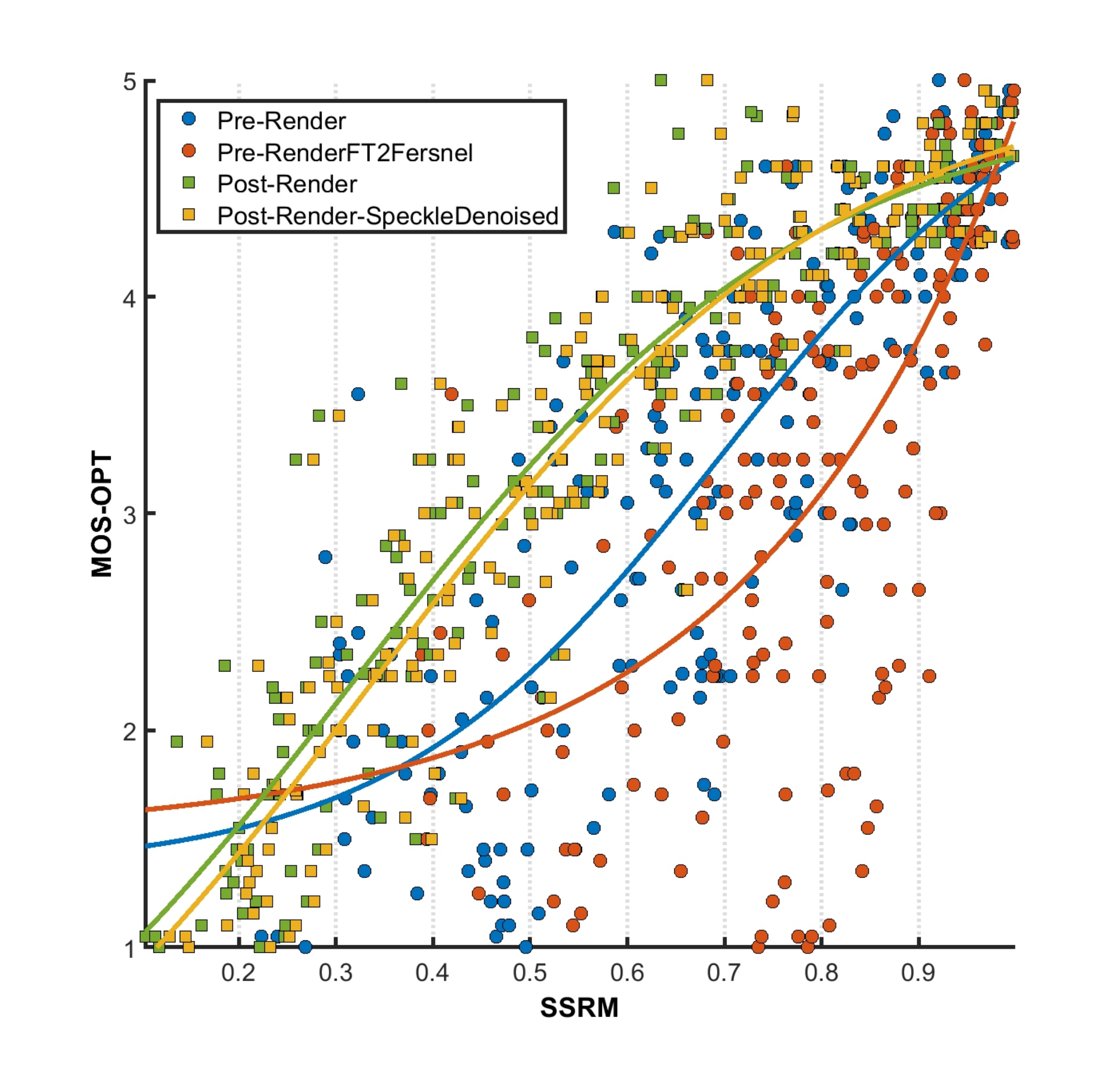}}
	    \subfloat[$SSRMt$\label{fig:Overall4s_SSRMt}]
	{\includegraphics[width=0.25\textwidth]{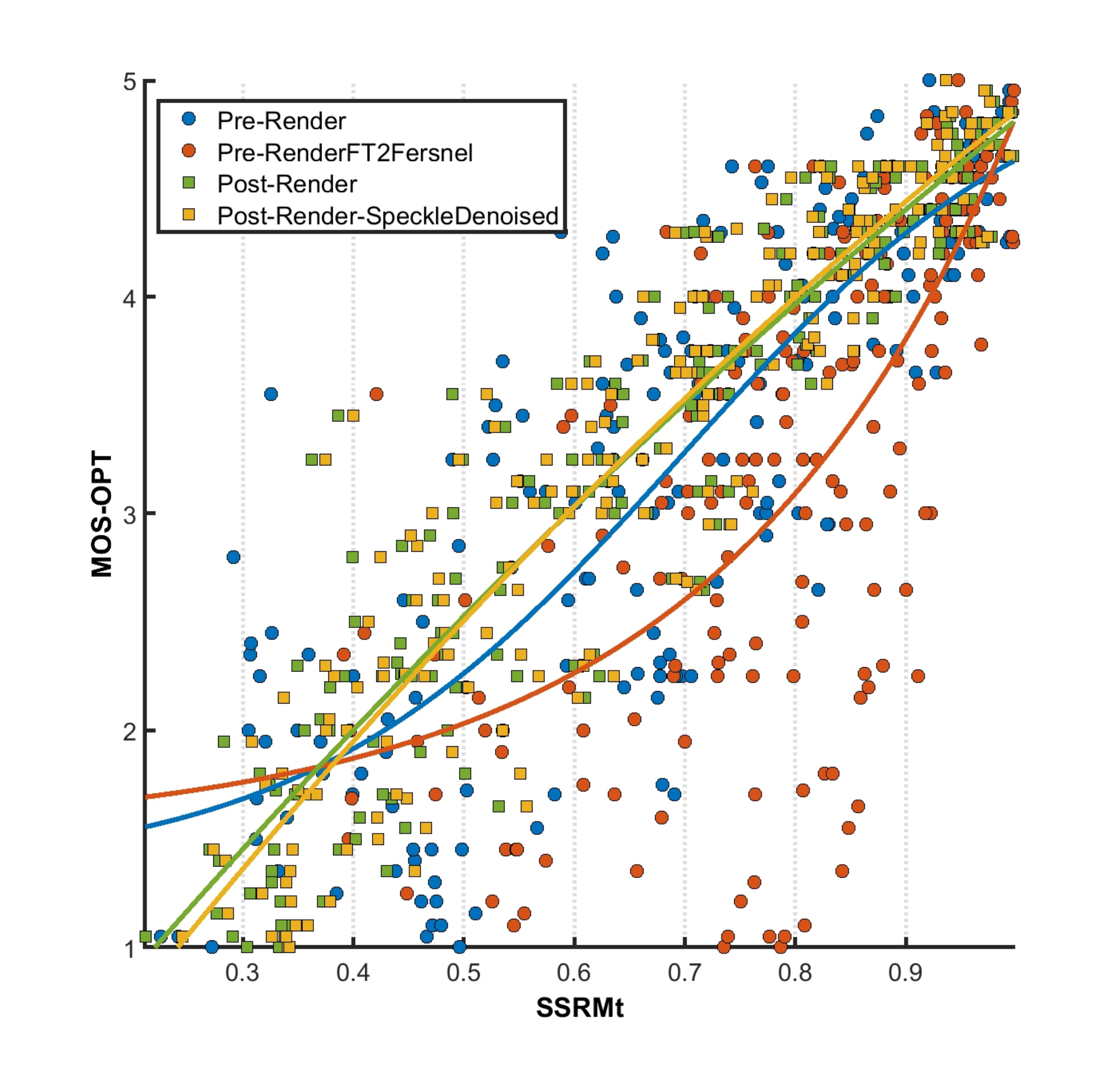}}
	    \subfloat[$UQI$\label{fig:Overall4s_UQI}]
	{\includegraphics[width=0.25\textwidth]{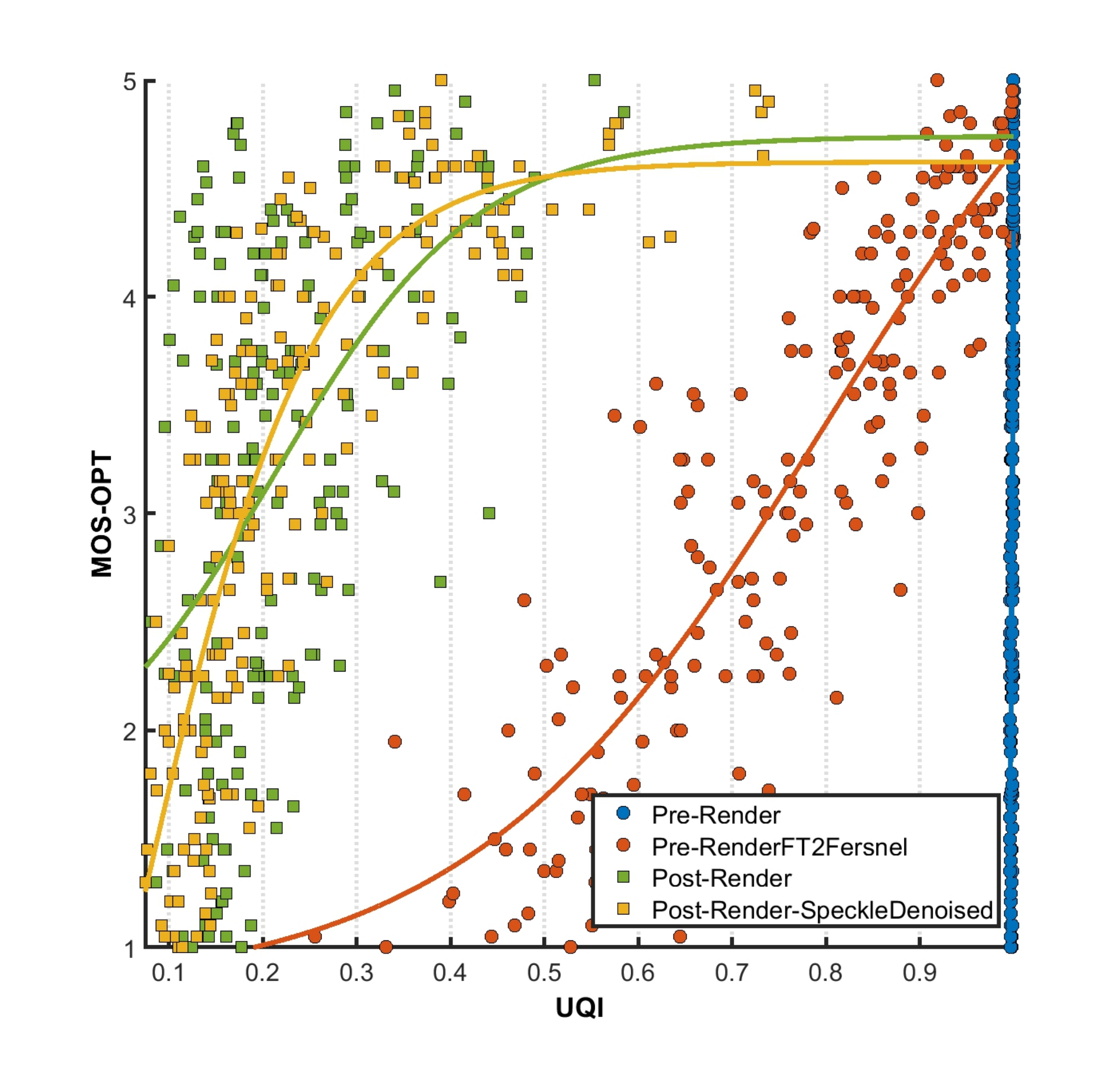}}
			
	\vspace*{-1em}
	    \subfloat[$VIFp$\label{fig:Overall4s_VIFp}]
	{\includegraphics[width=0.25\textwidth]{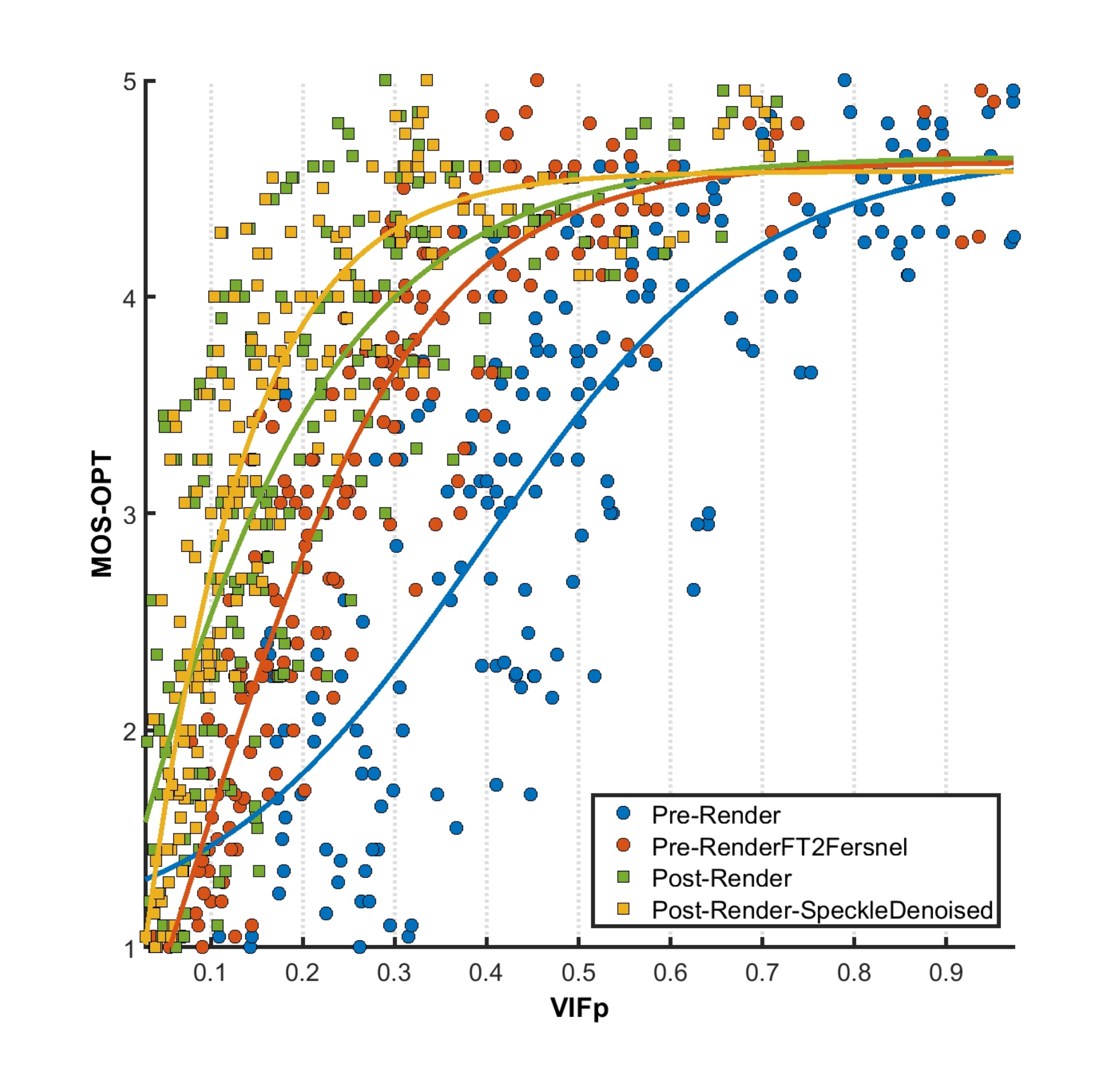}}
\caption{Overall scatter plots of the quality metric performances for each of the four experimental tracks based on the MOS obtained from the holographic display setup (MOS-OPT). The scores from the center view and the right corner view are combined and shown together per experimental track. Each data point for the two experimental tracks after reconstruction, represents the averaged MOS scores and predicted quality scores obtained from different focal distances per object. The results of the test tracks QA\_1 to QA\_4 are color coded with blue, orange, green, and yellow, respectively.}
\label{fig:Overall4s}%
\vspace*{-1em}
\end{figure*}

After reporting the results in each track, in this section a cross-track analysis of the IQM performances is provided. To do so, we present the overall scatter plots of quality predictions for each IQM in \figref{fig:Overall4s}. For each IQM, their quality predictions for QA\_1 to QA\_4 vs the MOS of the holographic setup are shown together. The scattered points from each test track are accompanied with a logistic fit curve calculated as explained in section \ref{sec:StatAnalysis}. The results from the QA\_1 to QA\_4 are colour coded with blue, orange, green, and yellow, respectively. 
Here, in the cases of MSE \protect\subref{fig:Overall4s_MSE}, NMSE \protect\subref{fig:Overall4s_NMSE}, PSNR \protect\subref{fig:Overall4s_PSNR}, and VIFp \protect\subref{fig:Overall4s_VIFp}, graphs reveal a distinctive shift when the measurements repeat in different steps of the holographic processing pipeline, \emph{i.e.} these IQMs show a similar behaviour overall but in a shifted score range for each test track. For the case of PSNR, MSE, and VIFp, these data-range shifts occur exactly with the same order of the test tracks progressions. While these plots show that measures like the MSE family in general exhibit a similar behaviour independent of the point in the processing pipeline, where they have been calculated, these range shifts imply that their values are strictly comparable only with the measurements performed in the same place. As an example, a PSNR of 30~dB does not necessarily correspond to the same visual quality when the measurements are done once on the hologram and once on the denoised reconstruction. For the IQMs which are designed to always provide a bounded measurement this is usually not an issue, though their overall behaviour compared to the MOS can drastically change depending on the QA step where they are used for quality prediction, e.g. in case of SSRM and SSIM. Overall, there seems to be a trade-off between generality and comparability. One should decide between having an unbounded measurement exclusively comparable with the measurements in the same test point but with stable behaviour across the processing pipeline, or having a bounded measurement but with varying behaviour dependent on where is used to predict the visual quality.

\begin{figure}
\centering
	    \subfloat[QA\_1: Fourier holograms prior to reconstruction\label{fig:OverallRanks1}]
	{\includegraphics[width=0.95\columnwidth]{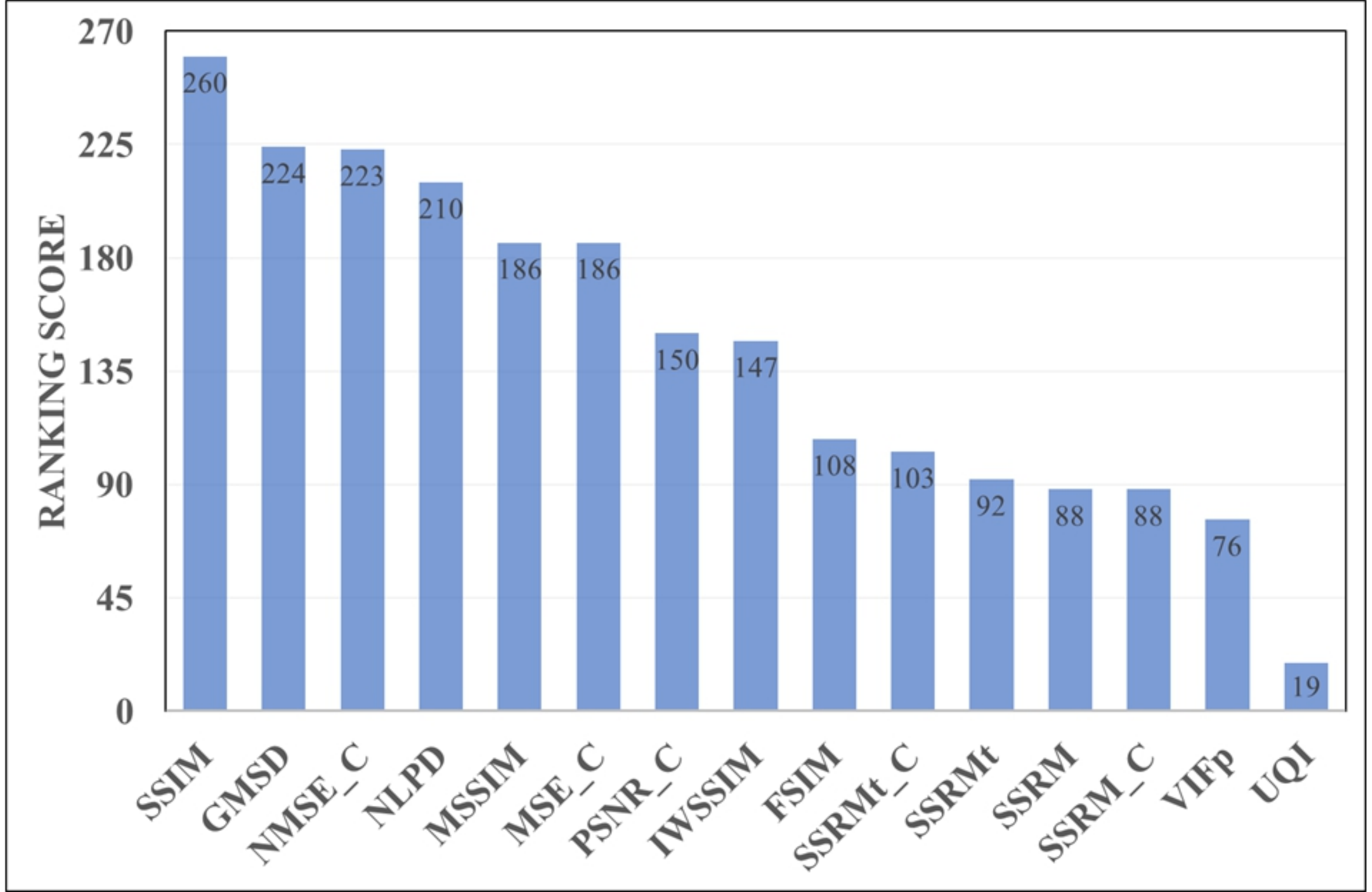}}

	\vspace*{-1em}
	    \subfloat[QA\_2: Fresnel holograms prior to reconstruction\label{fig:OverallRanks2}]
	{\includegraphics[width=0.95\columnwidth]{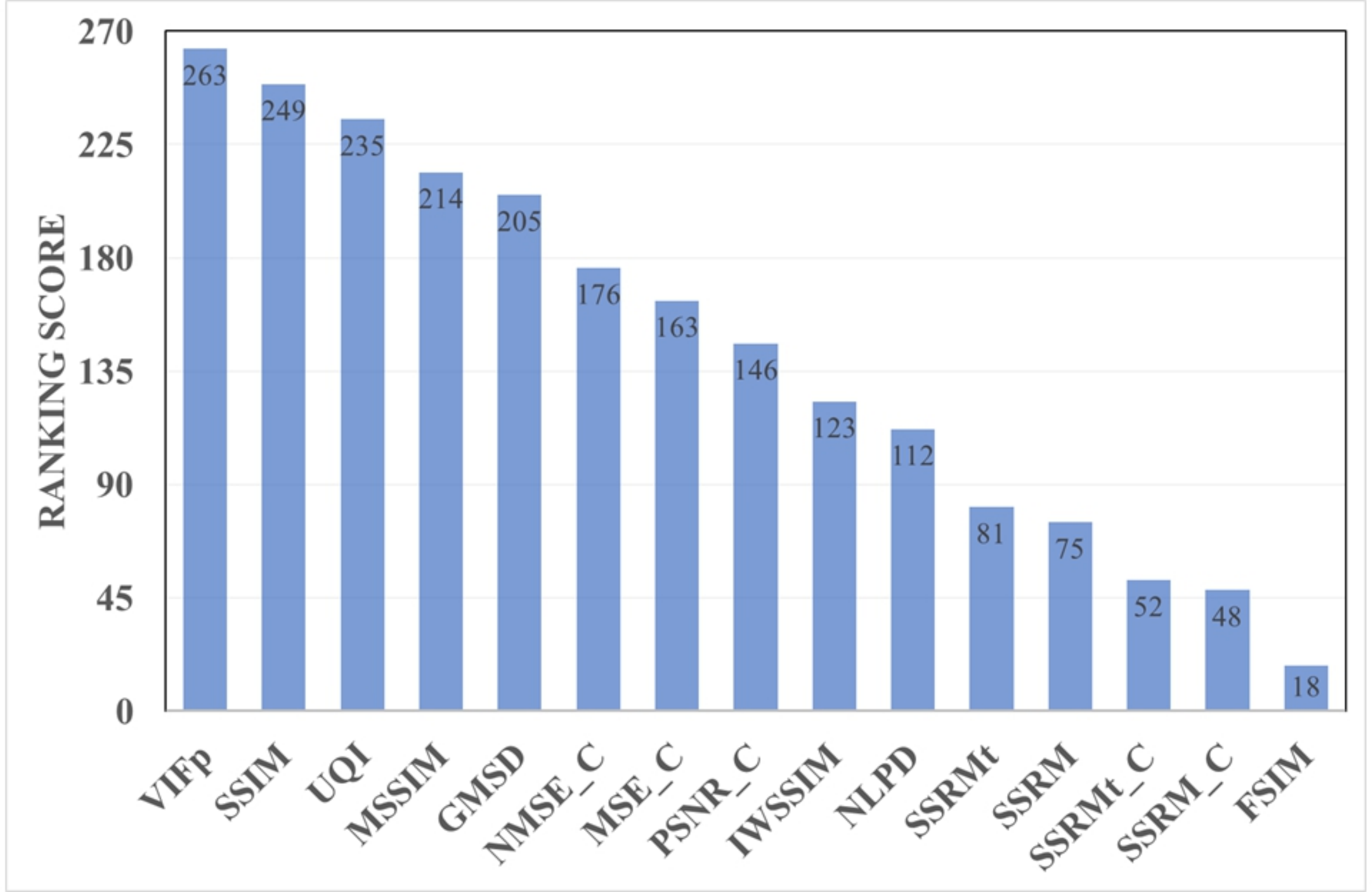}}
				
	\vspace*{-1em}
	    \subfloat[QA\_3: Fourier holograms after reconstruction\label{fig:OverallRanks3}]
	{\includegraphics[width=0.95\columnwidth]{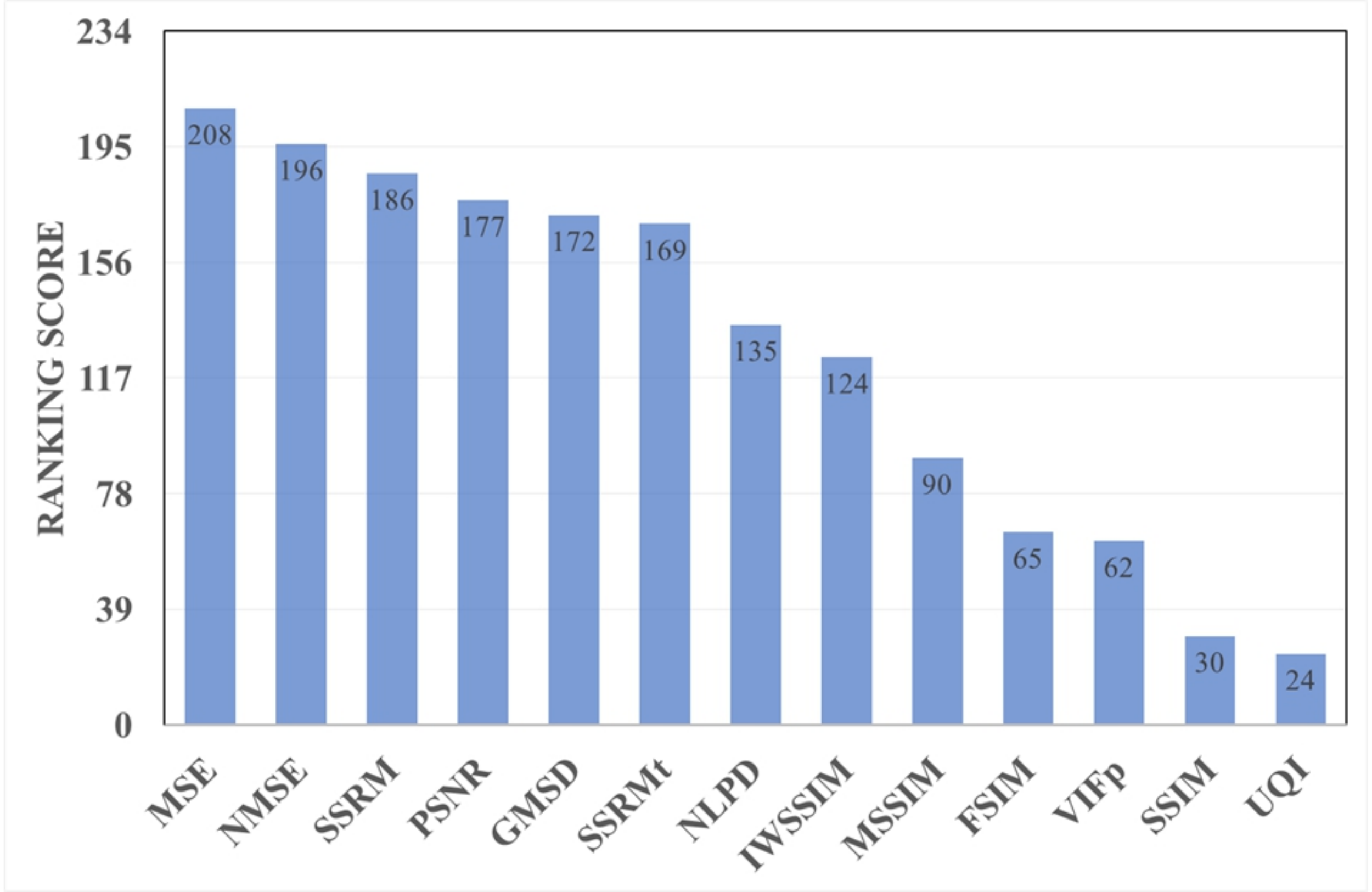}}
			
	\vspace*{-1em}
	    \subfloat[QA\_4: Fourier holograms after reconstruction and denoising\label{fig:OverallRanks4}]
	{\includegraphics[width=0.95\columnwidth]{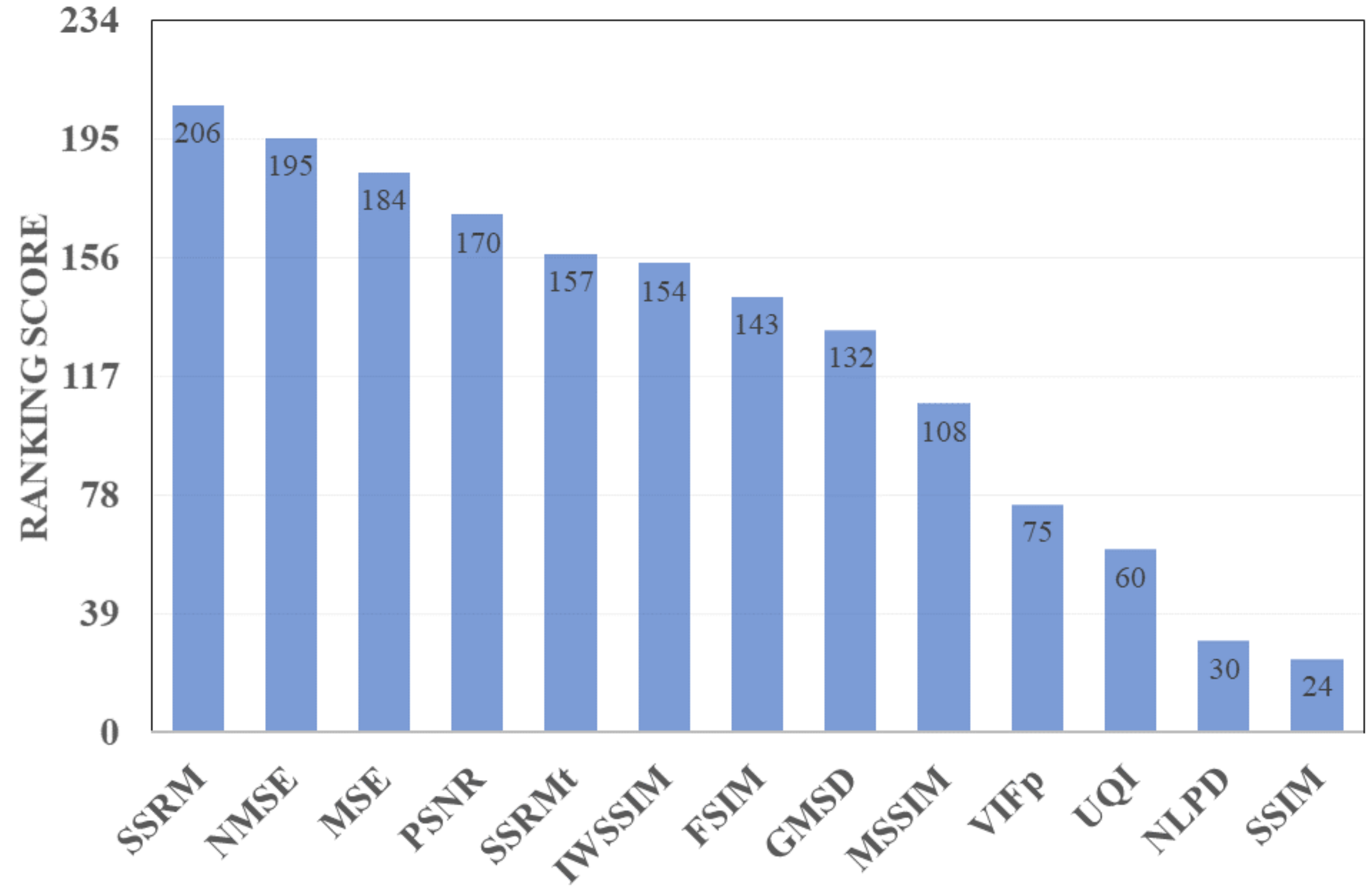}}
\caption{Total sum of the ranking scores of the IQMs ranked per row in the correlation tables \tabref{tab:CorrsPreReconFourier}-\ref{tab:CorrsPostReconNoSpeckle}. The overall performance IQMs in each test track is summarized based on all the 6 evaluation criteria and the 3 sets of MOS used for the benchmarking}
\label{fig:OverallRanks}%
\vspace*{-1em}
\end{figure}

Next, we want to have a simple quantitative evaluation of the performance of each IQM w.r.t. all of the benchmarked criteria and compared to all three sets of MOS. To do so, within each row of the \tabref{tab:CorrsPreReconFourier}-\ref{tab:CorrsPostReconNoSpeckle}, we ranked the IQMs (\emph{e.g} in \tabref{tab:CorrsPreReconFourier}, based on the SROCC criterion SSIM has the lowest rank so receives the highest score of 15 out of the 15 tested IQMs and on the other side UQI receives score of 1 for being the worst IQM w.r.t the SROCC). Thus for each table, every IQM receives 18 ranking scores for the 6 evaluation criteria (i.e. SROCC, KRCC, PCC\_NoFit, PCC\_Fitted, RMSE and Outlier Ratio) and all 3 sets of MOS. Then, we calculated a column-wise sum of these ranking scores per IQM. That way, we obtained a compact indicator of the IQM performance per test track for which the results are depicted in the bar charts of \figref{fig:OverallRanks}. We could continue summing the ranking scores across all 4 test tracks, though due to significant changes in the rankings of the IQMs for each test track compared to others, we prefer to provide the ranking scores for each test track separately. The bar charts of \figref{fig:OverallRanks} reveal the top performing IQMs in each track. Moreover, the ranking scores of the IQMs helps to see their relative performance compared to each other. For example, in \figref{fig:OverallRanks1}, the plateau of the ranking scores from FSIM till VIFp shows that these IQMs w.r.t all evaluation criteria are almost equally unreliable compared to the top three IQMs. 

\subsection{Final verdict}
\label{sec:verdict}
Our experiments for the first time provide a full view of the performance of the available quality measures for the case of digital holography w.r.t. MOS. We visualized the behaviour of each method when tested across the four test tracks (\figref{fig:Overall4s}) and quantified the performance of these quality measures based on several evaluation criteria (See \figref{fig:OverallRanks} ). Considering all aspects tested in this research, we would like to finalize the analysis of the tested IQMs by providing a usage guideline such that helps interested user to choose the most appropriate method for the holographic VQA tasks until more hologram-oriented quality metrics are designed. For each testing track, \tabref{tab:topIQMs} provides our recommendations where the recommendations are categorized in three groups colour coded with green: usage is advised, light-blue: Neutral or results were not conclusive to provide a strong recommendation and red: the application is discouraged.

According to our experimental results, MSE and NMSE which their measurements are purely based on signal fidelity and does not take into account any perceptual aspect, did not show any major drawback and in fact they are among the top three recommended methods especially after the reconstruction. To a lesser degree PSNR follows the same performance across all testing tracks. This means, its additional steps compared to the other two simpler methods (i.e the fractional measurement of error relative to the maximum possible signal value and the applied logarithmic scale) have a negative impact on the quality predictions w.r.t. the MOS. 

The SSIM and SSRM appear to demonstrate the most dramatic behaviour across all tracks. As mentioned earlier, while SSIM is found to be the top performing IQM prior to, and the worst after reconstruction, SSRM performs exactly the opposite. It performs poorly on holograms and jumps to the top of the list when tested on the reconstructed scenes. The SSRMt also demonstrates similar behaviour as the vanilla SSRM, which suggests a deeper connection between their identical operational core and their performance. Regarding the other members of the SSIM family (\emph{e.g.} MS-SSIM, IWSSIM) though, it appears that application of more complex perceptual models dropped their performance in the hologram domain. After the reconstruction, their multi-scale analysis based measurements could only create an improvement on their performance over the default SSIM but not adequate to make them competitive. The case of IWSSIM can also be considered in conjunction with the NLPD. Both of these methods benefit from the multi-resolution decomposition using Laplacian pyramid and both present a similar performance across our evaluation pipeline suggesting that such a decomposition may not necessarily be the most effective approach for VQA in digital holography. The other SSIM family member UQI, which does not benefit from the regularization in its operational core, is expected to perform worse than the more mathematically stable SSIM. Lack of regularization for this metric is known to cause instability especially in cases where the sum of squared means or sum of variances for the reference and distorted data gets close to zero. For this reason, we discourage its usage over the SSIM and in case of Fresnel holograms where it shows a competitive performance, we recommend utilizing it along with secondary quality prediction methods.     

Also, the image-gradient based methods -- which are normally able to make very accurate quality predictions in natural imagery -- do not show an outstanding performance (\emph{e.g}, FSIM, GMSD) suggesting that such approaches may not necessarily be effective for VQA in digital holography. However, the rather stronger performance of GMSD could be related to utilizing the standard deviation of the local gradient magnitude comparisons which enables a rather more stable prediction with more independence from the noisy nature of holograms and the speckle noise.

The VIFp which represents the information theoretic approach to the VQA and is been a long standing contender in the realm of IQA, also does not exhibit a reliable achievement across all of our experimental tracks. Only in one case when the Fresnel holograms are tested, outperforms others in most cases. This method exhibits a number of assumptions, which may or may not hold in case of holography causing large fluctuations in its performance. For example, it employs the natural image statistics by using the Gaussian Scale Mixture (GSM) to statistically model the wavelet coefficients after applying an steerable pyramid decomposition. This model is expected to substantially deviate from the statistics of the holographic signals. However, VIFp also models the distortion channel using an attenuating additive noise in wavelet domain which in this case potentially can relate well with the highly noisy nature of holograms. On top of them, in VIFp it is assumed that while passing through the HVS, the uncertainty level increases for perception of visual data. Thus the wavelet subbands for both the source and distortion channels are subject to an extra additive white Gaussian noise which models such increase in the uncertainty level. Now, wherever all these assumptions happen to fall in-line with the statistics of the tested holograms, a good prediction performance is foreseen. However, in most cases it is going to be the other way around. A re-adjustment on these assumptions based on the holographic statistical properties may potentially solve the unreliability issue for information theoretic based metrics like VIFp. Although, because of extremely noisy nature of holographic interference fringes and their strong dependence to the properties of the recorded scene, such statistical modelling may not be generalizable. We have recommended the usage of VIFp for the Fresnel holograms in \tabref{tab:topIQMs}. However, we strongly advise to use this metric along with other recommended IQMs due to its instability caused by the above explained reasons.

\begin{table}
\centering
\caption{Usage recommendation table of tested IQMs colour coded based on their overall performance in each test track. Green: advised to be utilized, Light blue: Neutral or results not conclusive within the bounds of this research and Red: usage is discouraged.}
\resizebox{\columnwidth}{!}{
\begin{tabular}{|p{1.75cm}|p{2cm}|p{2cm}|p{2cm}|p{2cm}|}

\hline
\textbf{IQM} & Fourier hologram (QA\_1) & Fresnel hologram (QA\_2) & Reconstructed hologram (QA\_3) & Reconstructed hologram + speckle denoising (QA\_4) \\ \hline
\textbf{MSE} & \cellcolor[HTML]{CBCEFB} & \cellcolor[HTML]{CBCEFB} & \cellcolor[HTML]{009901} & \cellcolor[HTML]{009901} \\ \hline
\textbf{NMSE} & \cellcolor[HTML]{009901} & \cellcolor[HTML]{CBCEFB} & \cellcolor[HTML]{009901} & \cellcolor[HTML]{009901} \\ \hline
\textbf{PSNR} & \cellcolor[HTML]{CBCEFB} & \cellcolor[HTML]{CBCEFB} & \cellcolor[HTML]{CBCEFB} & \cellcolor[HTML]{CBCEFB} \\ \hline
\textbf{SSRM} & \cellcolor[HTML]{E00000} & \cellcolor[HTML]{E00000} & \cellcolor[HTML]{009901} & \cellcolor[HTML]{009901} \\ \hline
\textbf{SSRMt} & \cellcolor[HTML]{E00000} & \cellcolor[HTML]{E00000} & \cellcolor[HTML]{CBCEFB} & \cellcolor[HTML]{CBCEFB} \\ \hline
\textbf{SSIM} & \cellcolor[HTML]{009901} & \cellcolor[HTML]{009901} & \cellcolor[HTML]{E00000} & \cellcolor[HTML]{E00000} \\ \hline
\textbf{IWSSIM} & \cellcolor[HTML]{CBCEFB} & \cellcolor[HTML]{CBCEFB} & \cellcolor[HTML]{CBCEFB} & \cellcolor[HTML]{CBCEFB} \\ \hline
\textbf{MS-SSIM} & \cellcolor[HTML]{CBCEFB} & \cellcolor[HTML]{CBCEFB} & \cellcolor[HTML]{E00000} & \cellcolor[HTML]{E00000} \\ \hline
\textbf{UQI} & \cellcolor[HTML]{E00000} & \cellcolor[HTML]{CBCEFB} & \cellcolor[HTML]{E00000} & \cellcolor[HTML]{E00000} \\ \hline
\textbf{GMSD} & \cellcolor[HTML]{009901} & \cellcolor[HTML]{CBCEFB} & \cellcolor[HTML]{CBCEFB} & \cellcolor[HTML]{CBCEFB} \\ \hline
\textbf{FSIM} & \cellcolor[HTML]{E00000} & \cellcolor[HTML]{E00000} & \cellcolor[HTML]{E00000} & \cellcolor[HTML]{CBCEFB} \\ \hline
\textbf{NLPD} & \cellcolor[HTML]{CBCEFB} & \cellcolor[HTML]{CBCEFB} & \cellcolor[HTML]{CBCEFB} & \cellcolor[HTML]{E00000} \\ \hline
\textbf{VIFp} & \cellcolor[HTML]{E00000} & \cellcolor[HTML]{009901} & \cellcolor[HTML]{E00000} & \cellcolor[HTML]{E00000} \\ \hline
\end{tabular}}
\label{tab:topIQMs}
\vspace*{-1em}
\end{table}

It should be noted that, these recommendations hold only within the bounds of current experiment (\emph{i.e.} as long as the degradations are limited to compression artifacts, and if the same procedure as this research is utilized for processing, rendering and quality assessment of the holograms). Moreover, being the best among themselves does not necessarily translate these IQMs into being reliably accurate in predicting the visual quality of holograms. This gets more clear, when the achieved correlations w.r.t. MOS in this research is compared to the ones of the same IQMs tested on digital images. Especially, the achieved correlations before the reconstruction (reported in \tabref{tab:CorrsPreReconFourier} and \tabref{tab:CorrsPreReconFresnel}), still have lots of room for improvement. Not to mention, our results showed that careful considerations should be taken into account for choosing the right quality prediction method among the available options based on the application and the characteristics of the holograms at the point where the measurement occurs. Therefore, in spite of the provided recommendations, we emphasize that none of the available options are at the level that confidently alleviate the strong need for design and development of especial quality assessment algorithms for the holographic data. 

\section{Conclusions}
\label{sec:conclusion}

In this research for the first time we utilized 96 holograms of HoloDB to make a systematic performance evaluation of the most viable options for the quality assessment of digital holograms. The results for 4 separate sets of evaluations are provided. We compared the reference holograms with the distorted versions: before rendering, on the real and imaginary parts of the complex-wavefield; after converting the Fourier holograms to regular Fresnel holograms; after rendering, on the quantized amplitude of the reconstructed data, and after speckle denoising. For every experimental track, the quality metric predictions are rigorously compared to three sets of MOS which were previously obtained via subjective experiments. Statistical analysis of IQM performances and a discussion on the behaviour of outstanding methods are presented. Finally their overall performance based on all of the utilized evaluation criteria and all three sets of MOS are summarized per test track to introduce the best performing quality metrics for each testing track. All aspects considered, turns out while for each test track, a couple of quality metrics present a significantly correlated performance compared to the multiple sets of available MOS, none of them show a consistently high-performance across all the four test-tracks. This emphasizes their sensitivity to the characteristics of the input and the position in the processing pipeline where the quality predictions are obtained. It also reveals once more the dire need to design efficient quality metrics for holographic content. We are hoping that this research provide a thorough understanding of the complexities involved in VQA of digital holograms and consequently act as the first systematic step toward designing advanced perceptual quality prediction methods for digital holograms.

\section*{Funding}
The research leading to these results has received funding from the European Research Council under the European Union's Seventh Framework Programme (FP7/2007-2013)/ERC Grant Agreement n.617779 (INTERFERE) and also the Cross-Ministry Giga KOREA Project (GigaKOREA GK20D0100) and the support from the WUT.
\section*{Acknowledgment}


\bibliography{refs}
\end{document}